\documentclass[twoside, twocolumn]{IEEEtran}
\ifCLASSINFOpdf
\else
\fi

\usepackage{epsfig,tikz,float}
\usepackage{cite}
\usepackage{amsmath}
\usepackage{amssymb}
\usepackage{amsfonts}
\usepackage{algorithmicx}
\usepackage{tikz}
\usetikzlibrary[arrows,graphs,quotes,shapes,chains,trees,positioning,graphs.standard]
\usepackage{algpseudocode}
\usepackage{graphicx}
\usepackage{textcomp}
\usepackage{xcolor}
\usepackage[ruled,vlined]{algorithm2e}
\usepackage{bm}
\usepackage{bbm}
\usepackage{dsfont}
\usepackage{bbold}
\usepackage{enumitem}
\usepackage{stmaryrd}
\usepackage[utf8]{inputenc} 
\usepackage[T1]{fontenc}
\usepackage{amsthm}

\usepackage{url}
\usepackage{ifthen}

\newtheorem{theorem}{Theorem}

\newtheorem{lemma}{Lemma}

\newtheorem{claim}{Claim}

\newcommand{\F}{\mathcal{F}}
\newcommand{\E}{\mathsf{E}}

\newcommand{\Bein}{\mathcal{B}^{(i,n)}_{\epsilon}}
\newcommand{\Yein}{\mathcal{Y}^{(i,n)}_{\epsilon}}
\newcommand{\Yei}{\mathcal{Y}^{(i)}_{\epsilon}}

\def\BibTeX{{\rm B\kern-.05em{\sc i\kern-.025em b}\kern-.08em
    T\kern-.1667em\lower.7ex\hbox{E}\kern-.125emX}}

\begin{document}
%
\title{Achievable Rates and Low-Complexity Encoding of Posterior Matching  for the BSC}
%
%
%

\author{Amaael Antonini,~\IEEEmembership{Student Member,~IEEE,}
        Rita Gimelshein,~\IEEEmembership{Student Member,~IEEE,}\\
        and~Richard Wesel,~\IEEEmembership{Fellow,~IEEE}
\thanks{This work was supported by the National Science Foundation (NSF) under Grant CCF-1955660. An earlier version of this paper was presented in part at the 2020 IEEE International Symposium on Information Theory (ISIT) \cite{9174232} [DOI: 10.1109/ISIT44484.2020.9174232 ]. (Corresponding author: Amaael Antonini.)

A.~Antonini is with the Department of Electrical and Computer Engineering, University of California, Los Angeles, Los Angeles, CA, 90095 USA (e-mail: amaael@ucla.edu).

R.~Gimelshein is with the Department of Electrical and Computer Engineering, University of California, Los Angeles, Los Angeles, CA, 90095 USA (e-mail: rgimel@ucla.edu).

Minghao Pan is with the California Institute of Technology.

R.~D.~Wesel is with the Department of Electrical and Computer Engineering, University of California, Los Angeles, Los Angeles, CA, 90095 USA (e-mail: wesel@ucla.edu).
    }
}

\maketitle

\begin{abstract}
Horstein, Burnashev, Shayevitz and Feder, Naghshvar \emph{et al.} and others have studied sequential transmission of a $k$-bit message over the binary symmetric channel (BSC) with full, noiseless feedback using posterior matching. 
Yang \emph{et al.} provide an improved lower bound on the achievable rate using martingale analysis that relies on the small-enough difference (SED) partitioning introduced by Naghshvar \emph{et al}.  SED requires a relatively complex encoder and decoder.  To reduce complexity, this paper replaces SED with relaxed constraints that admit the small enough {\em absolute} difference (SEAD) partitioning rule. The main analytical results show that achievable-rate bounds higher than those found by Yang \emph{et al.} \cite{Yang2021} are possible even under the new constraints, which are less restrictive than SED. The new analysis does not use martingale theory for the confirmation phase and applies a surrogate channel technique to tighten the results.  An initial systematic transmission further increases the achievable rate bound.  The simplified encoder associated with SEAD has a complexity below order $O(K^2)$ and allows simulations for message sizes of at least 1000 bits.  For example, simulations achieve $99$\% of of the channel's $0.50$-bit capacity with an average block size of 200 bits for a target codeword error rate of $10^{-3}$.

\end{abstract}


\IEEEpeerreviewmaketitle

\section{Introduction}

\label{sec: introduction}
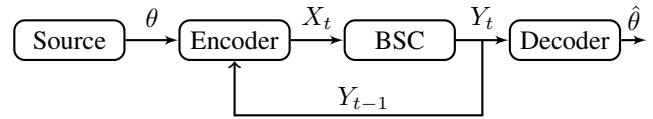
\begin{figure}
    \centering
\begin{tikzpicture}
[auto,
		point/.style={circle,inner sep=0pt, minimum size=0.1pt,fill=white},
		skip loop/.style={to path={-- ++(0,#1) -| (\tikztotarget)}},
		decision/.style={circle,draw=black,thick,fill=white,
								text width=0.7em,align=flush center,
								inner sep=0.01pt},
		block/.style={rectangle,draw=black,thick,fill=white,
							text width=3.5em,align=center,rounded corners,
							minimum height=1.5em},
		line/.style={draw,thick,-latex',shorten >=0.1pt},
		cloud/.style={circle,draw=red,thick,ellipse,fill=red!20,
							minimum height=1.5em}
		scale=0.5]
\matrix [column sep=3.5mm, row sep=6mm]
{
	\node 	[point]		(NULL0) 	{};
	&\node 	[block]		(D0)		{Source};
	&\node 	[point]	 	(NULL1)		{};
	&\node 	[block]		(D1)		{Encoder};
	&\node 	[point]	 	(NULL2)		{};
	&\node 	[block] 	(D2) 		{BSC};
	& \node [point]	 	(NULL3)		{};
	& \node [block] 	(D3)		{Decoder};
	& \node [point]	 	(NULL4)		{}; 
	\\
};
\begin{scope}[every path/.style=line]
\path (D0)--node[above of=NULL1,yshift=-2em]{$\theta$} (D1);
\path (D0);
\path (D1)--node[above of=D1,yshift=-2em]{$X_t$}(D2);
\path (D1);
\path (D2)--node[above of=D2,yshift=-2em]{$Y_t$} node[below of=D2, xshift=-4.5em, yshift=0.6em]{$Y_{t-1}$} (D3);
\path (D3)--node[above of=D3,yshift=-2em]{$\hat{\theta}$}  (NULL4);
\path (NULL3) edge [->,skip loop=-10mm] (D1);
\end{scope}
\end{tikzpicture}
\vspace{-1.0em}
\caption{System diagram of a BSC with full, noiseless feedback.}
\vspace{-1.5em}
\label{fig: system model}
\end{figure}

Consider sequential-transmission over the binary symmetric channel with full, noiseless feedback as depicted in Fig. \ref{fig: system model}.
The source data at the transmitter is a $K$-bit message $\theta$, uniformly sampled from $\{0,1\}^K \triangleq \Omega$.
At each time $t = 1, 2, \dots \tau$, input symbol $X_t$ is transmitted across the channel, and output symbol $Y_t$ is received, where $X_t, Y_t \in \{0,1\}$ and $\Pr(Y_t = 1 \mid X_t = 0) = \Pr(Y_t = 0 \mid X_t = 1) = p \ \forall t$. 
The received symbol $Y_t$ is available to the transmitter for encoding symbol $X_{t+1}$ (and subsequent symbols) via the noiseless feedback channel. 

\thispagestyle{empty}

The process terminates at stopping time $t=\tau$ when a reliability threshold is achieved, at which point the receiver computes an estimate $\hat{\theta} \in \Omega$ of $\theta$ from the received symbols $Y_1,Y_2,\dots,Y_\tau$. The communication problem consists of obtaining a decoding estimate of $\theta$ at the smallest possible time index $\tau$ while keeping the error probability $\Pr(\hat{\theta} \neq \theta)$ bounded by a small threshold $\epsilon$.

\subsection {Background} 
Shannon \cite{Shannon1956} showed that feedback cannot increase the capacity of discrete memoryless channels (DMC). However, when combined with variable-length coding, Burnashev \cite{Burnashev1976} showed that feedback can help increase the frame error rate's (FER) decay rate as a function of blocklength. One such variable length coding method was pioneered by Horstein \cite{Horstein1963}. Horstein's sequential transmission scheme  was presumed to achieve the capacity of the BSC, which was later proved by Shayevitz and Feder \cite{Shayevitz2011}
showing that it satisfies the criteria of a \emph{posterior matching scheme}.
A \emph{posterior matching (PM) scheme} was defined by Shayevitz and Feder as one that satisfies the two requirements of the \emph{posterior matching principle}:
\begin{enumerate} 
    \item The input symbol at time $t+1$, $X_{t+1}$, is a fixed function of a random variable $U$, that is independent of the received symbol history $Y^t \triangleq \{Y_1,Y_2,\dots,Y_t\} $; and 
    \item The transmitted message, $\theta$, can be uniquely recovered from $(U, Y^t)$ a.s. 
\end{enumerate}
Gorantla and Coleman \cite{Gorantla2010} used Lyapunov functions for an alternative proof that PM schemes achieve the channel capacity.
Later, Li and El-Gamal \cite{Li2015} proposed a capacity achieving ``posterior matching'' scheme with fixed block-length for DMC channels. Their scheme used a random cyclic shift that was later used by Shayevitz and Feder for a simpler proof that Horstein's scheme achieves capacity \cite{Shayevitz2016}.
Naghshvar \emph{et. al.}  \cite{Naghshvar2015} proposed a variable length, single phase ``posterior matching'' scheme for discrete DMC channels with feedback that exhibits Burnashev's optimal error exponent, and used a sub-martingale analysis to prove that it achieves the channel capacity. Bae and Anastasopoulos \cite{Bae2010} proposed a PM scheme that achieves the capacity of finite state channels with feedback.
Since then, other ``posterior matching'' algorithms have been developed, see \cite{Kostina2017, Kim2013, Sabag2018, Truong2014, Anastasopoulos2012}. Other variable length schemes that attain Burnashev's optimal error exponent have also been developed, and some can be found in \cite{Schalkwijk1971, Schalkwijk1973,Tchamkerten2002,Tchamkerten2006,Naghshvar2012}. 

Feedback communication over the BSC in particular has been the subject of extensive investigation. Capacity-approaching, fixed-length schemes have been developed such as \cite{Li2015}, but these schemes only achieve low frame error rates (FERs) at block sizes larger than 1000 bits. For shorter block lengths, capacity-approaching, variable-length schemes have  also been developed, e.g., \cite{Horstein1963}, \cite{Burnashev1976}, \cite{Naghshvar2015}. 
Recently, Yang \emph{et al.} \cite{Yang2021} provided the best currently available achievability bound for these variable-length schemes.  Yang \emph{et al.} derive an achievable rate using encoders that satisfy the small-enough-difference (SED) constraint.  However, the complexity of variable-length schemes satisfying that constraint can grow quickly with message size, becoming too complex for practical implementation even at block lengths significantly below those addressed by the fixed-length schemes such as in \cite{Li2015}.

\subsection{Contributions}
In our precursor conference paper \cite{9174232}, we simplified the implementation of an encoder that enforces the SED constraint both by initially sending systematic bits and by grouping the messages according to Hamming distance from the received systematic bits. The contributions of the current paper include the following:

\begin{itemize}
\item This paper provides a new analysis framework for posterior matching on the BSC that avoids martingale analysis in the communication phase in order to show that the achievable rate of \cite{Yang2021} can be achieved with a broader set of encoders that satisfy less restrictive criteria than the SED constraint. Thm. \ref{theorem: simple rule} provides an example of a constraint, the small-enough-{\em absolute}-difference (SEAD) constraint, that meets the new, relaxed criteria.

\item The relaxed criteria allow a significant reduction of encoder complexity.  Specifically, this paper shows that applying a new partitioning algorithm, thresholding of ordered posteriors (TOP), induces a partitioning that meets the SEAD constraints.  The TOP algorithm facilitates further complexity reduction by avoiding explicit computation of posterior updates for the majority of messages, since those posterior updates are not required to compute the threshold position. This low-complexity encoding algorithm achieves that same rate performance that has been previously established for SED encoders in, e.g., \cite{Yang2021}. 

\item Our new analysis further tightens the achievable rate bound provided in \cite{Yang2021}. This new achievable rate lower bound applies to both the SED encoder analyzed in \cite{Yang2021} and to our new, simpler, encoder. 

\item We also show that using systematic transmissions as in \cite{9174232} to initially send the message meets both the relaxed criteria including SEAD as well as the SED constraint. Complexity is reduced during the systematic transmission, with the required operations limited to simply storing the received sequence. 

\item  We generalize the concept of the ``surrogate process'' $U'_i(t)$,  used in Sec V-E of \cite{Yang2021}, to a broader class of processes that are not necessarily sub-martingales. The ability to construct such ``surrogate'' processes allows tighter bounds that also apply to the original process.

\item Taken together, these results demonstrate that variable-length coding with full noiseless feedback can closely approach capacity with modest complexity. 

\item Regarding complexity, the simplified encoder associated with SEAD has a complexity below order $O(K^2)$ and allows simulations for message sizes of at least $1000$ bits.  The simplified encoder organizes messages according to their type, i.e. their Hamming distance from the received word, orders messages according to their posterior, and partitions the messages with a simple threshold without requiring any swaps.

\item Regarding proximity to capacity, our achievable rate bounds show that with codeword error rate of $10^{-3}$ SEAD posterior matching can achieve $96$\% of  the channel's $0.50$-bit capacity for an average blocklength of $199.08$ bits corresponding to a message with $k=47$ bits.  Simulations with our simplified encoder achieve $99$\% of of the channel's $0.50$-bit capacity for a target codeword error rate of $10^{-3}$ with an average block size of $201.08$ bits corresponding to a message with $k=49$ bits.

\end{itemize}



\subsection{Organization}
\label{sec: organization}

The rest of the paper proceeds as follows. Sec. \ref{sec: Achievable Rate} describes the communication process, introduces the problem statement, and reviews the highest existing achievability bound, by Yang \emph{et al.} \cite{Yang2021}, as well as the scheme that achieves it, by Naghshvar \emph{et. al.}  \cite{Naghshvar2015}. 
Sec. \ref{sec: Main Section} introduces Thms. \ref{theorem: Main Theorem}, \ref{theorem: surrogate martingale} and \ref{theorem: simple rule} that together relax the sufficient constraints to guarantee a rate above Yang's lower bound and further tightens Yang's bound.
Sec. \ref{sec: helping lemmas and proof of main theorem} introduces Lemmas \ref{lemma: Phase I time}-\ref{lemma: final and initial values} and provides the proof of Thm. \ref{theorem: Main Theorem} via Lemmas \ref{lemma: Phase I time}-\ref{lemma: final and initial values}. 
Sec. \ref{sec: proof of lemmas and theorems 2 and 3} provides the proofs of Lemmas \ref{lemma: Phase I time}-\ref{lemma: final and initial values}, and the proof of Thm. \ref{theorem: surrogate martingale} and Thm. \ref{theorem: simple rule}.
Sec. \ref{sec: arbitrary distribution} generalizes the new rate lower bound to arbitrary input distributions and  derives an improved lower bound for the special case where a uniform input distribution is transformed into a binomial distribution through a systematic transmission phase. 
Sec. \ref{sec: algorithm and implementation} describes the TOP partitioning method and implements a simplified encoder that organizes messages according to
their type, applies TOP, and  employs initial systematic transmissions.
Sec. \ref{sec: simulation results} compares performance from simulations using the simplified encoder to the new achievability bounds. Sec. \ref{sec: conclusion} provides  our conclusions.  The  Appendix\ref{appendix: appendices} provides detailed proof of the second part of Thm. \ref{theorem: simple rule} and the proof of claim \ref{claim: j confirmation singleton}.


\section{Posterior Matching with SED Partitioning}
\label{sec: Achievable Rate}

\subsection{Communication Scheme}\label{sec: original scheme}
Our proposed communication scheme and simplified encoding algorithm are based on the single phase transmission scheme proposed by Naghshvar \emph{et. al.} \cite{Naghshvar2015}. Before each transmission, both the transmitter and the receiver partition the message set $\Omega = \{0,1\}^K$ into two sets, $S_0$ and $S_1$.  The partition is based on the received symbols $Y^t$ according to a specified deterministic algorithm known to both the transmitter and receiver. Then, the transmitter encodes $X_t = 0$ if $\theta \in S_0$ and $X_t = 1$ if $\theta \in S_1$, i.e.
\begin{equation}
    X_t=\text{enc}(\theta,Y^t) 
    = \mathbbm{1}_{i \in S_1}
\end{equation}
After receiving symbol $Y_t$, the receiver computes the posterior probabilities:
\begin{equation}
    \rho_i(y^t)\triangleq P(\theta = i \mid Y^t = y^t), \: \forall i \in \{0,1\}^K \label{eq: postierior defiition}.
\end{equation}
The transmitter also computes these posteriors, as it has access to the received symbol $Y_t$ via the noiseless feedback channel, which allows both transmitter and receiver to use the same deterministic partitioning algorithm. The process repeats until the first time $\tau$ that a single message $i$ attains a posterior $\rho_i(y^\tau) \ge 1 - \epsilon$. The receiver chooses this message $i$ as the estimate $\hat{\theta}$. 
Since $\theta$ is uniformly sampled, every possible message $j \in \{0,1\}^K$ has the same prior: $\Pr(\theta = j) = 2^{-K}$.

To prove that the SED scheme of Naghshvar \emph{et. al.} \cite{Naghshvar2015} is a posterior matching BSC scheme as described in \cite{Shayevitz2011}, it suffices to show that the the scheme uses the same encoding function as \cite{Shayevitz2011} applied to a permutation of the messages. Since 
the posteriors $\rho_i(y^t)$ are fully determined by the  history of received symbols $Y^t$, a permutation of the messages can be defined concatenating the messages in $S_0$ and $S_1$, each sorted by decreasing posterior. This permutation induces a c.d.f. on the corresponding posteriors. Then, to satisfy the \emph{posterior matching principle}, the random variable $U$ could just be the c.d.f. evaluated at the last message before $\theta$. The resulting encoding function is given by $X_{t+1} = 0$ if $U < 1/2$, otherwise $X_{t+1} = 1$.

Naghshvar \emph{et. al.} proposed two methods to construct the partitions $S_0$ and $S_1$. The simplest one, described as the small enough difference encoder (SED) \cite{Yang2021}, consists of an algorithm that terminates when the SED constraint bellow is met:
\begin{equation}
    0\le \sum_{i \in S_0}\rho_i(y^t) - \sum_{i \in S_1}\rho_i(y^t)  < \min_{i \in S_0} \rho_i(y^t) \, .
    \label{eq: sed rule}
\end{equation}
The algorithm starts with all messages in $S_0$ and a vector of posteriors $\bm{\rho}_t \triangleq [\rho_1(y^t),\dots,\rho_{2^{K}}(y^t)]$ of the messages $\{1,\dots,2^K\}$. The items are moved to $S_1$ one by one, from smallest to largest posterior. The process ends at any point where rule \eqref{eq: sed rule} is met. If the accumulated probability in $S_0$ falls below $\frac{1}{2}$, then the labelings of $S_0$ and $S_1$ are swapped, after which the process resumes.

The worst case scenario complexity of this algorithm is of order $O(M^2)$, where $M=2^K$ is the number of posteriors. The $M$ is squared because part of the process repeats after every swap, and in the worst case scenario the number of swaps is proportional to $M$. However, a likely scenario is that the process ends after very few swaps, in which case the complexity is of order $O(M)=O(2^K)$. 

The second method by which Naghshvar \emph{et al.} proposed to construct $S_0$ and $S_1$  consists of an exhaustive search over all possible partitions, i.e., the power set $2^\Omega$, and a metric to determine the optimal partition. This search would clearly include the partitioning of the first method, and therefore, also provide the guarantees of equations \eqref{eq: submartingale C} and \eqref{eq: average llr guarantee}.

\subsection{Yang's Achievable Rate}\label{sec: SED Rate}
Yang \emph{et. al.} \cite{Yang2021} developed the upper bound \eqref{eq: general bound} on the expected block length $\tau$ of the SED encoder that, to the best of our knowledge, is the best upper bound that has been developed for the model. 

The analysis by Yang \emph{et al.} consists of two steps. The first step, in \cite{Yang2021} Thm. $7$, consists of splitting the single phase process from Naghshvar \emph{et. al.} \cite{Naghshvar2015} into a two phase process: the communication phase, with stopping time $T$, where $\rho_i(y^t) < \frac{1}{2}$ and a confirmation phase where $\rho_i(y^t) \ge \frac{1}{2}$, when the transmitted message $\theta$ is the message $i$. This is a method first used by Burnahsev in \cite{Burnashev1976}.
With the first step alone, the following upper bound on the expected blocklength can be constructed:
\ifCLASSOPTIONonecolumn
\begin{align}
    \E[\tau] &\le \frac{\log_2(M-1)+C_2}{C}
    +\left\lceil\frac{\log_2(\frac{1-\epsilon}{\epsilon})}{C_2}\right\rceil\frac{C_2}{C_1} 
    + 2^{-C_2}\left(\frac{2 C_2}{C}-\frac{C_2}{C_1} \right)\frac{1 - \frac{\epsilon}{1-\epsilon}2^{-C_2}}{1 - 2^{-C_2}} \, .
  \label{eq: original bound}
\end{align}
\else
\begin{align}
    \E[\tau] &\le \frac{\log_2(M-1)+C_2}{C}
    +\left\lceil\frac{\log_2(\frac{1-\epsilon}{\epsilon})}{C_2}\right\rceil\frac{C_2}{C_1} \nonumber \\ 
  &\phantom{=\,} + 2^{-C_2}\left(\frac{2 C_2}{C}-\frac{C_2}{C_1} \right)\frac{1 - \frac{\epsilon}{1-\epsilon}2^{-C_2}}{1 - 2^{-C_2}} \, .
  \label{eq: original bound}
\end{align}
\fi
where $C$ is the channel capacity, defined by $C\triangleq 1-H(p)$, and $H(p) \triangleq -p\log_2(p) - (1-p) \log_2(1-p)$, and the constants $C_2$ and $C_1$ from \cite{Yang2021} are given by:
\begin{align}
    C_2 &\triangleq \log_2\left ( \frac{q}{p}\right) \label{eq: C2 def}
    \\
    C_1 &\triangleq q \log_2\left(\frac{q}{p}\right) + p \log_2\left(\frac{p}{q}\right) \, .
\end{align}
The second step, in \cite{Yang2021} Lemma $4$, consists of synthesizing a surrogate martingale $U'_i(t)$ with stopping time $T'$ that upper bounds $T$, which is a degraded version of the sub-martingale $U_i(t)$. The martingale $U'_i(t)$ guarantees that whenever $U'_i(t) < 0$, then $U'_i(t+1) \le \frac{1}{q}\log_2(2q)$ and while still satisfying the constraints needed to guarantee the bound \eqref{eq: original bound}. An achievability bound on the expected blocklength for the surrogate process, $U'_i(t)$, is constructed from \eqref{eq: original bound} by replacing some of the $C_2$ values by $\frac{1}{q}\log_2(2q)$. The new bound from  \cite{Yang2021} Lemma $4$ is given by:  
\ifCLASSOPTIONonecolumn
\begin{align}
    \E[\tau] &\le \frac{\log_2(M-1)}{C}
    +\frac{\log_2(2q)}{q\cdot C}
    +\left\lceil\frac{\log_2(\frac{1-\epsilon}{\epsilon})}{C_2}\right\rceil\frac{C_2}{C_1} 
  =
  + 2^{-C_2}\left(\frac{C_2 + \frac{\log_2(2q)}{q} }{C}-\frac{C_2}{C_1} \right)\frac{1 - \frac{\epsilon}{1-\epsilon}2^{-C_2}}{1 - 2^{-C_2}} \, .
  \label{eq: general bound}
\end{align}
\else
\begin{align}
    \E[\tau] &\le \frac{\log_2(M-1)}{C}
    +\frac{\log_2(2q)}{q\cdot C}
    +\left\lceil\frac{\log_2(\frac{1-\epsilon}{\epsilon})}{C_2}\right\rceil\frac{C_2}{C_1} \nonumber \\ 
  &\phantom{=\,} + 2^{-C_2}\left(\frac{C_2 + \frac{\log_2(2q)}{q} }{C}-\frac{C_2}{C_1} \right)\frac{1 - \frac{\epsilon}{1-\epsilon}2^{-C_2}}{1 - 2^{-C_2}} \, .
  \label{eq: general bound}
\end{align}
\fi
This bound also applies to the original process $U_i(t)$, since the blocklength of the process $U'_i(t)$ upper bounds that of $U_i(t)$. The bound \eqref{eq: general bound} is lower because $\frac{1}{q}\log_2(2q)$ is smaller than $C_2$. The improvement is more significant as $p \rightarrow 0$ because $\frac{1}{q}\log_2(2q)$ grows from $0$ to $1$ as $p \rightarrow 0$, while $C_2$, instead, grows from $0$ to infinity.
The rate lower bound is given by $\frac{K}{\E[\tau]}$, where $\E[\tau]$ is upper bounded by \eqref{eq: general bound} from Thm. $7$ \cite{Yang2021}.
\subsection{Original Constraints that Ensure Yang's Achievable Rate}\label{sec: SED Requirements}

Let $\mathcal{F}_t \triangleq \sigma(Y^t)$, the $\sigma$-algebra generated by the sequence of received symbols up to time $t$, where $Y^t = [Y_1, Y_2,\dots, Y_t]$. 
For each $i = 1,\dots,M$, let the processes $U_i(Y^t)$ by:
\begin{align}
    U_i(t) = U_i(Y^t) &\triangleq \log_2 \left( \frac{\rho_i(Y^t)}{1-\rho_i(Y^t)}\right)
    \label{eq: U process} \, .
\end{align}

Yang \emph{et. al.} show that the SED encoder from Naghshvar \emph{et. al.} \cite{Naghshvar2015} guarantees that the following constraints \eqref{eq: submartingale C}-\eqref{eq: Markov step size} are met:
 \begin{align}
      \E[U_i(t+1)|\F_t,\theta=i] &\ge U_i(t) + C,&
      \quad &\text{if } U_i(t) < 0, \label{eq: submartingale C}
      \\
      |U_i(t+1) - U_i(t)| &\le C_2.&& \label{eq: Max U step}
      \\
      \E[U_i(t+1)|\F_t,\theta=i] &= U_i(t) + C_1, &\quad &\text{if } U_i(t) \ge 0, \label{eq: martingale C1}
      \\
      |U_i(t+1) - U_i(t)| &= C_2,&
      \quad &\text{if } U_i(t) \ge 0,  \label{eq: Markov step size}
\end{align}
Meanwhile, Naghshvar \emph{et. al.} show that the SED encoder also satisfied the more strict constraint that the average log likelihood ratio $\mathbf{U}(t)$, as defined in equation \eqref{eq: average llr}, is also a submartingale that satisfies equation \eqref{eq: average llr guarantee}, which is equivalent to \eqref{eq: average llr guarantee 2}:
\begin{align}
    &\mathbf{U}(Y^t) \triangleq \sum_{i=1}^M \rho_i(Y^t) U_i(Y^t)
    \label{eq: average llr}
    \\
    &\E[\mathbf{U}(Y^{t+1})\mid \mathcal{F}_t] \ge \mathbf{U}(Y^t) + C
    \label{eq: average llr guarantee}
    \\
    &\E\left[\sum_{i=1}^M \left(\rho_i(y^{t \! + \! 1}) U_i(t \! + \! 1) \! - \!  \rho_i(y^t) U_i(t)\right) \Bigg | \mathcal{F}_t\right] \ge C 
    \,.
    \label{eq: average llr guarantee 2}
\end{align}
The process $\mathbf{U}(t)$ is a weighted average of values $U_i(t)$, some of which increase and some of which decrease after the next transmission $t+1$.

To derive the bound \eqref{eq: original bound}, Yang \emph{et al.} split
the decoding time $\tau$ into $T$ and $\tau-T$, where $T$ is an intermediate stopping time defined by the first crossing into the confirmation phase. The expectation $\E[T]$ is analyzed in \cite{Yang2021} as a martingale stopping time, and requires that if $\theta = i$, then $U_i(t)$ be a strict submartingale that satisfies the inequalities \eqref{eq: submartingale C} and \eqref{eq: Max U step}. The expectation $\E[\tau-T]$ is analyzed using a Markov Chain
that exploits the larger and fixed magnitude step size \eqref{eq: Markov step size} and inequality \eqref{eq: martingale C1}. Since $T$ is the time of the first crossing into the confirmation phase, the Markov Chain model, needs to include in the time $\tau-T$ the time that message $i$ takes to return to the confirmation phase if it has fallen back to the communication phase, that is: $\rho_i(y^t) < \frac{1}{2}$ for some $t > T$.




\section{A New Bound and Relaxed Partitioning}\label{sec: Main Section}
In the following section, we introduce relaxed conditions that are still sufficient to allow a sequential encoder over the BSC with full feedback to attain the performance of Yang's bound \eqref{eq: general bound}. Specifically, we replace the requirement in \eqref{eq: submartingale C} that applies separately to each message with a new requirement in \eqref{eq: C step size} that applies to an average over all possible messages.  For each individual message, we require in \eqref{eq: V ge 0} that each step size is larger than the same positive $\textit{a}$.  
The relaxed conditions are easier to enforce than \eqref{eq: submartingale C},  e.g. by the SEAD partitioning constraint introduced in Thm. \ref{theorem: simple rule}.

\subsection{Relaxed Constraints that Also Guarantee Bound \eqref{eq: original bound}}
We begin with a theorem that introduces relaxed conditions and shows that they guarantee the performance \eqref{eq: original bound}, corresponding to the first step of Yang's analysis.

\begin{theorem}\label{theorem: Main Theorem}
Let $\tau$ be the stopping time of a sequential transmission system over the BSC. At each time $t$ let the posteriors $\rho_1(Y^t), \rho_2(Y^t),\dots,\rho_M(Y^t)$ be as defined in \eqref{eq: postierior defiition} and the log likelihood ratios $U_1(t),\dots,U_M(t)$ be as defined in \eqref{eq: U process}. 
Suppose that for all times $t$ for all received symbols $y^t$, and for each $j\in \Omega$, the constraints \eqref{eq: V ge 0}-\eqref{eq: phase II step size}
are satisfied:
\begin{alignat}{4}
    &\E[U_j(t+1)-U_j(t)|\mathcal{F}_t,\theta = j] &&\ge a \, ,  &&\text{where } a &&> 0\, , \label{eq: V ge 0}
    \\
    &U_j(t+1)-U_j(t)   &&\le  C_2 \, , \quad  &&\text{if } U_j(t) &&\le 0 \, , \label{eq: phase 1 max}
    \\
    &\E[U_j(t+1)-U_j(t)|\mathcal{F}_t,\theta = j] &&= C_1 \, , &&\text{if } U_j(t) &&\ge 0 \, , \label{eq: phase II V}
    \\
    &\mid U_j(t+1)-U_j(t) \mid  &&= C_2 \, , &&\text{if } U_j(t) &&\ge 0 \, . \label{eq: phase II step size}
\end{alignat}
Suppose further that for all $t$ and $y^t$ the following condition is satisfied:
\begin{alignat}{4}
     &\sum_{j=1}^M\E[\rho_j(y^t)\left(U_j(t \! + \! 1) \! -  \! U_j(t)\right)|&& Y^t = y^t , && \theta = j]) &&\ge C .
    \label{eq: C step size}
\end{alignat}
Then, expected stopping time $\E[\tau]$ is upper bounded by \eqref{eq: stopping time bound}.
\ifCLASSOPTIONonecolumn
\begin{align}
    \E[\tau] &\le \frac{\log_2(M \! - \! 1)+C_2}{C}
    +\left\lceil\frac{\log_2(\frac{1-\epsilon}{\epsilon})}{C_2}\right\rceil\frac{C_2}{C_1} 
    + 2^{-C_2}\left(\frac{ C_2 }{C}-\frac{C_2}{C_1} \right)\frac{1 - \frac{\epsilon}{1-\epsilon}2^{-C_2}}{1 - 2^{-C_2}} \, . 
  \quad \boxempty
  \label{eq: stopping time bound}
\end{align}
\else
\begin{align}
    \E[\tau] &\le \frac{\log_2(M \! - \! 1)+C_2}{C}
    +\left\lceil\frac{\log_2(\frac{1-\epsilon}{\epsilon})}{C_2}\right\rceil\frac{C_2}{C_1} \nonumber \\ 
  &\phantom{+\,} + 2^{-C_2}\left(\frac{ C_2 }{C}-\frac{C_2}{C_1} \right)\frac{1 - \frac{\epsilon}{1-\epsilon}2^{-C_2}}{1 - 2^{-C_2}} \, . 
  \quad \boxempty
  \label{eq: stopping time bound}
\end{align}
\fi
The proof is provided in Sec \ref{sec: proof of Thm 1}. 
\end{theorem}
In equation \eqref{eq: C step size} the values of $\rho_j(y^t)$ and $U_j(t)$ are fixed since they are functions of the $y^t$ specified by the conditioning. Then, the expectation $\E[\rho_j(y^t)|Y^t = y^t , \theta = j])$ is simply the constant $\rho_j(y^t)$, and we can also write \eqref{eq: C step size} as the weighted sum of expectations:
\begin{alignat}{4}
     &\sum_{j=1}^M \rho_j(y^t) \E[\left(U_j(t \! + \! 1) \! -  \! U_j(t)\right)|&& Y^t = y^t , && \theta = j]) &&\ge C .
    \label{eq: C step size 2}
\end{alignat}
Meanwhile the value of $U_j(t \! + \! 1)$ for each $j$ is a random variable that takes on two possible values depending on the value of $Y_{t+1}$.



The sequential transmission process begins by randomly selecting a message $\theta$ from $\Omega$.  Using that selected message, at each time $t$ until the decoding process terminates, the process computes an $X_{t}=x_{t}$, which induces a $Y_{t}=y_{t}$ at the receiver.  The original constraint \eqref{eq: submartingale C} dictates that $\{U_i(t), \theta = i\}$ is a sub-martingale and allows for a bound on $U_i(t)$ at any future time $t+s$ for any possible selected message $i$, i.e. $\E[U_i(t+s)\mid \mathcal{F}_t, \theta = i] \ge U_i(t) + s C$. This is no longer the case with the new constraints in Thm. \ref{theorem: Main Theorem}. While equation \eqref{eq: V ge 0} of the new constraints  makes the process $U_i(t)$ a sub-martingale, it only guarantees that $\E[U_i(t+s)\mid \mathcal{F}_t, \theta = i] \ge U_i(t) + s a$ and $a$ could be any small positive constant. The left side of equation \eqref{eq: C step size} is a sum that includes all $M$ realizations of the message, it is a constraint for each fixed time $t$ and each fixed event $Y^t = y^t$ that governs the behavior across the entire message set and does not define a sub-martingale.
For this reason, the martingale analysis used by Naghshvar \emph{et. al.} in \cite{Naghshvar2015} and Yang \emph{et al.} in \cite{Yang2021} no longer be applies.  

A new analysis is needed to derive \eqref{eq: stopping time bound}, the bound on the expected stopping time $\tau$, using only the constraints of Thm. \ref{theorem: Main Theorem}. This new analysis needs to exploit the property that the expected stopping time is over all messages, that is: $\E[\tau] = \sum_{i=1}^M \Pr(\theta = i) \E[\tau \mid \theta = i]$ which the original does not use because it guarantees that the bound \eqref{eq: general bound} holds for each message, i.e.,  $\E[\tau \mid \theta = i], i = 1,\dots,M$. Note, however, that the original constraint \eqref{eq: submartingale C} does imply that the new constraints are satisfied, so that the results we derive below also apply to the setting of Naghshvar \emph{et. al.} in \cite{Naghshvar2015} and Yang \emph{et al.} in \cite{Yang2021}.
The new constraints allow for a much simpler encoder and decoder design. This simpler design motivates our new analysis that forgoes the simplicity afforded by modeling the process $\{U_i(t), \theta = i\}$ as a martingale.

The new analysis seeks to accumulate the entire time that a message  $i$ is not in its confirmation phase, i.e. the time during which the encoder is either in the communication phase or in some other message's confirmation phase.  For each $n=1,2,3,\dots$, let $T_n$ be the time at which the confirmation phase for message $i$ starts for the $n$th time (or the process terminates) and let $t^{(n)}_0$ be the time the encoder exits the confirmation phase for message $i$ for the $(n-1)^{th}$ time (or the process terminates). That is, for each $n=1,2,3,\dots$, let $t^{(n)}_0$ and $T_n$ be defined recursively by $t_0^{(1)}=0$ and :
\begin{alignat}{4}
    &T_n& &= \min\{t \ge t^{(n)}_0& &: 
    U_i(t) \ge 0 \, \text{or } t=\tau\}
    \label{eq: T_n definition}
    \\
    &t^{(n+1)}_0& &= \min\{t \ge T_n& &: 
    U_i(t) <  0 \, \text{or } t=\tau\} \, .
    \label{eq: t_0 definition}
\end{alignat}
Thus, the total time the process $U_i(t)$ is not in its confirmation phase is given by:
\begin{equation}
    T \triangleq \sum_{n=1}^{\infty} \left(T_n \! - \! t^{(n)}_0\right) \label{eq: T definition} \, .
\end{equation}

\subsection{A ``Surrogate Process'' that can Tighten the Bound}
\label{sec: surrogate process}


First we want to note that the bound \eqref{eq: stopping time bound} is loose compared to \eqref{eq: general bound}. It is loose because when the expectation $\E[\tau]$ is split in two parts, $\E[T]$ and $\E[\tau-T]$, to analyze them separately, a sub-optimal factor is introduced in the expression for $\E[T]$, which is $\frac{1}{C}(\log_2(M-1)+C_2)$.
The sub-optimality is derived from the term $C_2$, which is the largest value that $U_i(t)$ can take at the start of the confirmation phase and makes the term $\E[T]$ large. However, this large $C_2$ is not needed to satisfy any of the constraints in Thm. \ref{theorem: Main Theorem}.
To overcome this sub-optimality, we use a surrogate process that is a degraded version of the process $U_i(t)$, where the value at the start of the confirmation phase is bounded by a constant smaller than $C_2$. The surrogate process is a degradation in the sense that it is always below the value of the original process $U_i(t)$. 

\begin{figure}[t]
\centering
\includegraphics[width=0.5\textwidth]{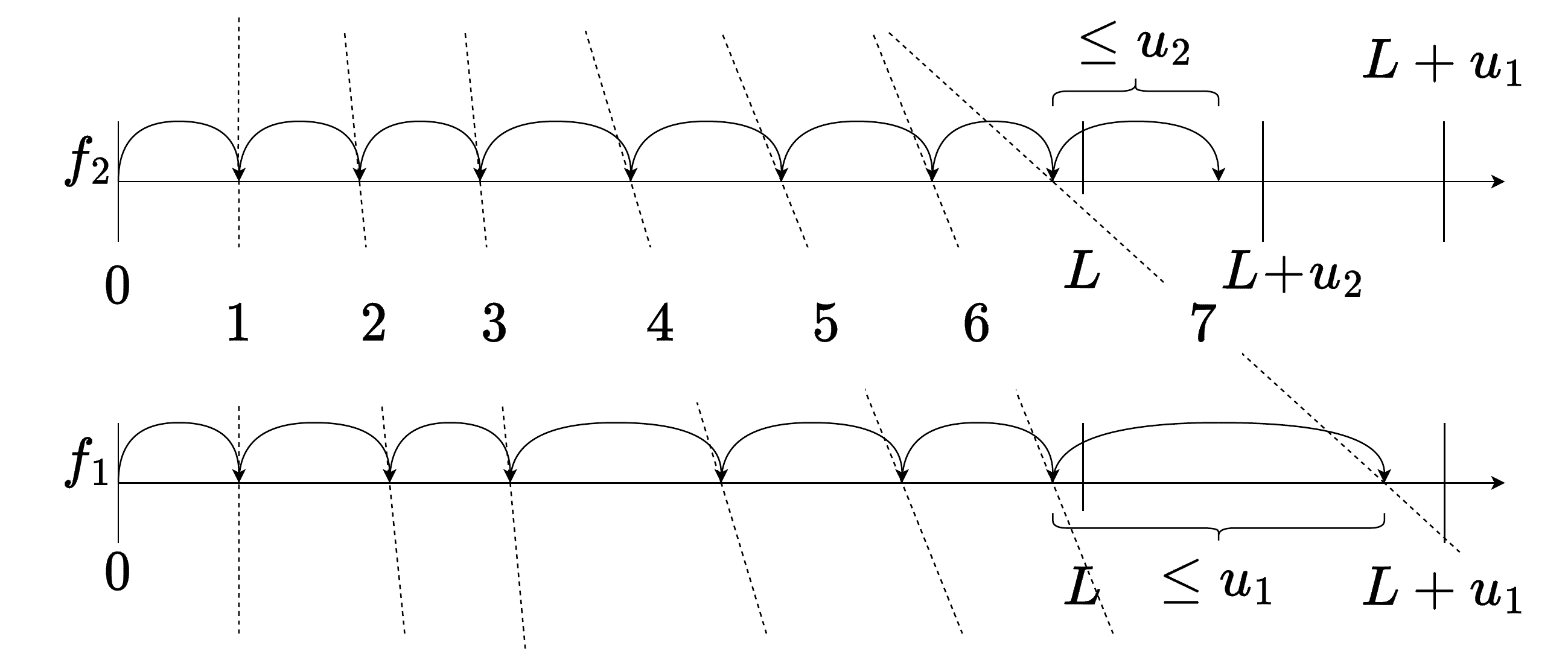}
\caption{Example: frogs $f_1$ and $f_2$ jumping from $0$ to $L$. The length of a single jump by $f_1$ is at most $u_1$. Frog $f_2$ jumps at the same times as $f_1$, however, the length of a single jump by $f_2$ is at most $u_2 < u_1$. This restriction forces frog $f_2$ to be always behind $f_1$ and thus reach $L$ no sooner than frog $f_1$.
}
\label{fig: frog race}
\end{figure}

Perhaps the utility of the surrogate process can be better understood through the following frog-race analogy illustrated in Fig. \ref{fig: frog race}.
A frog $f_1$ traverses a race track of length $L$ jumping from one point to the next. The distance traveled by frog $f_1$ in a single jump is upper bounded by $u_1$. The jumps are not necessarily IID, but we know that the expected length of each jump is lower bounded by $l$. It is also possible that $f_1$ takes some jumps backwards. With only this information, we want to determine an upper bound on the average number of steps frog $f_1$ takes to reach the end of the track. This could be done using Doob's optional stopping theorem \cite{almaOST2019} to compute the upper bound as $\frac{L+u_1}{l}$, the maximum distance $L + u_1$  traveled from the origin to the last jump divided by the lower bound on average distance  $l$ of a single jump. 

Perhaps this bound can be improved. The final point is located between $L$ and $L + u_1$ and is reached in a single jump from a point between $L - u_1$ and $L$. If for instance, the frog was restricted to only forward jumps, we could replace $u_1/l$ by just $1$, but the process $U_i(t)$ actually can take steps backwards. Instead we exploit another property of $U_i(t)$, which is that maximum step size $C_2$ is not needed to guarantee the lower bound $C$ on the average step size. 

Suppose now that a surrogate frog $f_2$ participates in the race along $f_1$ but with the following restrictions: 
\begin{enumerate}
    \item $f_1$ and $f_2$ start in the same place and always jump at the same time.
    \item  $f_2$ is never ahead of $f_1$, i.e. when $f_1$ jumps forward, $f_2$ jumps at most as far, and when $f_1$ jumps backwards, $f_2$ jumps at least as far.  \item Moreover, the forward distance traveled by frog $f_2$ in a single jump is upper bounded by $u_2<u_1$.
     \item Despite its slower progress, the surrogate frog $f_2$ still satisfies the property that the expected length of each jump is lower bounded by $l$.
\end{enumerate}
The average number of steps taken by $f_2$ will be upper bounded by $\frac{L+u_2}{l}$, also by Doob's optional stopping theorem. Since $f_2$ is never ahead of $f_1$ then $f_2$ crossing the finish line implies that $f_1$ has as well.  Thus, $\frac{L+u_2}{l}$ is also an upper bound on the average number of jumps required for frog $f_1$ reach across $L$.

The equivalent to the surrogate frog $f_2$ is what we proceed to define in Thm. \ref{theorem: surrogate martingale}, where the length $L=\log(M-1)$,  $u_1=C_2$, $u_2 = B$ and $l=C$.



\begin{theorem}[Surrogate Process Theorem]
\label{theorem: surrogate martingale}
Let the surrogate process $U'_i(t)$ be a degraded version of $U_i(t)$ that still satisfies the constraints of Thm. \ref{theorem: Main Theorem}.
Initialize the surrogate process as $U'_i(0)=U_i(0)$ and reset $U'_i(t)$ to $U_i(t)$ at every $t=t^{(n)}_0$, that is at each $t$ that the encoder exits a confirmation phase round for message $i$. 
Define $T'_n \triangleq \min\{t \ge t_0^{(n)}: U'_i(t) \ge 0  \, \text{or } t = t_0^{(n+1)}\}$.
Suppose that for some $B < C_2$, the process $U'_i(t)$ also satisfies the following constraints:
\begin{align}
    &U_i(t) < 0 \implies U'_i(t \! + \! 1) \! - \! U'_i(t)& &\le U_i(t \! + \! 1) \! - \! U_i(t) \label{eq: U' smaller}
    \\
    &U'_i(t) < 0 \implies U'_i(t \! + \! 1)&
    &\le B
    \\
    &U'_i \left ( T'_n \right ) - \frac{p}{q}\left ( U_i(T_n) - C_2\right )& 
    &\le B
    \label{eq: U' B bound}.
\end{align}
Then the total time $U'_i(t)$ is not in its confirmation phase is given by $T' \triangleq \sum_{n=1}^{\infty} \left(T'_n \! - \! t^{(n)}_0\right)$, and $E[T]$ is bounded by:
\begin{equation}
    \E[T] \le \E[T']
    \le \frac{B}{C}\left(1 \! + \! 2^{-C_2}\frac{1 \! - \! 2^{-NC_2}}{1 \! - \! 2^{-C_2}}\right)-\frac{\E[U_i(0)]}{C} \, . \label{eq: surrogate time}
\end{equation}
Note that $T_n \le T'_n$ for all $n$ because $U_i(t) \ge U'_i(t)$ from the definition of $U'_i(t)$ and constraint \eqref{eq: U' smaller}, Therefore $T \le T'$. 
Also note that after the process terminates at the stopping time $\tau$, both $T_n$ and $t^{(n)}_0$ are equal to $\tau$, which makes their difference $0$. Then the communication phase times $T$ and $T'$ are a sum of finitely many non-zero values. \quad $\boxempty$

The proof is provided in Sec. \ref{sec: proof of theorems 2 and 3} 

\end{theorem}

\subsection{Relaxed Constraints that Achieve a Tighter Bound}
\label{sec: SEAD constraint}
The following theorem introduces partitioning  constraints that guarantee that the constraints in Thm. \ref{theorem: surrogate martingale} are satisfied with a value of $B=\log_2(2q)/q$ for the surrogate process. The new constraints are looser than the original SED constraint, and therefore are satisfied by an encoder that enforces the SED constraint. Using a new analysis we show that this encoder guarantees an achievability bound tighter than the bound \eqref{eq: general bound} obtained in the second step of Yang's analysis described in Section \ref{sec: Achievable Rate}. The value $B=\log_2(2q)/q$ is the lowest possible $B$ value that satisfies the constraints \eqref{eq: U' smaller}-\eqref{eq: U' B bound} of Thm. \ref{theorem: surrogate martingale} for a system that enforces the original SED constraint \eqref{eq: sed rule}. 
The new achievability applies to an encoder that satisfies the new relaxed constraint as well as one that satisfies the SED constraint. 
\begin{theorem}
\label{theorem: simple rule}
Consider sequential transmission over the BSC with noiseless feedback as described in Sec. \ref{sec: introduction} with an encoder that enforces the Small Enough Absolute Difference (SEAD) encoding constraints, equations \eqref{eq: simple partitioning} and \eqref{eq: simple partitioning 2} bellow:
 \begin{align}
         &  \left |\underset{i \in S_0}{\sum}\rho_i(y^t) - \underset{i \in S_1}{\sum}\rho_i(y^t) \right | \le \underset{i \in S_0}{\min}\rho_i(y^t)
         \label{eq: simple partitioning} 
         \\
         & \rho_i(y^t) \ge \frac{1}{2} \implies S_0 = \{i\} \; \text{or } S_1 = \{i\}
         \,.
         \label{eq: simple partitioning 2} 
    \end{align}
Then, the constraints \eqref{eq: phase II step size}-\eqref{eq: C step size} in Thm. \ref{theorem: Main Theorem} are satisfied and a process $U'_i(t)$, $i=1,\dots,M$ described in Thm. \ref{theorem: surrogate martingale} can be constructed with $B = \frac{1}{q}\log_2(2q)$. 
The resulting upper bound on $\E[\tau]$ is given by: 
\ifCLASSOPTIONonecolumn
\begin{align}
    \E[\tau] \le&  \frac{\log_2(M-1) +\frac{\log_2(2q)}{q}}{C} + \frac{ C_2}{C_1}\left \lceil \frac{\log_2\left(\frac{1-\epsilon}{\epsilon}\right)}{C_2}\right \rceil
    + 2^{-C_2}\frac{1-\frac{\epsilon}{1-\epsilon}2^{- C_2}}{1-2^{-C_2}}
    \left(\frac{\log_2(2q)}{q C}-\frac{C_2}{C_1}\right) \, ,\label{eq: stopping time optimized bound}
\end{align} 
\else
\begin{align}
    \E[\tau] \le&  \frac{\log_2(M-1) +\frac{\log_2(2q)}{q}}{C} + \frac{ C_2}{C_1}\left \lceil \frac{\log_2\left(\frac{1-\epsilon}{\epsilon}\right)}{C_2}\right \rceil
    \nonumber
    \\
    &+ 2^{-C_2}\frac{1-\frac{\epsilon}{1-\epsilon}2^{- C_2}}{1-2^{-C_2}}
    \left(\frac{\log_2(2q)}{q C}-\frac{C_2}{C_1}\right) \, ,\label{eq: stopping time optimized bound}
\end{align} 

\fi
which is lower than \eqref{eq: general bound} from \cite{Yang2021}.
Note that meeting the SEAD constraints guarantees that both sets $S_0$ and $S_1$ are non empty. This is because if either set is empty, the the other one is the whole space $\Omega$ and the difference in \eqref{eq: simple partitioning} is $1$, which is greater than any posterior in a space with more than one element. 
\quad $\boxempty$

The proof is provided in Sec. \ref{sec: proof of theorems 2 and 3}
\end{theorem}

The requirement $\rho_i(y^t) \ge \frac{1}{2} \implies S_0 = \{i\} \; \text{or } S_1 = \{i\}$, is needed to satisfy constraint \eqref{eq: phase II step size} and guarantees constraint \eqref{eq: phase II V}. This requirement is also enforced by the SED partitioning constraints in \cite{Naghshvar2015} and \cite{Yang2021}. 

The SEAD partitioning constraint is satisfied whenever the SED constraint \eqref{eq: sed rule} is. However, the SEAD partitioning constraint allows for constructions of $S_0$, $S_1$ that do not meet either of the SED constraints in \cite{Naghshvar2015} and \cite{Yang2021}, and is therefore looser. Particularly, SEAD partitioning allows for the case where $P_1 - P_0 > \underset{j \in S_1}{\max}\rho_j(y^t)$ that often arises in the implementation shown in Sec \ref{sec: set construction}. This case is not allowed under either of the SED constraints because they both demand that:
\begin{equation}
    -\underset{j \in S_1}{\min}\rho_j(y^t) \le P_0-P_1 \le \underset{j \in S_0}{\min}\rho_j(y^t) \, . 
\end{equation}



\section{Supporting Lemmas and Proof of Theorem \ref{theorem: Main Theorem}}
\label{sec: helping lemmas and proof of main theorem}

This section presents some supporting Lemmas and the full proofs of Thm. \ref{theorem: Main Theorem}. 
Let $T$ be the time the transmitted message $\theta$ spends in the communication phase or on an incorrect confirmation phase, that is, for $\theta = i$, $U_i(t) < 0$ as defined equation \eqref{eq: T definition}. Note that this definition is different from the stopping time used in \cite{Yang2021} described in Sec. \ref{sec: Achievable Rate}.
The proof of Thm. \ref{theorem: Main Theorem} consists of bounding $\E[T]$ and $\E[\tau - T]$ by expressions that derive from the constraints \eqref{eq: V ge 0}-\eqref{eq: C step size}.


Since $\rho_i(y^t) \ge 1-\epsilon \iff \frac{\rho_i(y^t)}{1-\rho_i(y^t)}\ge \frac{1-\epsilon}{\epsilon} \iff U_i(t) \ge \log_2\left(\frac{1-\epsilon}{\epsilon}\right)$, the stopping rule described in Sec. \ref{sec: Achievable Rate} could be expressed by: $\tau \triangleq \{\min \; t:  \underset{i}{\max}\{ U_i(t)\} \ge \log_2\left(\frac{1-\epsilon}{\epsilon}\right)\}$.
To prove Thm. \ref{theorem: Main Theorem}, we will instead use the stopping rule introduced by Yang \emph{et. al.} \cite{Yang2021}, defined by:
\begin{equation}
    \tau \triangleq \{\min t:  \underset{i}{\max}\{ U_i(t)\} \ge N C_2\} \, ,\label{eq: stopping rule ceiling}
\end{equation}
where $N \triangleq \left \lceil \frac{\log_2\left(\frac{1 - \epsilon}{\epsilon}\right)}{C_2}\right \rceil$.
This rule models the confirmation phase as a fixed Markov Chain with exactly $N+1$ states. Since $N C_2 \ge \log_2 \left(\frac{1-\epsilon}{\epsilon}\right)$, the stopping time under the new rule is larger than or equal to that of the original rule without the ceiling as explained in \cite{Yang2021}.

\subsection{Five Helping Lemmas to Aid the Proof of Thms. \ref{theorem: Main Theorem}-\ref{theorem: simple rule}}
\label{sec: lemmas}
The expression to bound the expectation expectation $\E[T]$ is constructed via five  inequalities (or equalities) each of which derives from one of the following five Lemmas. The proofs of the Lemmas will be provided in Sec. \ref{sec: proof of lemmas}.
\begin{lemma}\label{lemma: Phase I time}
Let the total time the transmitted message spends in the communication (or an incorrect confirmation phase) be $T$ and let $T_n$ and $t^{(n)}_0$ be as defined in \eqref{eq: T_n definition} and \eqref{eq: t_0 definition}. Define $T^{(n)} \triangleq T_n - t^{(n)}_0$ and let
\begin{equation}
    \mathcal{Y}_\epsilon^{(i)} \triangleq \{y^t: \rho_i(y^t) < \frac{1}{2}, \rho_j(s) < 1 \! - \!\epsilon \; \forall s \le t, j \in  \Omega\}
\end{equation}
then:
\begin{align}
    \E[T] 
    &= \sum_{i=1}^M \Pr(\theta = i) \sum_{t=1}^\infty \sum_{y^t \in \mathcal{Y}_\epsilon^{(i)}}\Pr(Y^t = y^t \mid \theta = i) \, . \label{eq: Phase I time}
\end{align}

Note that $T^{(1)}=T_1$ is the time before entering the correct confirmation phase for the first time, that is, the time
spent in the communication phase (or an incorrect confirmation phase) before the posterior $\rho_i(y^t)$ of the transmitted message ever crosses $\frac{1}{2}$.
If the decoder stops (in error) before ever entering the correct confirmation phase, then $T^{(1)}$  is the time until the decoder stops. For $n>1$, $T^{(n)}$ is the time between falling back from the correct confirmation for the $(n-1)^{th}$ time and either stopping (in error) or reentering the correct confirmation phase for the $(n)^{th}$ time.  
Thus, the total time the transmitted message $\theta=i$ has $U_i(t) < 0$ is also given by $T = \sum_{n=1}^\infty T^{(n)}$.
Also note that that if the decoder stops before entering the correct confirmation phase for the $n^{th}$ time, then $T^{(n+m)}=0$ for all $m \ge 1$.  \quad $\boxempty$
\end{lemma}

\begin{lemma}\label{lemma: sum C bound}
Suppose constraints \eqref{eq: V ge 0}, \eqref{eq: phase II step size}, and \eqref{eq: C step size} of Thm. \ref{theorem: Main Theorem} are satisfied and let:
\begin{equation}
    V_i(y^t) \triangleq
    \E[U_i(t \! + \! 1) \! - \! U_i(t) 
    \!\mid \! Y^t = y^t, \theta = i] \, ,
    \label{eq: Viyt definition}
\end{equation}
then:
\ifCLASSOPTIONonecolumn
\begin{align}
    \sum_{i = 1}^M \Pr(\theta = i) \underset{y^t \in \mathcal{Y}_\epsilon^{(i)}}{\sum }
    V_i(y^t)\Pr(Y^t = y^t \mid \theta = i) 
    \ge C \sum_{i = 1}^M \Pr(\theta = i) \underset{y^t \in \mathcal{Y}_\epsilon^{(i)}}{\sum }
    \Pr(Y^t = y^t \mid \theta = i) \, . \label{eq: Vi M wide sum}  \quad \boxempty
\end{align}
\else
\begin{align}
    \sum_{i = 1}^M &\Pr(\theta = i) \underset{y^t \in \mathcal{Y}_\epsilon^{(i)}}{\sum }
    V_i(y^t)\Pr(Y^t = y^t \mid \theta = i) 
    \nonumber
    \\
    &\ge C \sum_{i = 1}^M \Pr(\theta = i) \underset{y^t \in \mathcal{Y}_\epsilon^{(i)}}{\sum }
    \Pr(Y^t = y^t \mid \theta = i) \, . \label{eq: Vi M wide sum}  \quad \boxempty
\end{align}
\fi
\end{lemma}

\begin{lemma}\label{lemma: stopping time bound}
Let $T_n$ and $t_0^{(n)}$ be the times defined in \eqref{eq: T_n definition} and \eqref{eq: t_0 definition}. Then, for the left side of sum \eqref{eq: Vi M wide sum} in Lemma \eqref{lemma: sum C bound} the following equality holds:
\ifCLASSOPTIONonecolumn
\begin{align}
    \sum_{i = 1}^M \Pr(\theta = i) \sum_{y^t \in \mathcal{Y}_\epsilon^{(i)}} V_i(y^t) \Pr(Y^t = y^t \mid \theta = i) 
    = \sum_{i=1}^M \! \Pr(\theta \! = \! i) \sum_{n=1}^\infty  \E[U_i(T_n) \! - \! U_i(t_0^{(n)}) \! \mid \! \theta \! = \! i]  \, . \label{eq: telescopic equality}  \quad \boxempty
\end{align}
\else
\begin{flalign}
    \sum_{i = 1}^M & \Pr(\theta = i) \sum_{y^t \in \mathcal{Y}_\epsilon^{(i)}} V_i(y^t) \Pr(Y^t = y^t \mid \theta = i) 
    \nonumber&
    \\
    = & \sum_{i=1}^M \! \Pr(\theta \! = \! i) \sum_{n=1}^\infty  \E[U_i(T_n) \! - \! U_i(t_0^{(n)}) \! \mid \! \theta \! = \! i]  \, . \label{eq: telescopic equality}  \quad \boxempty&
\end{flalign}
\fi
\end{lemma}

\begin{lemma}\label{lemma: fall back probability}
Let $\epsilon$ be the decoding threshold and let the decoding rule be \eqref{eq: stopping rule ceiling}. Define the fallback probability as the probability that a subsequent round of communication phase occurs, computed at the start of a confirmation phase. Then, this fallback probability is a constant $p_f$ independent of the message $i = 1,\dots,M$, independent of the number of previous confirmation phase rounds $n$, and is given by: 
\begin{align}
    p_f &= 2^{-C_2}\frac{1-2^{-NC_2}}{1-2^{-(N+1)C_2}} \, .  \quad \boxempty
    \label{eq: fall back probability}
\end{align}
\end{lemma}

\begin{lemma}
\label{lemma: final and initial values}
Let $p_f$ be the fallback probability in Lemma \ref{lemma: fall back probability} and suppose that $U_i(0) < 0 \; \forall i = 1,\dots,M$. Then the expectation \eqref{eq: telescopic equality} in Lemma \ref{lemma: stopping time bound} is upper bounded by:
\ifCLASSOPTIONonecolumn
\begin{align}
      \sum_{i=1}^M \Pr(\theta = i) \sum_{n=1}^\infty  \E[U_i(T_n) \! - \! U_i(t_0^{(n)}) \mid \theta = i]  
      &\le \sum_{i=1}^M \Pr(\theta = i) \left( \frac{p_f}{1-p_f}C_2 + C_2-U_i(0)\right)\label{eq: mid lemma 5 pf}
      \\
      &\le 2^{-C_2}\frac{1-2^{-NC_2}}{1-2^{-C_2}}C_2+C_2-\E[U_i(0)] \, .  \quad \boxempty
\end{align}
\else
\begin{align}
      \sum_{i=1}^M& \Pr(\theta = i) \sum_{n=1}^\infty  \E[U_i(T_n) \! - \! U_i(t_0^{(n)}) \mid \theta = i]  
      \nonumber
      \\
      &\le \sum_{i=1}^M \Pr(\theta = i) \left( \frac{p_f}{1-p_f}C_2 + C_2-U_i(0)\right)\label{eq: mid lemma 5 pf}
      \\
      &\le 2^{-C_2}\frac{1-2^{-NC_2}}{1-2^{-C_2}}C_2+C_2-\E[U_i(0)] \, .  \quad \boxempty
\end{align}
\fi
\end{lemma}

\subsection{Proof of Thm. \ref{theorem: Main Theorem} Using Lemmas \ref{lemma: Phase I time}-\ref{lemma: final and initial values}:}
\label{sec: proof of Thm 1}
\begin{IEEEproof} 
Using Lemmas  \ref{lemma: Phase I time} and \ref{lemma: sum C bound}, the expectation $\E[T]$ is bounded as follows:
\ifCLASSOPTIONonecolumn
\begin{align}
    \E[T] &= \sum_{i = 1}^M \underset{y^t \in \mathcal{Y}_\epsilon^{(i)}}{\sum }\Pr(Y^t = y^t, \theta = i) \label{eq: start proof thm 1}&
    \\
    &\le \frac{1}{C}\sum_{i = 1}^M \underset{y^t \in \mathcal{Y}_\epsilon^{(i)}}{\sum }V_i(y^t)\Pr(Y^t = y^t, \theta = i) \, . \label{eq: ET le sum V}&
\end{align}
\else
\begin{align}
    \E[T] &=  \sum_{i = 1}^M \underset{y^t \in \mathcal{Y}_\epsilon^{(i)}}{\sum }\Pr(Y^t = y^t, \theta = i) \label{eq: start proof thm 1}&
    \\
    &\le \frac{1}{C}\sum_{i = 1}^M \underset{y^t \in \mathcal{Y}_\epsilon^{(i)}}{\sum }V_i(y^t)\Pr(Y^t = y^t, \theta = i) \, . \label{eq: ET le sum V}&
\end{align}
\fi
By Lemma \eqref{lemma: stopping time bound}, expression \eqref{eq: ET le sum V} is equal to the left side of inequality \eqref{eq: mid lemma 5 pf}, which is bounded by \eqref{eq: C2 pf and EU} according to Lemma \ref{lemma: final and initial values}:
\ifCLASSOPTIONonecolumn
\begin{align}
    \frac{1}{C}\sum_{i=1}^M  \Pr(\theta \! = \! i) \sum_{n=1}^\infty  \E[U_i(T_n) \! - \! U_i(t_0^{(n)}) \! \mid \! \theta \! = \! i] 
    \le \left(1+2^{-C_2}\frac{1-2^{-NC_2}}{1-2^{-C_2}}\right)\frac{C_2}{C}-\frac{\mathbf{U}(Y^0)}{C} \, , \label{eq: C2 pf and EU}
\end{align}
\else
\begin{align}
    \frac{1}{C}\sum_{i=1}^M  &\Pr(\theta \! = \! i) \sum_{n=1}^\infty  \E[U_i(T_n) \! - \! U_i(t_0^{(n)}) \! \mid \! \theta \! = \! i]
    \nonumber
    &
    \\
    &\le \left(1+2^{-C_2}\frac{1-2^{-NC_2}}{1-2^{-C_2}}\right)\frac{C_2}{C}-\frac{\mathbf{U}(Y^0)}{C} \, ,& \label{eq: C2 pf and EU}
\end{align}
\fi
where $\mathbf{U}(Y^0)$ is the expected value of the log likelihood ratio of the true message according to the {\em a-priori} message distribution, i.e. from the perspective of the receiver before any symbols have been received.  Note that $\mathbf{U}(Y^0)$ is $-\log(M-1)$ for a uniform {\em a-priori} input distribution. Then, equations \eqref{eq: start proof thm 1}-\eqref{eq: C2 pf and EU} yield the following bound on $\E[T]$:
\begin{align}
    \E[T] &\le  2^{-C_2}\frac{1-2^{-NC_2}}{1-2^{-C_2}}\frac{C_2}{C}+\frac{C_2-\mathbf{U}(Y^0) }{C} \, .
    \label{eq: T upper bound}
\end{align}

A bound on the expectation $\E[\tau-T]$ can be obtained using the Markov Analysis in \cite{Yang2021}, Section V. \textit{F}. However, our analysis of  $\E[T]$ already accounts for all time spent in the communication phase, including the additional communication phases that occur after the system falls back from the confirmation phase. Accordingly, we reduce the self loop weight $\Delta_0$ in \cite{Yang2021} Sec. V \textit{F} from $\Delta_0 = 1 +\frac{C_2}{C}+\frac{\log_2(2q)}{qC}$ to $\Delta_0 = 1$. 
The resulting bound is given by:
\begin{align}
    \E[\tau-T] &\le \left(N - 2^{-C_2}\frac{1-2^{-NC_2}}{1-2^{-C_2}}\right)\frac{C_2}{C_1} \, . \label{eq: Etau-ET}
\end{align}
The inequality in \eqref{eq: Etau-ET} is not equality because, in our analysis, the transmission ends if any message $j$, other than the transmitted message $\theta$, attains $U_j(t) \ge N C_2$. However, in \cite{Yang2021} the transmission only terminates when $U_i(t)\ge N C_2$.
The upper bound on the expected stopping time $\E[\tau]$ is obtained by adding the bounds in equations \eqref{eq: T upper bound} and \eqref{eq: Etau-ET} and replacing $N$ by its definition in equation \eqref{eq: stopping rule ceiling}:
\ifCLASSOPTIONonecolumn
\begin{align}
    \E[\tau] \le&  \frac{\log_2(M-1) + C_2}{C} + \frac{ C_2}{C_1}\left \lceil \frac{\log_2\left(\frac{1-\epsilon}{\epsilon}\right)}{C_2}\right \rceil 
    + 2^{-C_2}\frac{1-\frac{\epsilon}{1-\epsilon}2^{- C_2}}{1-2^{-C_2}}
    \left(\frac{C_2}{C}-\frac{C_2}{C_1}\right) \, .
    \label{eq: stopping time C2 bound}
\end{align} 
\else
\begin{align}
    \E[\tau] \le&  \frac{\log_2(M-1) + C_2}{C} + \frac{ C_2}{C_1}\left \lceil \frac{\log_2\left(\frac{1-\epsilon}{\epsilon}\right)}{C_2}\right \rceil \nonumber
    \\
    &+ 2^{-C_2}\frac{1-\frac{\epsilon}{1-\epsilon}2^{- C_2}}{1-2^{-C_2}}
    \left(\frac{C_2}{C}-\frac{C_2}{C_1}\right) \, .
    \label{eq: stopping time C2 bound}
\end{align} 
\fi
\ifCLASSOPTIONonecolumn
Note that $\frac{C_2}{C} -\frac{C_2}{C_1} \ge 0$ and since $N = \left\lceil \frac{1}{C_2} \log_2\left(\frac{1-\epsilon}{\epsilon}\right)\right\rceil \le \frac{1}{C_2}\log_2\left(\frac{1-\epsilon}{\epsilon}\right)+1$, then:
$2^{-N C_2} \ge 2^{-\log_2\left(\frac{1-\epsilon}{\epsilon}\right)-C_2}=\frac{\epsilon}{1-\epsilon}2^{-C_2}$ which is also used in \cite{Yang2021} for a more compact upper bound expression.
\else
Note that $\frac{C_2}{C} -\frac{C_2}{C_1} \ge 0$ and since \\ $N = \left\lceil \frac{1}{C_2} \log_2\left(\frac{1-\epsilon}{\epsilon}\right)\right\rceil \le \frac{1}{C_2}\log_2\left(\frac{1-\epsilon}{\epsilon}\right)+1$, then \\
$2^{-N C_2} \ge 2^{-\log_2\left(\frac{1-\epsilon}{\epsilon}\right)-C_2}=\frac{\epsilon}{1-\epsilon}2^{-C_2}$ which is also used in \cite{Yang2021} for a more compact upper bound expression.
\fi

\end{IEEEproof}
\section{Proof of Lemmas \ref{lemma: Phase I time}-\ref{lemma: final and initial values}, Thm. \ref{theorem: surrogate martingale}, and Thm. \ref{theorem: simple rule}}
\label{sec: proof of lemmas and theorems 2 and 3}
Before proceeding to prove Lemmas  \ref{lemma: Phase I time}-\ref{lemma: final and initial values}, we will introduce a claim that will aid in some of the proofs.
\begin{claim}\label{claim: j confirmation singleton}
For the communication scheme described in Sec. \ref{sec: Achievable Rate}, the following are equivalent:
\begin{enumerate}[label=(\roman*)]
    \item $ \mid U_j(t+1)-U_j(t) \mid  = C_2$
    \item $S_0 = \{j\}$ or $S_1 = \{j\}$
\end{enumerate}
This claim implies that for constraint \eqref{eq: phase II step size} to hold, the set containing item $j$, with $U_j(t)\ge 0$, must be a singleton.

\begin{IEEEproof}
See appendix \ref{sec: proof of singleton claim}
\end{IEEEproof}
\end{claim}

\subsection{Proof of Lemmas \ref{lemma: Phase I time}-\ref{lemma: final and initial values}}
\label{sec: proof of lemmas}
\begin{IEEEproof}
[Proof of Lemma \ref{lemma: Phase I time}]
We begin by defining sets that are used in the proof.  First we define $\mathcal{E}_t^\epsilon$,  the set of sequences of length $t$ where the process has not stopped:
\begin{alignat}{3}
    &\mathcal{E}_{t}^\epsilon \triangleq  && \{y^t \! \in \{0,1\}^t \! \mid \rho_j(s) < 1 \! - \! \epsilon, \: \forall j \in \Omega, s \le t\} \, , \label{eq: E length t}
\end{alignat}
and let $\mathcal{E}^\epsilon \triangleq \cup_{t=0}^\infty \mathcal{E}_{t}^\epsilon$.

For each sequence  $y^t \in \mathcal{E}^\epsilon$, $N_i(y^t)$ is the set of time values $t^{(1)}_0, t^{(2)}_0, \dots, t^{(n)}_0 \le t$ where message $i$ begins an interval with $U_i(t) <0$.  This includes time zero and all the times $s$ where from time $s-1$ to $s$, message $i$ transitions from $U_i(s-1) \ge 0$ to $U_i(s) <0$, i.e. the decoder falls back from confirmation phase to communication phase.
\begin{alignat}{3}   
    &N_i(y^t) &&\triangleq  \{ 0\} \cup \{s \le t: U_i(s) < 0, U_i(s-1) \ge 0 \}
    \, .\label{eq: T vals}
 \end{alignat}
Now we define the set $\Yein$ of sequences $y^t$ for which the the following are all true: 1) the decoder has not stopped, 2) the decoder has entered the confirmation phase for message $i$ $n$ times, and 3) the decoder is not in the confirmation phase for message $i$ at time $t$, where the sequence ends.
\begin{alignat}{3}
   &\Yein &&\triangleq \{y^t \! \in \mathcal{E}^\epsilon \!: \! \left| N_i(y^t) \right| = n,  U_i(t)  <  0  \} \, . \label{eq: Beta round n}
\end{alignat}
For $t \ge s$, let $y^{s:t} = [y_{s+1},\dots, y_t]$ and let $y^s y^{s:t} = \text{cat}([y_1,\dots,y_s],[y_{s+1},\dots, y_t]) = [y_1,\dots,y_s,y_{s+1},\dots,y_t]$, the concatenation of the strings $y^s$ and $y^{s:t}$.
Now we define the set $\Yein(y^s)$, which is the subset sequences in $\Yein$ that have the sequence $y^s$ as a prefix.
\begin{alignat}{3}
   &\Yein(y^s) &&\triangleq \{y^t \in \mathcal{Y}_
    \epsilon^{(i,n)}: y^t = y^{s} y^{s:t}\} \label{eq: beta y round n}
\end{alignat}
As our final definition, let $\Bein$ be the set
containing only the sequences where the final received symbol $y_t$ is the symbol for which the decoder resumes the communication phase for message $i$ for the $n$th time, or the empty string, that is:
\begin{alignat}{3}
&\Bein &&\triangleq \{y^t \in \Yein \big | \; t \in N_i(y^t)\} \, .
   \label{eq: Y initials}
\end{alignat}
Each $y^t \in \Bein$, sets an initial condition for the communication phase where $U_i(t) < 0$, so that $T^{(n)} \ge 1$, that is $t$ is of the form $t^{(n)}_0$ defined in \eqref{eq: t_0 definition}.
By the property of conditional expectation, $\E[T]$ is given by:
\begin{equation}
    \E[T] = \sum_{i=1}^M \Pr(\theta = i) \E[T\mid \theta = i] \, .
\end{equation}
Now we explicitly write this expression as a function of all the possible initial conditions for each of the communication phase rounds $n$, that is, the set $\Bein$:
\ifCLASSOPTIONonecolumn
\begin{align}
    \sum_{i=1}^M \Pr(\theta = i) \E[T\mid \theta = i] \label{eq: expand T mid theta}
    &= \sum_{i=1}^M \Pr(\theta = i) \E\left[\left(\sum_{n=1}^\infty T^{(n)}\right)\bigg | \theta = i\right]
    \\
    &= \sum_{i=1}^M \Pr(\theta = i) \sum_{n=1}^\infty \E\left[T^{(n)}  \Big| \theta = i\right]  \label{eq: E T n given i}
    \\
    &= \sum_{i=1}^M \Pr(\theta \! = \! i)
    \sum_{n=1}^\infty
    \underset{y^s \in  \Bein}{\sum \Pr(Y^s} \! = \! y^s \! \mid \! \theta \! = \! i) 
    \E \Big[T^{(n)}  \Big| Y^s \! = \! y^s, \theta \! = \! i\Big] 
    \label{eq: before tail sum} \, .&
\end{align}
\else
\begin{flalign}
    \sum_{i=1}^M &\Pr(\theta = i) \E[T\mid \theta = i] \nonumber&
    \\
    &= \sum_{i=1}^M \Pr(\theta = i) \E\left[\left(\sum_{n=1}^\infty T^{(n)}\right)\bigg | \theta = i\right] 
    \label{eq: expand T mid theta}
    & 
    \\
    &= \sum_{i=1}^M \Pr(\theta = i) \sum_{n=1}^\infty \E\left[T^{(n)}  \Big| \theta = i\right]  \label{eq: E T n given i}&
    \\
    &= \sum_{i=1}^M \Pr(\theta \! = \! i)
    \sum_{n=1}^\infty
    \underset{y^s \in  \Bein}{\sum \Pr(Y^s} \! = \! y^s \! \mid \! \theta \! = \! i) 
    \nonumber
    & 
    \\
    & \quad \quad \quad \quad \quad \quad \quad \quad \quad \quad
    \cdot \E \Big[T^{(n)}  \Big| Y^s \! = \! y^s, \theta \! = \! i\Big] \label{eq: before tail sum} \, .&
\end{flalign}
\fi
Now we proceed to write the last expectation \eqref{eq: before tail sum} using the tail sum formula for expectations in \eqref{eq: P T^n} and then as an expectation of the indicator of $\{T^{(n)} > 0 \}$ in \eqref{eq: E indicator T^n}. Then, since $T^{(n)}$ is a random function of $Y^t = Y^s Y^r$, where $Y^r \in \{0,1\}^r$, given by $\mathbbm{1}_{T^{(n)} > r} =\mathbbm{1}_{Y^s Y^r \in \Yein}$,  \eqref{eq: E indicator Yein} follows:
\ifCLASSOPTIONonecolumn
\begin{align}
    \E \Big[ T^{(n)}  \Big| Y^s \! = \! y^s, \theta \! = \! i\Big] 
    &= \sum_{r=0}^\infty 
    \Pr(T^{(n)} > r | Y^s = y^s, \theta = i) \label{eq: P T^n}
    \\
    &=
     \sum_{r=0}^\infty 
    \E[\mathbbm{1}_{T^{(n)} > r}  |  Y^s = y^s, \theta = i] \label{eq: E indicator T^n}
    \\
    &=
    \sum_{r=0}^\infty 
    \E[\mathbbm{1}_{Y^{s \! + \! r} \in \Yein(y^s)} |  Y^s = y^s, \theta = i] \label{eq: E indicator Yein} \, .
\end{align}
\else
\begin{align}
    \E \Big[ & T^{(n)}  \Big| Y^s \! = \! y^s, \theta \! = \! i\Big] 
    \nonumber
    \\
    &= \sum_{r=0}^\infty 
    \Pr(T^{(n)} > r | Y^s = y^s, \theta = i) \label{eq: P T^n}
    \\
    &=
     \sum_{r=0}^\infty 
    \E[\mathbbm{1}_{T^{(n)} > r}  |  Y^s = y^s, \theta = i] \label{eq: E indicator T^n}
    \\
    &=
    \sum_{r=0}^\infty 
    \E[\mathbbm{1}_{Y^{s \! + \! r} \in \Yein(y^s)} |  Y^s = y^s, \theta = i] \label{eq: E indicator Yein} \,.
\end{align}
\fi
Expanding the expectation in \eqref{eq: E indicator Yein} we obtain \eqref{eq: expand E 1 Yein}. Since the indicator in \eqref{eq: expand E 1 Yein}  is $0$ outside $\Yein$ and $1$ inside, it is omitted in \eqref{eq: union over r}, where we have only considered values of $y^s z^r$ that intersect with $\Yein$.
Since $\cup_{r=1}^\infty \{ \{0,1\}^r \cap \Yein (y^s)\} = \Yein(y^s)$ \eqref{eq: union over r} follows. 
\ifCLASSOPTIONonecolumn
\begin{align}
    \sum_{r=0}^\infty 
    \E[\mathbbm{1}_{Y^{s \! + \! r} \in \Yein(y^s)} |  Y^s = y^s, \theta = i]
    &= \sum_{r=0}^\infty \sum_{z^r \in \{0,1\}^r}  \mathbbm{1}_{Y^{s \! + \! r} \in \Yein(y^s)} 
    \Pr(Y^{s+r} \! = \! y^s z^r 
    \mid Y^s \! = \! y^s, \theta \! = 
    \! i) \label{eq: expand E 1 Yein}&
    \\
    &= \underset{y^s z^r \in \cup_{r=1}^\infty \{ \{0,1\}^r \cap \Yein (y^s)\} }{\sum
    \Pr(Y^{s+r} \! = \! y^s z^r }
    \mid Y^s \! = \! y^s, \theta \! = 
    \! i) \label{eq: union over r}&
    \\
    &=
    \underset{y^{s+r} \in \Yein(y^s)}
    {\quad \sum
    \Pr(Y^{s+r}} = y^s z^r 
    \mid Y^s = y^s, \theta = i)
    \,. \label{eq: restrict to Yein ys}
\end{align}
\else
\begin{flalign}
    \sum_{r=0}^\infty 
    &\E[\mathbbm{1}_{Y^{s \! + \! r} \in \Yein(y^s)} |  Y^s = y^s, \theta = i]
    \nonumber
    &
    \\
    &= \sum_{r=0}^\infty \sum_{z^r \in \{0,1\}^r}  \mathbbm{1}_{Y^{s \! + \! r} \in \Yein(y^s)}& \nonumber
    \\
    & \quad \quad \quad \quad \quad \quad \cdot
    \Pr(Y^{s+r} \! = \! y^s z^r 
    \mid Y^s \! = \! y^s, \theta \! = 
    \! i) \label{eq: expand E 1 Yein}&
    \\
    &= \underset{y^s z^r \in \cup_{r=1}^\infty \{ \{0,1\}^r \cap \Yein (y^s)\} }{\sum
    \Pr(Y^{s+r} \! = \! y^s z^r }
    \mid Y^s \! = \! y^s, \theta \! = 
    \! i) \label{eq: union over r}&
    \\
    &=
    \underset{y^{s+r} \in \Yein(y^s)}
    {\quad \sum
    \Pr(Y^{s+r}} = y^s z^r 
    \mid Y^s = y^s, \theta = i)
    \,. \label{eq: restrict to Yein ys}
    &
\end{flalign}
\fi
The product of the conditional probabilities $\Pr(Y^s \! = \! y^s \mid \theta \! = \! i)$ in \eqref{eq: before tail sum} and $\Pr(Y^{s+r} = y^s z^r 
\mid Y^s = y^s, \theta = i)$ in \eqref{eq: restrict to Yein ys} is given by $\Pr(Y^{s+r} = y^s z^r 
\mid \theta = i)$.
Replacing the expectation in \eqref{eq: before tail sum} by \eqref{eq: restrict to Yein ys} the inner-most sum in \eqref{eq: before tail sum} becomes \eqref{eq: ys and Yein ys}. 
The summation in \eqref{eq: ys and Yein ys} is over $\Yein(y^s)$ for each $y^s$ in $\Bein$, which is equivalent to the sum over $\underset{y^s \in  \Bein}{ \cup}\Yein(y^s) = \Yein$ and \eqref{eq: over all Yein} follows:
\ifCLASSOPTIONonecolumn
\begin{align}
    \underset{y^s \in  \Bein}{ \sum} \Pr(Y^s \! = \! y^s \! \mid \! \theta \! = \! i) 
    \E \Big[T^{(n)}  \Big| Y^s \! = \! y^s, \theta \! = \! i\Big] 
    &= \! \underset{y^s \in  \Bein}{ \sum}
    \underset{y^{s+r} \in \Yein(y^s)}
    {\sum
    \Pr(Y^{s+r} \! = \! } y^s z^r 
    \mid Y^s \! = \! y^s, \theta \! = \! i)
    \label{eq: ys and Yein ys}
    \\
    &=
    \underset{y^t \in  \Yein}{ \sum
    \Pr(Y^t} = y^t 
    \mid \theta = i) \label{eq: over all Yein}
    \,. 
\end{align}
\else
\begin{flalign}
    &\underset{y^s \in  \Bein}{ \sum} \Pr(Y^s \! = \! y^s \! \mid \! \theta \! = \! i) 
    \E \Big[T^{(n)}  \Big| Y^s \! = \! y^s, \theta \! = \! i\Big] &
    \nonumber
    \\
    & \quad = \! \underset{y^s \in  \Bein}{ \sum}
    \underset{y^{s+r} \in \Yein(y^s)}
    {\sum
    \Pr(Y^{s+r} \! = \! } y^s z^r 
    \mid Y^s \! = \! y^s, \theta \! = \! i)&
    \label{eq: ys and Yein ys}
    \\
    & \quad =
    \underset{y^t \in  \Yein}{ \sum
    \Pr(Y^t} = y^t 
    \mid \theta = i) \label{eq: over all Yein}
    \,. &
\end{flalign}
\fi
We can now rewrite \eqref{eq: expand T mid theta} by replacing the expectation in \eqref{eq: E T n given i} by \eqref{eq: over all Yein} to obtain \eqref{eq: n and Yein}. In \eqref{eq: Yi from union Yein} the two summations are consolidated into a single sum over union over all $n$ of each $\Yein$:
\ifCLASSOPTIONonecolumn
\begin{align}
    \sum_{i=1}^M \Pr(\theta \! = \! i)
    \sum_{n=1}^\infty\sum_{r=0}^\infty \E[\mathbbm{1}_{T^{(n)} > r}  | \theta = i]
    &=
     \sum_{i=1}^M \Pr(\theta \! = \! i) \sum_{n=1}^\infty
     \underset{y^t \in  \Yein}{ \sum
    \Pr(Y^t} = y^t 
    \mid \theta = i) \label{eq: n and Yein}
     \\
     &= \sum_{i=1}^M\underset{y^t \in \cup_{n=0}^\infty \Yein}{\Pr(\theta = i) \sum \Pr(Y^t = y^t}   \mid \theta = i) \, . \label{eq: Yi from union Yein}
\end{align}
\else
\begin{align}
    \sum_{i=1}^M &\Pr(\theta \! = \! i)
    \sum_{n=1}^\infty\sum_{r=0}^\infty \E[\mathbbm{1}_{T^{(n)} > r}  | \theta = i]
    \nonumber
    \\
    &=
     \sum_{i=1}^M \Pr(\theta \! = \! i) \sum_{n=1}^\infty
     \underset{y^t \in  \Yein}{ \sum
    \Pr(Y^t} = y^t 
    \mid \theta = i) \label{eq: n and Yein}
     \\
     &= \sum_{i=1}^M\underset{y^t \in \cup_{n=0}^\infty \Yein}{\Pr(\theta = i) \sum \Pr(Y^t = y^t}   \mid \theta = i) \, . \label{eq: Yi from union Yein}
\end{align}
\fi
To conclude the proof, note that the union $\cup_{n=0}^\infty \Yein$ is the set $\Yei$ defined in the statement of the Lemma \ref{lemma: Phase I time}.
\end{IEEEproof}



\begin{IEEEproof}[Proof of Lemma \ref{lemma: sum C bound}]
Define the set $\mathcal{A}_\epsilon$ by:
\begin{equation}
    \mathcal{A}_\epsilon \triangleq \{y^t \in \mathcal{Y}_\epsilon^{(i)}: \rho_j(y^t) < \frac{1}{2} \; \forall j = 1,\dots,M\} \, , \label{eq: set A definition}
\end{equation}
where $\mathcal{A}_\epsilon$ does not depend on $i$.
Let the set $\mathcal{Y}_\epsilon^{(i)}$ be partitioned into $\mathcal{A}_\epsilon$ and $\mathcal{Y}_\epsilon^{(i)}\setminus \mathcal{A}_\epsilon$.
Then, we can split the sum in the left side of \eqref{eq: inside A}, which is the left side of \eqref{eq: Vi M wide sum} in Lemma \ref{lemma: sum C bound}, into a sum over $\mathcal{A}_\epsilon$, right side of \eqref{eq: inside A}, and a sum over the sets $\mathcal{Y}_\epsilon^{(i)}\setminus \mathcal{A}_\epsilon$, expression \eqref{eq: outside A} as follows:
\ifCLASSOPTIONonecolumn
\begin{align}
    \sum_{i = 1}^M \Pr(\theta = i)\sum_{y^t \in \mathcal{Y}_\epsilon^{(i)}}\Pr(Y^t = y^t \mid \theta = i)V_i(y^t)
    =&
    \sum_{i = 1}^M \Pr(\theta = i) \sum_{y^t \in \mathcal{A}_\epsilon} \Pr(Y^t = y^t \mid \theta = i)V_i(y^t)
    \label{eq: inside A}
    \\
    & \quad \quad +
    \sum_{i = 1}^M \underset{y^t \in \mathcal{Y}_\epsilon^{(i)}\setminus \mathcal{A}_\epsilon}{\Pr(\theta = i)\sum\Pr(Y^t = y^t} \mid \theta = i)V_i(y^t) \, .
    \label{eq: outside A}&
\end{align}
\else
\begin{flalign}
    \sum_{i = 1}^M \Pr&(\theta = i)\sum_{y^t \in \mathcal{Y}_\epsilon^{(i)}}\Pr(Y^t = y^t \mid \theta = i)V_i(y^t) 
    \nonumber&
    \\
    =&
    \sum_{i = 1}^M \Pr(\theta = i) \sum_{y^t \in \mathcal{A}_\epsilon} \Pr(Y^t = y^t \mid \theta = i)V_i(y^t)
    \label{eq: inside A}&
    \\
    +&
    \sum_{i = 1}^M \underset{y^t \in \mathcal{Y}_\epsilon^{(i)}\setminus \mathcal{A}_\epsilon}{\Pr(\theta = i)\sum\Pr(Y^t = y^t} \mid \theta = i)V_i(y^t) \, .
    \label{eq: outside A}&
\end{flalign}
\fi
For $y^t \in \mathcal{Y}_\epsilon^{(i)}\setminus \mathcal{A}_\epsilon: \; \exists j\neq i$ s.t. $U_j(t) \ge 0$ and $U_i(t) < 0$. Then $S_0 = \{j\}$ by constraint \eqref{eq: phase II step size} and \emph{Claim \eqref{claim: j confirmation singleton}}, and therefore $\Delta > 0$, (see the proof of Thm. \ref{theorem: simple rule}, for definition of $\Delta$). By equation \eqref{eq: jensen over S0} with $\Delta \ge 0$, this results in $\E[U_i(t+1)-U_i(t) \mid Y^t = y^t, \theta = i] \ge C$ for all $i$. 
Note that $\{U_j(t) \ge 0\} \cap \{ S_0 = \{j\}\}$ means that in this case the SED constraint \eqref{eq: sed rule} is satisfied.
It suffices to show the bound holds also for \eqref{eq: inside A}. The product of conditional probabilities: $\Pr(\theta =i)$ and $\Pr(Y^t = y^t \mid \theta = i)$ in \eqref{eq: inside A} is equal to $\Pr(Y^t = y^t, \theta \! = \! i)$ and can be factored into $\Pr(Y^t = y^t)\Pr(\theta = i \mid Y^t = y^t)$. Since $0 < V_i(y^t) \le C_2$ and $\mathcal{A}_\epsilon$ does not depend on $i$, then the summation order in  \eqref{eq: inside A} can be reversed to obtain:
\ifCLASSOPTIONonecolumn
\begin{align}
    \sum_{i = 1}^M \sum_{y^t \in \mathcal{A}_\epsilon}  \! \Pr(Y^t = y^t)\Pr(\theta = i \mid Y^t = y^t)V_i(y^t)
    &\underset{y^t \in \mathcal{A}_\epsilon}{=\sum} \! \Pr(Y^t = y^t)
    \sum_{i = 1}^M \! \Pr(\theta = i \mid Y^t = y^t)V_i(y^t) \label{eq: isolate y}
\end{align}
\else
\begin{align}
    \sum_{i = 1}^M & \sum_{y^t \in \mathcal{A}_\epsilon}  \! \Pr(Y^t = y^t)\Pr(\theta = i \mid Y^t = y^t)V_i(y^t)
    \nonumber
    \\
    &\underset{y^t \in \mathcal{A}_\epsilon}{=\sum} \! \Pr(Y^t = y^t)
    \sum_{i = 1}^M \! \Pr(\theta = i \mid Y^t = y^t)V_i(y^t) \label{eq: isolate y}
\end{align}
\fi
The probability $\Pr(\theta = i \mid Y^t = y^t)$ in \eqref{eq: isolate y} is just $\rho_i(y^t)$ and using the definition of $V_i(y^t)$ \eqref{eq: Vi by def} follows. In \eqref{eq: form theorem 1} $\rho_i(y^t)$ is moved inside the expectation, to obtain the form in constraint \eqref{eq: C step size} of Thm. \ref{theorem: Main Theorem}: 
\ifCLASSOPTIONonecolumn
\begin{align}
    \sum_{i = 1}^M \Pr(\theta = i \mid Y^t = y^t)V_i(y^t)
    &=
    \sum_{i = 1}^M \! \rho_i( y^t)\E[U_i(t \! + \! 1) \! - \! U_i(t) 
    \!\mid \! Y^t = y^t, \theta = i]
    \label{eq: Vi by def}
    \\
    &=
    \sum_{i = 1}^M \E[\rho_i(y^t)(U_i(t \! + \! 1) \! - \! U_i(t) )
    \!\mid \! Y^t = y^t, \theta = i]
    \label{eq: form theorem 1}
    \\
    &\ge C
\end{align}
\else
\begin{align}
    \sum_{i = 1}^M & \Pr(\theta = i \mid Y^t = y^t)V_i(y^t) \nonumber
    \\
    &=
    \sum_{i = 1}^M \! \rho_i( y^t)\E[U_i(t \! + \! 1) \! - \! U_i(t) 
    \!\mid \! Y^t = y^t, \theta = i]
    \label{eq: Vi by def}
    \\
    &=
    \sum_{i = 1}^M \E[\rho_i(y^t)(U_i(t \! + \! 1) \! - \! U_i(t) )
    \!\mid \! Y^t = y^t, \theta = i]
    \label{eq: form theorem 1}
    \\
    &\ge C
\end{align}
\fi
Constraint \eqref{eq: C step size} dictates that \eqref{eq: form theorem 1} is lower bounded by $C$ and \eqref{eq: apply theorem 1} follows. Then we multiply by $1 = \sum_{i =1}^M { \rho_i(y^t)}$ to produce \eqref{eq: multiply by 1 or sum rho}. In \eqref{eq: ro to P} note that $\rho_i(y^t) = \Pr(\theta = i\mid Y^t = y^t)$ and the product $\Pr(\theta = i\mid Y^t = y^t)\Pr(Y^t=y^t)$ is given $\Pr(Y^t=y^t, \theta = i) = \Pr(Y^t=y^t \mid \theta = i)\Pr(\theta = i)$. This is used to obtain \eqref{eq: ro to P} and then \eqref{eq: end lemma 2}:
\ifCLASSOPTIONonecolumn
\begin{align}
   \underset{y^t \in \mathcal{A}_\epsilon}{\sum} \! \Pr(Y^t = y^t)
    \sum_{i = 1}^M \! \Pr(\theta = i \mid Y^t = y^t)V_i(y^t)
    &\ge \sum_{y^t \in \mathcal{A}_\epsilon}\Pr(Y^t = y^t) C
    \label{eq: apply theorem 1}
    \\ 
    &= C\sum_{y^t \in \mathcal{A}_\epsilon}\Pr(Y^t = y^t)\sum_{i =1}^M { \rho_i(y^t)} \label{eq: multiply by 1 or sum rho}
    \\
    &= C\sum_{i = 1}^M \sum_{y^t \in \mathcal{A}_\epsilon} \Pr(Y^t = y^t, \theta = i)
    \label{eq: ro to P}
    \\
    &=C\sum_{i = 1}^M \Pr(\theta = i) \sum_{y^t \in \mathcal{A}_\epsilon} \Pr(Y^t = y^t \mid \theta = i) \, . \label{eq: end lemma 2}
\end{align}
\else
\begin{flalign}
   \underset{y^t \in \mathcal{A}_\epsilon}{\sum} \! &\Pr(Y^t = y^t)
    \sum_{i = 1}^M \! \Pr(\theta = i \mid Y^t = y^t)V_i(y^t)&
   \nonumber
    \\
    &\ge \sum_{y^t \in \mathcal{A}_\epsilon}\Pr(Y^t = y^t) C&
    \label{eq: apply theorem 1}
    \\ 
    &= C\sum_{y^t \in \mathcal{A}_\epsilon}\Pr(Y^t = y^t)\sum_{i =1}^M { \rho_i(y^t)}& \label{eq: multiply by 1 or sum rho}
    \\
    &= C\sum_{i = 1}^M \sum_{y^t \in \mathcal{A}_\epsilon} \Pr(Y^t = y^t, \theta = i)&
    \label{eq: ro to P}
    \\
    &=C\sum_{i = 1}^M \Pr(\theta = i) \sum_{y^t \in \mathcal{A}_\epsilon} \Pr(Y^t = y^t \mid \theta = i) \label{eq: end lemma 2}
    \, . &
\end{flalign}
\fi
In both \eqref{eq: inside A} and \eqref{eq: outside A} replacing $V_i(y^t)$ by $C$ provide and upper bound on the original expression. Combining the two upper bounds we recover the Lemma.
\end{IEEEproof}


\begin{IEEEproof}[Proof of Lemma \ref{lemma: stopping time bound}]
We start writing, in the left side of \eqref{eq: split n and Yein}, the sum in the left side of equation \eqref{eq: telescopic equality} of Lemma \ref{lemma: stopping time bound},
using an equivalent form for $\Yei$, which is $\cup_{n=0}^\infty \Yein$. This equivalent form was also used in the proof of Lemma \ref{lemma: Phase I time}. Then in \eqref{eq: split n and Yein} we break it into two summations, first over $n$ and then over $\Yein$:
\ifCLASSOPTIONonecolumn
\begin{align}
    \sum_{i = 1}^M \Pr(\theta \! = \! i) \underset{y^t \in \cup_{n=0}^\infty \Yein}{\sum V_i(y^t)
    }\Pr(Y^t \! = \! y^t \mid \theta = i) 
    =&\sum_{i = 1}^M \Pr(\theta \! = \! i) \sum_{n=1}^\infty \underset{y^t \in \Yein}{\sum V_i}(y^t) \Pr(Y^t \! = \! y^t \mid \theta \! = \! i) \label{eq: split n and Yein}
    \, .
\end{align}
\else
\begin{flalign}
    \sum_{i = 1}^M &\Pr(\theta \! = \! i) \underset{y^t \in \cup_{n=0}^\infty \Yein}{\sum V_i(y^t)
    }\Pr(Y^t \! = \! y^t \mid \theta = i) 
    \nonumber
    & 
    \\
    =&\sum_{i = 1}^M \Pr(\theta \! = \! i) \sum_{n=1}^\infty \underset{y^t \in \Yein}{\sum V_i}(y^t) \Pr(Y^t \! = \! y^t \mid \theta \! = \! i) \label{eq: split n and Yein}
    \, .&
\end{flalign}
\fi
The set $\Yein$ is a subset of $\cup_{t=0}^\infty \{0,1\}^t$ and therefore can be expressed a union of all the intersections over $n$: $\Yein=\cup_{t=0}^\infty \{\Yein \cap \{0,1\}^t\}$. We use this new form 
to rewrite the inner sum in \eqref{eq: split n and Yein} as the left side of \eqref{eq: indicator of Yein}. Then, we remove the intersections with $\Yein$ by using its indicator in the right side of \eqref{eq: indicator of Yein}:
\ifCLASSOPTIONonecolumn
\begin{align}
    \sum_{t=0}^\infty  \underset{y^t \in \Yein \cap \{0,1\}^t}{\sum} V_i(y^t) \Pr(Y^t = y^t \mid \theta = i)
    =&\sum_{t=0}^\infty 
    \sum_{y^t \in \{0,1\}^t} \mathbbm{1}_{y^t \in \Yein}V_i(y^t) \Pr(Y^t \! = \! y^t \mid \theta \! = \! i)
    \label{eq: indicator of Yein}
    \, .
\end{align}
\else
\begin{flalign}
    \sum_{t=0}^\infty & \underset{y^t \in \Yein \cap \{0,1\}^t}{\sum} V_i(y^t) \Pr(Y^t = y^t \mid \theta = i)& \nonumber
    \\
    =&\sum_{t=0}^\infty 
    \sum_{y^t \in \{0,1\}^t} \mathbbm{1}_{y^t \in \Yein}V_i(y^t) \Pr(Y^t \! = \! y^t \mid \theta \! = \! i)
    \label{eq: indicator of Yein}
    \, .&
\end{flalign}
\fi
Recall that $V_i(y^t)=\E\left[ U_i(t+1) \! - \! U_i(t) \mid \! Y^t=y^t, \theta \! = \! i\right]$ from \eqref{eq: Viyt definition}. Also recall from \eqref{eq: U process} that $U_i(t)=U_i(Y^t)$, a random function of $Y^t$. Let $D_i(Y^{t+1})\triangleq U_i(Y^{t+1})-U_i(Y^t)$, then 
we expand $V_i(y^t)$ as:
\ifCLASSOPTIONonecolumn
\begin{align}
    V_i(y^t) &= \E[U_i(t+1)-U_i(t) \mid Y^t = y^t, \theta = i]
    \underset{z \in \{0,1\}}
    {=\sum D_i}(Y^{t+1})\Pr(Y_{t+1} \! = \! z \mid Y^t \! = \! y^t, \theta \! = \! i)
    \, .& \label{eq: Diyt introduction}
\end{align}
\else
\begin{flalign}
    V_i(y^t) &= \E[U_i(t+1)-U_i(t) \mid Y^t = y^t, \theta = i] \nonumber
    &
    \\
    & \underset{z \in \{0,1\}}
    {=\sum D_i}(Y^{t+1})\Pr(Y_{t+1} \! = \! z \mid Y^t \! = \! y^t, \theta \! = \! i)
    \, .& \label{eq: Diyt introduction}
\end{flalign}
\fi
The product of the probabilities in \eqref{eq: indicator of Yein} and \eqref{eq: Diyt introduction} is given by $\Pr(Y^{t+1}= y^{t}z \mid \theta = i)$. Replacing $V_i(y^t)$ in \eqref{eq: indicator of Yein} using \eqref{eq: Diyt introduction} we obtain the left side of \eqref{eq: E Di}. The equality in \eqref{eq: E Di} follows by definition of expectation:
\ifCLASSOPTIONonecolumn
\begin{align}
    \sum_{t=0}^\infty
    \underset{y^{t+1} \in \{0,1\}^{t+1}}{ \quad \sum  D_i(y^{t+1})}\mathbbm{1}_{y^t \in \Yein} \Pr(Y^{t+1} \! = \! y^{t+1} \mid \theta \! = \! i)
    &=
    \sum_{t=0}^\infty \E[ D_i(Y^{t+1})\mathbbm{1}_{Y^t \in \Yein}\mid \theta \! = \! i]
     \label{eq: E Di} 
    \, .
\end{align}
\else
\begin{flalign}
    \sum_{t=0}^\infty &
    \underset{y^{t+1} \in \{0,1\}^{t+1}}{ \quad \sum  D_i(y^{t+1})}\mathbbm{1}_{y^t \in \Yein} \Pr(Y^{t+1} \! = \! y^{t+1} \mid \theta \! = \! i)& \nonumber
    \\
    &=
    \sum_{t=0}^\infty \E[ D_i(Y^{t+1})\mathbbm{1}_{Y^t \in \Yein}\mid \theta \! = \! i]
     \label{eq: E Di} 
    \, .&
\end{flalign}
\fi
We expand $D_i(Y^t)$ using its definition to write  \eqref{eq: E Di} as the left side of \eqref{eq: expectation of sum E Vt} and use linearity of expectations to the equality \eqref{eq: expectation of sum E Vt}. The indicator $\mathbbm{1}_{Y^t \in \Yein}$ is zero before time $t=t^{(n)}_0$ and after time $t=t^{(n)}_0+T^{(n)}-1$, and is one in between. Accordingly, in \eqref{eq: telescopic sum} we adjust the limits of summation and remove the indicator function.
Note that the times $t_0^{(n)}$ and $T^{(n)}$ are themselves random variables. Lastly, observe that \eqref{eq: telescopic sum} is a telescopic sum that is replaced by the end points in \eqref{eq: Tn expectation}:
\ifCLASSOPTIONonecolumn
\begin{align}
    \sum_{t=0}^\infty \E
    \left[ \left(U_i\left(Y^{t+1}\right) \! - \! U_i\left(Y^t\right)\right)
    \mathbbm{1}_{Y^t \in \Yein} 
    \big | \theta \! = \! i\right]
    &= \E\left[\sum_{t=0}^\infty
    \left(U_i\left(Y^{t+1}\right) \! - \! U_i\left(Y^t\right)\right)
    \mathbbm{1}_{Y^t \in \Yein}
    \Bigg | \theta = i \right ]
    \label{eq: expectation of sum E Vt}
    \\
    &= \E\left[\sum_{t = t_0^{(n)}}^{T^{(n)}+t_0^{(n)}-1}(U_i(Y^{t+1}) \! - \! U_i(Y^t)) 
    \Bigg | \theta = i \right ] \label{eq: telescopic sum}
    \\
    &= \E\left [ U_i \left ( t_0^{(n)}+T^{(n)} \right ) - U_i \left ( t_0^{(n)}\right )  \big | \theta = i \right ] 
    \label{eq: Tn expectation}
    \,.
\end{align}
\else
\begin{flalign}
    \sum_{t=0}^\infty &\E
    \left[ \left(U_i\left(Y^{t+1}\right) \! - \! U_i\left(Y^t\right)\right)
    \mathbbm{1}_{Y^t \in \Yein} 
    \big | \theta \! = \! i\right]
    \nonumber&
    \\
    &= \E\left[\sum_{t=0}^\infty
    \left(U_i \! \left(Y^{t+1}\right) \! - \! U_i\left(Y^t\right)\right)
    \! \mathbbm{1}_{Y^t \in \Yein}
    \Bigg | \theta \! = \! i \right ]
    \label{eq: expectation of sum E Vt}&
    \\
    &= \E\left[\sum_{t = t_0^{(n)}}^{T^{(n)}+t_0^{(n)}-1}(U_i(Y^{t+1}) \! - \! U_i(Y^t)) 
    \Bigg | \theta = i \right ] \label{eq: telescopic sum}&
    \\
    &= \E\left [ U_i \left ( t_0^{(n)}+T^{(n)} \right ) - U_i \left ( t_0^{(n)}\right )  \big | \theta = i \right ] 
    \label{eq: Tn expectation}
    \,.&
\end{flalign}
\fi
 To conclude the proof, we replace the inner most summation in \eqref{eq: split n and Yein} with \eqref{eq: Tn expectation}:
\ifCLASSOPTIONonecolumn
\begin{align}
    \sum_{i = 1}^M \Pr(\theta \! = \! i) \sum_{n=1}^\infty \underset{y^t \in \Yein}{\sum V_i}(y^t) \Pr(Y^t \! = \! y^t \mid \theta \! = \! i)
    \label{eq: original inner sum lemma 3}
    =& \sum_{i = 1}^M \Pr(\theta \! = \! i) \! \sum_{n=1}^\infty \!
    \E\left [ U_i \left ( \! t_0^{(n)} \! + \! T^{(n)} \! \right ) \! - \! U_i \left ( t_0^{(n)}\right ) \! \mid \! \theta \! = \! i \right ] 
    \, .
\end{align}
\else
\begin{flalign}
    \sum_{i = 1}^M &\Pr(\theta \! = \! i) \sum_{n=1}^\infty \underset{y^t \in \Yein}{\sum V_i}(y^t) \Pr(Y^t \! = \! y^t \mid \theta \! = \! i)
    \label{eq: original inner sum lemma 3}& 
    \\
    =& \sum_{i = 1}^M \Pr(\theta \! = \! i) \! \sum_{n=1}^\infty \!
    \E\left [ U_i \left ( \! t_0^{(n)} \! + \! T^{(n)} \! \right ) \! - \! U_i \left ( t_0^{(n)}\right ) \! \mid \! \theta \! = \! i \right ] 
    \nonumber
    \, .&
\end{flalign}
\fi
\end{IEEEproof}


\begin{IEEEproof}[Proof of Lemma \ref{lemma: fall back probability}] The confirmation phase starts at a time $t$ of the form $T_n$ defined in \eqref{eq: T_n definition}, at which the transmitted message $i$ attains $U_i(T_n) \ge 0$ and $U_i(T_n-1) < 0$. 
Then, like the product martingale in  \cite{almaEM2019}, the process $\zeta_i(t)$, $t \ge T_n$, is a martingale respect to $\mathcal{F}_t = \sigma(Y^t)$, where:
\begin{align}
    \zeta_i(t) = \left(\frac{p}{q}\right)^{\frac{U_i(t)}{C_2}} \, .
\end{align}
Note that $U_i(t)$ is a biased random walk, see the Markov Chain in \cite{Yang2021}, with $U_i(t) = U_i(T_n) + \sum_{m=T_n}^t\xi_m$, where $\xi_m$ is an R.V. distributed according to:
\begin{align}
    \xi_m = 
    \begin{cases}
    +C_2 \quad \text{w.p. } q
    \\
    -C_2 \quad \text{w.p. } p
    \end{cases}
    \, ,
\end{align}
We verify that $\E[\zeta_i(t+1)\mid \mathcal{F}_t] = \zeta_i(t)$ as follows:
\ifCLASSOPTIONonecolumn
\begin{align}
    \E[\zeta_i(t+1)\mid \mathcal{F}_t] 
    &= \zeta_i(t)\left(p \left(\frac{p}{q}\right)^{-1}+q\left(\frac{p}{q}\right)^{1}\right)
    = \zeta_i(t)\left(p+q\right)= \zeta_i(t) 
    \, .
\end{align}
\else
\begin{align}
    \E[\zeta_i(t+1)\mid \mathcal{F}_t] 
    &= \zeta_i(t)\left(p \left(\frac{p}{q}\right)^{-1}+q\left(\frac{p}{q}\right)^{1}\right)
    \nonumber
    \\
    &= \zeta_i(t)\left(p+q\right)= \zeta_i(t) 
    \, .
\end{align}
\fi
Let $S_n$ be the time at which decoding either terminates at $U_i(t) = U_i(T_n)+N C_2$, or a fall back occurs, when $U_i(t) = U_i(T_n)-C_2 < 0$, that is $S_n \triangleq \min\{t \ge T_n: U_i(t) \in \{U_i(T_n)-C_2,  U_i(T_n)+ N C_2\}\}$. Then, the process $\zeta_i(t\wedge S_n)$ is a two side bounded martingale and:
\begin{align}
    \E[\zeta_i(S_n)] &= p_f\left(\frac{p}{q}\right)^{\frac{U_i(T_n)}{C_2}-1} \! \! + (1 \! - \! p_f) \left(\frac{p}{q}\right)^{\frac{U_i(T_n)}{C_2}+N} \label{eq: Ezeta Sn from Tn-1 and Tn+N}
    \\
    \E[\zeta_i(T_n)]&= \left(\frac{p}{q}\right)^{\frac{U_i(T_n)}{C_2}}
    \label{eq: Ezeta Tn}
\end{align}
By Doob's optional stopping theorem \cite{almaOST2019}, $\E[\zeta_i(S_n)]$ is equal to $\E[\zeta_i(T_n)]$ . Let the fall back probability be $p_f \triangleq \Pr(U_i(S_n) = U_i(T_n)-C_2\mid t = T_n)$, then we can solve for $p_f$ using equations \eqref{eq: Ezeta Sn from Tn-1 and Tn+N} and \eqref{eq: Ezeta Tn} by setting both right sides equal. In \eqref{eq: factor Tn C2} we factor out and cancel $\left(p/q\right)^{U_i(T_n)/C_2}$. In \eqref{eq: collect pf} we collect the terms with factor $p_f$ and in \eqref{eq: solve pf} we solve for $p_f$.
\begin{align}
    1 &= p_f \frac{q}{p}+(1-p_f)\left(\frac{p}{q}\right)^N 
     \label{eq: factor Tn C2}
    \\
    0 &= p_f \frac{q}{p}\left(1-\left(\frac{p}{q}\right)^{N+1}\right) - \left(1 -\left(\frac{p}{q}\right)^N\right)
    \label{eq: collect pf}
    \\
    p_f &= \frac{p}{q}\frac{1-\left(\frac{p}{q}\right)^N}{1-\left(\frac{p}{q}\right)^{N+1}}
    \label{eq: solve pf}
    \, .
\end{align}
Since $p_f$ is just a function of $N$ and $p$, then it is the same constant across all messages $i = 1,\dots,M$ and indexes $n = 1,\dots$. 
To complete the proof we use the definition of $C_2$ from equation \eqref{eq: C2 def}, which is $C_2=\log_2\left(\frac{q}{p}\right)$ and express $p_f$ in terms of $C_2$:
\ifCLASSOPTIONonecolumn
\begin{align}
    p_f &= 2^{-\log_2(\frac{q}{p})}\frac{1-2^{-N\log_2(\frac{q}{p})}}{1-2^{-(N+1)\log_2(\frac{q}{p})}}
    = 2^{-C_2}\frac{1-2^{-NC_2}}{1-2^{-(N+1)C_2}} 
\end{align}
\else
\begin{align}
    p_f &= 2^{-\log_2(\frac{q}{p})}\frac{1-2^{-N\log_2(\frac{q}{p})}}{1-2^{-(N+1)\log_2(\frac{q}{p})}} \nonumber
    \\
    &= 2^{-C_2}\frac{1-2^{-NC_2}}{1-2^{-(N+1)C_2}} 
\end{align}
\fi
\end{IEEEproof}



\begin{IEEEproof}[Proof of Lemma \ref{lemma: final and initial values}]
We start by conditioning the expectation in the left side of equation \eqref{eq: mid lemma 5 pf}, in Lemma \ref{lemma: final and initial values}, on the events $\{T^{(n)} > 0, \theta = i\}$, $\{T^{(n)} = 0, \theta = i\}$, and  $\{T^{(n)} < 0, \theta = i\}$, whose union results in the original conditioning event, $\{\theta = i\}$, to express the original conditional probability using Bayes rule:
\ifCLASSOPTIONonecolumn
\begin{align}
    \E[&U_i(T_{n})  - U_i(t_0^{(n)}) \mid \theta = i]    
    =\E[U_i(T_n) \!  - \!  U_i(t_0^{(n)}) \! \mid \! T^{(n)} \! > \! 0, \theta \! = \! i] \Pr(T^{(n)} \! > \! 0 \mid \! \theta \! = \! i)
     \label{eq: T^n cases}
    \\
    & +\E[U_i(T_n) \!  - \!  U_i(t_0^{(n)}) \! \mid \! T^{(n)} \! = \! 0, \theta \! = \! i] \Pr(T^{(n)} \! = \! 0 \mid \! \theta \! = \! i)
     +\E[U_i(T_n) \!  - \!  U_i(t_0^{(n)}) \! \mid \! T^{(n)} \! < \! 0, \theta \! = \! i] \Pr(T^{(n)} \! < \! 0 \mid \! \theta \! = \! i) \, . \nonumber
\end{align}
\else
\begin{flalign}
    &\E[U_i(T_{n})  - U_i(t_0^{(n)}) \mid \theta = i]     \label{eq: T^n cases}&
    \\
    &=\E[U_i(T_n) \!  - \!  U_i(t_0^{(n)}) \! \mid \! T^{(n)} \! > \! 0, \theta \! = \! i] \Pr(T^{(n)} \! > \! 0 \mid \! \theta \! = \! i)&
    \nonumber
    \\
    &+\E[U_i(T_n) \!  - \!  U_i(t_0^{(n)}) \! \mid \! T^{(n)} \! = \! 0, \theta \! = \! i] \Pr(T^{(n)} \! = \! 0 \mid \! \theta \! = \! i)& \nonumber
    \\
    &+\E[U_i(T_n) \!  - \!  U_i(t_0^{(n)}) \! \mid \! T^{(n)} \! < \! 0, \theta \! = \! i] \Pr(T^{(n)} \! < \! 0 \mid \! \theta \! = \! i) \, . & \nonumber
\end{flalign}
\fi
Note that $T^{(n)}$ is non-negative, thus  the last term in the right side of \eqref{eq: T^n cases} vanishes as $\Pr(T^{(n)} < 0 ) = 0$. When $T^{(n)}=0$, then $T_n=t_0^{(n)}+T^{(n)} = t_0^{(n)}$ so that $U_i(T_n) = U_i(t_0^{(n)})$. Therefore, the second term in the right side of \eqref{eq: T^n cases} also vanishes, leaving only the first term conditioned on $\{T^{(n)} > 0, \theta = i\}$. Let $\mathcal{C}(t^{(n)}_0)$ be the event that message $i$ enters confirmation after time $t^{(n)}_0$, rather than another message $j\neq i$ ending the process by attaining $U_j(t) \ge \log_2(1-\epsilon)-\log_2(\epsilon)$, that is: $\mathcal{C}(t^{(n)}_0)\triangleq \{\exists t > t^{(n)}_0: U_i(t) \ge 0\}$. Then, the probability $\Pr(T^{(n+1)} \ge 0\mid \theta = i)$ can be expressed as:
\ifCLASSOPTIONonecolumn
\begin{align}
    \Pr(T^{(n+1)} \ge 0\mid \theta = i)
    =\Pr(T^{(n+1)} \ge 0 \mid T^{(n)} > 0, \mathcal{C}(t^{(n)}_0), \theta = i) 
    \label{eq: Tn given n-1, C}
    \Pr( T^{(n)} > 0, \mathcal{C}(t^{(n)}_0) \mid \theta = i)
    \,.
\end{align}
\else
\begin{flalign}
    \Pr(& T^{(n+1)} \ge 0\mid \theta = i)& \nonumber
    \\
    &=\Pr(T^{(n+1)} \ge 0 \mid T^{(n)} > 0, \mathcal{C}(t^{(n)}_0), \theta = i) 
    \label{eq: Tn given n-1, C}&
    \\
    &\quad \quad \quad \quad \quad \quad \, \cdot \Pr( T^{(n)} > 0, \mathcal{C}(t^{(n)}_0) \mid \theta = i)
    \,. & \nonumber
\end{flalign}
\fi
Note that the first probability in the right side of \eqref{eq: Tn given n-1, C} is just the fall back probability $p_f$ computed in Lemma \ref{lemma: fall back probability}. The last probability in \eqref{eq: Tn given n-1, C} can be also expressed as a product of conditional probabilities, see \eqref{eq: Tn given CTn and P CTn}. In \eqref{eq: Tn given CTn and P CTn} note that event $\mathcal{C}(t^{(n)}_0)$ is the event that an $n$-th confirmation phase phase occurs, which implies that a preceding $n$-th communication phase round occurs. Then, $\mathcal{C}(t^{(n)}_0) \implies T^{(n)} > 0$ and the first factor in the product of probabilities in \eqref{eq: Tn given CTn and P CTn} vanishes:
\ifCLASSOPTIONonecolumn
\begin{align}
    \Pr ( T^{(n)} > 0, \mathcal{C}(t^{(n)}_0) \mid \theta = i)
    &= \Pr( T^{(n)} \! > \! 0 \mid \mathcal{C}(t^{(n)}_0), \theta \! = \! i) \Pr(\mathcal{C}(t^{(n)}_0) \mid \theta \! = \! i) \label{eq: Tn given CTn and P CTn}
    =\Pr(\mathcal{C}(t^{(n)}_0) \mid \theta = i)
    \,.
\end{align}
\else
\begin{flalign}
    \Pr &( T^{(n)} > 0, \mathcal{C}(t^{(n)}_0) \mid \theta = i) 
    \nonumber&
    \\
    &= \Pr( T^{(n)} \! > \! 0 \mid \mathcal{C}(t^{(n)}_0), \theta \! = \! i) \Pr(\mathcal{C}(t^{(n)}_0) \mid \theta \! = \! i)& \nonumber
    \\
    &=\Pr(\mathcal{C}(t^{(n)}_0) \mid \theta = i)
    \,.
    \label{eq: Tn given CTn and P CTn}&
\end{flalign}
\fi
Combining \eqref{eq: Tn given n-1, C} and \eqref{eq: Tn given CTn and P CTn} we obtain:
\begin{align}
    \Pr( T^{(n+1)} \! > \! 0 \mid \theta \! = \! i)=\Pr(\mathcal{C}(t^{(n)}_0) \mid \theta = i) p_f 
    \label{eq: P Tn C Tn}\,.
\end{align}
We can also bound $\Pr(\mathcal{C}(t^{(n)}_0) \mid \theta = i)$ as follows:
\ifCLASSOPTIONonecolumn
\begin{align}
    \Pr(\mathcal{C}(t^{(n)}_0) \mid \theta = i) 
    &=\Pr(\mathcal{C}(t^{(n)}_0) \mid T^{(n)} \! > \! 0, \theta \! = \! i)\Pr(T^{(n)} \!  > \! 0 \mid \theta \! = \! i)
    \le \Pr(T^{(n)} \!  > \! 0 \! \mid \theta \! = \! i) \, . \label{eq: C Tn bound}
\end{align}
\else
\begin{flalign}
    \Pr&(\mathcal{C}(t^{(n)}_0) \mid \theta = i) 
    \nonumber&
    \\
    &=\Pr(\mathcal{C}(t^{(n)}_0) \mid T^{(n)} \! > \! 0, \theta \! = \! i)\Pr(T^{(n)} \!  > \! 0 \mid \theta \! = \! i)
    \nonumber&
    \\
    &\le \Pr(T^{(n)} \!  > \! 0 \! \mid \theta \! = \! i) \, . \label{eq: C Tn bound}&
\end{flalign}
\fi
Then, we can recursively bound  $\Pr( T^{(n+1)} \! > \! 0 \mid \theta \! = \! i)$ by $\Pr( T^{(n)} \! > \! 0 \mid \theta \! = \! i)p_f$ using \eqref{eq: P Tn C Tn} and \eqref{eq: C Tn bound}. For $n \ge 1$, this results in the general bound:
\begin{equation}
    \Pr(T^{(n)} \!  \ge \!  0 \mid \theta \! = \! i) \le p_f^{n \! - \! 1} 
    \label{eq: recursive P Tn bound}\,.
\end{equation}
Using \eqref{eq: recursive P Tn bound} we can bound the expectation $\E[U_i(T_{n})  - U_i(t_0^{(n)}) \mid \theta = i]$ in the left side of \eqref{eq: T^n cases} by:
\ifCLASSOPTIONonecolumn
\begin{align}
    \E[U_i(T_{n}) - U_i(t_0^{(n)}) \mid \theta = i]
    & \le \E[U_i(T_n) \!  - \!  U_i(t_0^{(n)}) \! \mid \! T^{(n)} \! > \! 0, \theta \! = \! i] p^{n-1}_f
    \label{eq: E dif U bound}
    \, .
\end{align}
\else
\begin{align}
    \E[U_i(&T_{n}) - U_i(t_0^{(n)}) \mid \theta = i] \nonumber
    \\
    & \le \E[U_i(T_n) \!  - \!  U_i(t_0^{(n)}) \! \mid \! T^{(n)} \! > \! 0, \theta \! = \! i] p^{n-1}_f
    \label{eq: E dif U bound}
    \, .
\end{align}
\fi
The value of $U_i(t^{(n)}_0)$ at $n=1$, when $t^{(1)}_0=0$ is a constant that can be directly computed for every $i$ from the initial distribution. Using \eqref{eq: E dif U bound}, we can then bound the second sum in the left side of equation \eqref{eq: mid lemma 5 pf} by:
\ifCLASSOPTIONonecolumn
\begin{align}
    \sum_{n=1}^\infty \E [U_i(T_n)  - &U_i(t_0^{(n)}) \mid \theta = i]
    \label{eq: lemma 5 inner sum}
    \\
    &\le  \E[U_i(T^{(1)})-U_i(0) \mid
    \! T^{(1)} \! > \! 0, \theta = i]p_f^0
    +\sum_{n=2}^\infty  \E[U_i(T_n) \!  - \! U_i(t_0^{(n)}) \mid \! T^{(n)} \! > \! 0, \theta \! = \! i] p_f^{n \! - \! 1}
    \label{eq: T1 and Tn}
    \\
    &= - U_i(0) + \sum_{n=1}^\infty  \E[U_i(T_n) \mid T^{(n)} > 0, \theta = i] p_f^{n-1}
    -\sum_{n=2}^\infty  \E[U_i(t_0^{(n)}) \mid T^{(n)} > 0, \theta = i] p_f^{n-1} \label{eq: sum U T n=2}
    \,.
\end{align}
\else
\begin{flalign}
    \sum_{n=1}^\infty & \E [U_i(T_n)  - U_i(t_0^{(n)}) \mid \theta = i]
    \label{eq: lemma 5 inner sum}&
    \\
    &\le   \E[U_i(T^{(1)})-U_i(0) \mid
    \! T^{(1)} \! > \! 0, \theta = i]p_f^0
    \nonumber&
    \\
    &+\sum_{n=2}^\infty  \E[U_i(T_n) \!  - \! U_i(t_0^{(n)}) \mid \! T^{(n)} \! > \! 0, \theta \! = \! i] p_f^{n \! - \! 1}
    \label{eq: T1 and Tn}
    &
    \\
    =&\sum_{n=1}^\infty  \E[U_i(T_n) \mid T^{(n)} > 0, \theta = i] p_f^{n-1} - U_i(0) 
    \nonumber&
    \\
    &-\sum_{n=2}^\infty  \E[U_i(t_0^{(n)}) \mid T^{(n)} > 0, \theta = i] p_f^{n-1} \label{eq: sum U T n=2}
    \,.&
\end{flalign}
\fi
The conditioning event $\{T^{(n)}>0\}$ implies events $\{T^{(m)}>0\}$ for $m = 1,\dots, n$ because if $T^{(m)}=0$ means the process has stopped and no further communication rounds occur. Event $\{T^{(n)}>0\}$ also implies that the $n$-th round of communication occurs, and therefore $U_i(t_0^{(n)})$ is given by the previous crossing value $U_i(T_{n-1})$ minus $C_2$  by constraint \eqref{eq: phase II step size}, then:
\ifCLASSOPTIONonecolumn
\begin{align}
    \sum_{n=2}^\infty \E[U_i(t_0^{(n)}) & \mid T^{(n)} > 0, \theta = i]p_f^{n-1} \label{eq: last t0n}
    =
    \sum_{n=2}^\infty \E[U_i(T_{n-1}) \! - \! C_2 \mid T^{(n)} > 0, \theta \! = \! i]p_f^{n-1}
    \\
    &=
    \sum_{n=1}^\infty \E[U_i(T_n) \! - \! C_2\mid T^{(n+1)} > 0, \theta \! = \! i]p_f^{n}
    \ge
    \sum_{n=1}^\infty \E[U_i(T_n) \! - \! C_2\mid T^{(n)} \! > \! 0, \theta \! = \! i]p_f^{n}\, .
    \label{eq: sum T n+1 conditioning bound}
\end{align}
\else
\begin{align}
    \sum_{n=2}^\infty &\E[U_i(t_0^{(n)}) \mid T^{(n)} > 0, \theta = i]p_f^{n-1} 
    \nonumber
    \\
    &=
    \sum_{n=2}^\infty \E[U_i(T_{n-1}) \! - \! C_2 \mid T^{(n)} > 0, \theta \! = \! i]p_f^{n-1} \label{eq: last t0n}
    \\
    &=
    \sum_{n=1}^\infty \E[U_i(T_n) \! - \! C_2\mid T^{(n+1)} > 0, \theta \! = \! i]p_f^{n} \nonumber
    \\
    &\ge
    \sum_{n=1}^\infty \E[U_i(T_n) \! - \! C_2\mid T^{(n)} \! > \! 0, \theta \! = \! i]p_f^{n}\, .
    \label{eq: sum T n+1 conditioning bound}
\end{align}
\fi
The inequality in \eqref{eq: sum T n+1 conditioning bound} follows from the following inequality:
\ifCLASSOPTIONonecolumn
\begin{align}
    \E[U_i(T_n) \mid T^{(n+1)} \! > \! 0, \theta \! = \! i]
    \ge \E[U_i(T_n)\mid T^{(n)} \! > \! 0, \theta \! = \! i] \, . \label{eq: T n+1 inequality}
\end{align}
\else
\begin{flalign}
    &\E[U_i(T_n) \mid T^{(n+1)} \! > \! 0, \theta \! = \! i]
    \ge \E[U_i(T_n)\mid T^{(n)} \! > \! 0, \theta \! = \! i] \, .& \label{eq: T n+1 inequality}
\end{flalign}
\fi
For the proof of inequality \eqref{eq: T n+1 inequality} see Appendix \ref{appendix: proof of T n+1 inequality}. From \eqref{eq: sum T n+1 conditioning bound} it follows that:
\ifCLASSOPTIONonecolumn
\begin{align}
    - \sum_{n=1}^\infty \E[U_i(T_n) \! - \! C_2\mid T^{(n+1)} \! > \! 0 \theta \! = \! i]p_f^{n} \label{eq: inner sum telescope}
    \le&
    -\sum_{n=1}^\infty \E[U_i(T_n) \! - \! C_2\mid T^{(n)} \! > \! 0, \theta \! = \! i]p_f^{n}&
    \\
    =&
    -\sum_{n=1}^\infty \E[p_f(U_i(T_n) \! - \! C_2) \mid T^{(n)} \! > \! 0, \theta \! = \! i]p_f^{n-1}
    \label{eq: E U C2 factor pf}
    \, .&
\end{align}
\else
\begin{flalign}
    - \sum_{n=1}^\infty &\E[U_i(T_n) \! - \! C_2\mid T^{(n+1)} \! > \! 0 \theta \! = \! i]p_f^{n}
    \nonumber& 
    \\
    \le&
    -\sum_{n=1}^\infty \E[U_i(T_n) \! - \! C_2\mid T^{(n)} \! > \! 0, \theta \! = \! i]p_f^{n}&
    \label{eq: inner sum telescope} \\
    =& \! - \!
    \sum_{n=1}^\infty \E[p_f(U_i(T_n) \! - \! C_2) \! \mid \! T^{(n)} \! > \! 0, \theta \! = \! i]p_f^{n \! - \! 1}
    \label{eq: E U C2 factor pf}
    \, .&
\end{flalign}
\fi
In \eqref{eq: E U C2 factor pf} we have factored one $p_f$ inside the expectation. We can now replace \eqref{eq: sum U T n=2} by \eqref{eq: E U C2 factor pf} to upper bound \eqref{eq: lemma 5 inner sum}:
\ifCLASSOPTIONonecolumn
\begin{align}
    \sum_{n=1}^\infty \E[U_i(T_n) & - U_i(t_0^{(n)}) \mid \theta = i]
    \nonumber
    \\
    &\le -U_i(0)
    +\sum_{n=1}^\infty  \E[U_i(T_n) \mid T^{(n)} > 0, \theta = i] p_f^{n-1}
    -\sum_{n=1}^\infty \E[p_f(U_i(T_n) \! - \! C_2) \! \mid T^{(n)} \! > \! 0, \theta \! = \! i]p_f^{n \! - \! 1}
    \label{eq: sum U Tn 2}
    \\
    &=\sum_{n=1}^\infty \E[U_i(T_n) \! - \! p_f(U_i(T_n) \! - \! C_2) \! \mid \! T^{(n)} \! > \! 0, \theta \! = \! i]p_f^{n \! - \! 1}
    -U_i(0)\, . 
    \label{eq: key lemma 5}
\end{align}
\else
\begin{flalign}
    &\sum_{n=1}^\infty \E[U_i(t_0^{(n)}+T^{(n)})  - U_i(t_0^{(n)}) \mid \theta = i]& \nonumber
    \\
    &\le -U_i(0)
    +\sum_{n=1}^\infty  \E[U_i(T_n) \mid T^{(n)} > 0, \theta = i] p_f^{n-1}&  \label{eq: sum U Tn 2}
    \\
    & \quad \quad -\sum_{n=1}^\infty \E[p_f(U_i(T_n) \! - \! C_2) \! \mid T^{(n)} \! > \! 0, \theta \! = \! i]p_f^{n \! - \! 1}&
    \nonumber
    \\
    &=\sum_{n=1}^\infty \E[U_i(T_n) \! - \! p_f(U_i(T_n) \! - \! C_2) \! \mid \! T^{(n)} \! > \! 0, \theta \! = \! i]p_f^{n \! - \! 1} & \nonumber \\
    & \quad \quad -U_i(0)\, . &
    \label{eq: key lemma 5}
\end{flalign}
\fi
The expectation in \eqref{eq: key lemma 5} combines the two sums in from \eqref{eq: sum U Tn 2} by subtracting $p_f(U_i(T_n)-C_2)$ from $U_i(T_n)$. The first term $U_i(T_n)$ is the value of $U_i(t)$ at the communication-phase stopping time $t=T_n$. In the second term $p_f(U_i(T_n)-C_2)$, the difference $U_i(T_n)-C_2$ is the unique value that $U_i(t_0^{(n+1)})$ can take once the $n$-th confirmation phase round starts at a point $U_i(T_n)$. Equation \eqref{eq: key lemma 5} is an important intermediate result in the proof of Thm. \ref{theorem: surrogate martingale}. This is because when considering the process $U'_i(t)$, the starting value of each communication-phase round $U'_i(t_0^{(n+1)})$ is still that of the original process $U_i(T_n)-C_2$, and therefore the argument of the expectation would change to $U'_i(T'_n) -p_f(U_i(T_n)-C_2)$. For the proof of Lemma \ref{lemma: final and initial values}, we just need to bound \eqref{eq: key lemma 5}, so we write the sum in \eqref{eq: key lemma 5} as:
\ifCLASSOPTIONonecolumn
\begin{align}
    \sum_{n=1}^\infty  \E[U_i(T_n)(1-p_f) \! + \! p_f C_2) \! \mid T^{(n)} \! > \! 0, \theta \! = \! i]p_f^{n \! - \! 1} 
    \label{eq: gather UTn} 
    =
    \sum_{n=1}^\infty \E[U_i(T_n) \mid T^{(n)} \! > \! 0, \theta \! = \! i](1 \! - \! p_f)p_f^{n \! - \! 1} 
    \! + \! \sum_{n=1}^\infty C_2 p_f^{n}
    \, . 
\end{align}
\else
\begin{align}
    \sum_{n=1}^\infty  \E[&U_i(T_n)(1-p_f) \! + \! p_f C_2) \! \mid T^{(n)} \! > \! 0, \theta \! = \! i]p_f^{n \! - \! 1} 
    \label{eq: gather UTn}
    \\
    =&
    \sum_{n=1}^\infty \E[U_i(T_n) \mid T^{(n)} \! > \! 0, \theta \! = \! i](1 \! - \! p_f)p_f^{n \! - \! 1} 
    \! + \! \sum_{n=1}^\infty C_2 p_f^{n} \nonumber
    . 
\end{align}
\fi
By constraint \eqref{eq: phase 1 max}, $U_i(T_n) \le U_i(T_n-1)+C_2$, and since $U_i(T_n-1) < 0$, then, $U_i(T_n)$ is bounded by $C_2$. Thus, the expectation $\E[U_i(T_n) \mid T^{(n)} \! > \! 0, \theta \! = \! i]$ is also bounded by $C_2$. Then \eqref{eq: gather UTn} is bounded by:
\begin{align}
    \sum_{n=1}^\infty C_2(1 \! - \! p_f)p_f^{n \! - \! 1} 
    + \sum_{n=1}^\infty C_2 p_f^{n} =  C_2 + \frac{p_f}{1-p_f}C_2 \label{eq: inner sum bound}
\end{align}
Finally, the left side of equation \eqref{eq: mid lemma 5 pf} in Lemma \ref{lemma: final and initial values}, (the left side of \eqref{eq: end lemma 5 with pf}), is upper bounded using the bounds \eqref{eq: key lemma 5} and \eqref{eq: inner sum bound} on the inner sum \eqref{eq: inner sum telescope}  as follows: 
\ifCLASSOPTIONonecolumn
\begin{align}
      \sum_{i=1}^M \Pr(\theta = i) \sum_{n=1}^\infty  \E[U_i(T_n) \! - \! U_i(t_0^{(n)}) \mid \theta = i]  
      &\le \sum_{i=1}^M \Pr(\theta = i) \left( \frac{p_f}{1-p_f}C_2 + C_2-U_i(0)\right)\label{eq: end lemma 5 with pf}&
      \\
      &= 2^{-C_2}\frac{1-2^{-NC_2}}{1-2^{-C_2}}C_2+C_2-\E[U_i(0)] 
      \label{eq: replace pf}
      \, .&
\end{align}
\else
\begin{flalign}
      \sum_{i=1}^M& \Pr(\theta = i) \sum_{n=1}^\infty  \E[U_i(T_n) \! - \! U_i(t_0^{(n)}) \mid \theta = i]  
      \nonumber&
      \\
      &\le \sum_{i=1}^M \Pr(\theta = i) \left( \frac{p_f}{1-p_f}C_2 + C_2-U_i(0)\right)\label{eq: end lemma 5 with pf}&
      \\
      &= 2^{-C_2}\frac{1-2^{-NC_2}}{1-2^{-C_2}}C_2+C_2-\E[U_i(0)] 
      \label{eq: replace pf}
      \, .&
\end{flalign}
\fi
To transition from \eqref{eq: end lemma 5 with pf} to \eqref{eq: replace pf}
we have used the definition of $p_f$ from Lemma \ref{lemma: fall back probability}. The proof of Lemma \ref{lemma: final and initial values} is complete.
\end{IEEEproof}

\subsection{Proof of Theorems \ref{theorem: surrogate martingale} and \ref{theorem: simple rule}}
\label{sec: proof of theorems 2 and 3}
\begin{IEEEproof}[Proof of Thm. \ref{theorem: surrogate martingale}]
Suppose $U'_i(t)$ is a process that satisfies the constraints \eqref{eq: V ge 0}-\eqref{eq: C step size} in Thm. \ref{theorem: Main Theorem} and constraints \eqref{eq: U' smaller}-\eqref{eq: U' B bound} of Thm. \ref{theorem: surrogate martingale} for some $B< C_2$. 
Because the constraints of  Thm. \ref{theorem: Main Theorem}  are satisfied, Lemmas 1-5 all hold for the process $U'_i(t)$. To bound $\E[T']$  we begin by bounding the sum on the right side of Lemma \ref{lemma: stopping time bound}, which is \eqref{eq: telescopic equality}, but using the new process $U'_i(t)$.  Dividing the new bound by $C$ produces the desired result. We follow the procedure in the proof of Lemma \ref{lemma: final and initial values}, but replacing $U_i(t)$ by $U'_i(t)$, up to the equation \eqref{eq: sum U T n=2}. By the definition of $U'_i(t)$ we have that $U'_i(t_0^{(n)}) = U_i(t_0^{(n)})$ and from equation \eqref{eq: phase II step size} it follows that, for $n > 1$, $T^{(n)} > 0$ implies $U_i(t_0^{(n)}) =  U_i(T_{n-1})-C_2$. Then, from equation \eqref{eq: sum U T n=2} to \eqref{eq: key lemma 5}, we replace $U'_i(t_0^{(n)})$ by  $U_i(T_{n-1})-C_2$ instead. Using equation \eqref{eq: key lemma 5} we have that:
\ifCLASSOPTIONonecolumn
\begin{align}
    \sum_{n=1}^\infty \E[U'_i(T'_n)  - U'_i(t_0^{(n)}) \mid \theta = i] 
    \le -U'_i(0)
    +\sum_{n=1}^\infty \E[U'_i(T'_n) \! - \! p_f (U_i(T_n)-C_2) \! \mid T^{(n)} \! > \! 0, \theta \! = \! i]p_f^{n \! - \! 1}
\end{align}
\else
\begin{flalign}
    \sum_{n=1}^\infty &\E[U'_i(T'_n)  - U'_i(t_0^{(n)}) \mid \theta = i] \le -U'_i(0)&
    \\
    +&\sum_{n=1}^\infty \E[U'_i(T'_n) \! - \! p_f (U_i(T_n)-C_2) \! \mid T^{(n)} \! > \! 0, \theta \! = \! i]p_f^{n \! - \! 1}& \nonumber
\end{flalign}
\fi
We can replace $U'_i(t)$ by $U'_i(0) = U_i(0)$ using the definition of $U'_i(t)$. We further claim that the constraints of Thm. \ref{theorem: surrogate martingale} guarantee that $U'_i(T'_n) \! - \! p_f (U_i(T_n)-C_2) \le B$. This is derived from constraint \eqref{eq: U' B bound}: $U'_i(T'_n)-\frac{p}{q}(U_i(T_n) - C_2) \le B$ by replacing $p_f$ by $\frac{p}{q}$. The replacement is possible because $p_f < \frac{p}{q}$, see \eqref{eq: solve pf}, and $U_i(T_n) - C_2 < 0$ by constraint \eqref{eq: phase 1 max}. Therefore, the expectation in the last sum can be replaced with $B$ for an upper bound to obtain: 
\ifCLASSOPTIONonecolumn
\begin{align}
    \sum_{n=1}^\infty \E[U'_i(T'_n) \! - \! U_i(t_0^{(n)}) \! \mid \! \theta = i]
    \le -U_i(0) \! + \!
    \sum_{n=1}^\infty B p_f^{n \! - \! 1}
    =&
    B \! + \! \frac{B p_f }{1 \! - \! p_f} \! - \! U_i(0)
    = B \! + \! 2^{-C_2}\frac{1 \! - \! 2^{-NC_2}}{1 \! - \! 2^{-C_2}}B -U_i(0)
    \label{eq: U' and U} 
    \, .
\end{align}
\else
\begin{flalign}
    \sum_{n=1}^\infty &\E[U'_i(T'_n) \! - \! U_i(t_0^{(n)}) \! \mid \! \theta = i]
    \le -U_i(0) \! + \!
    \sum_{n=1}^\infty B p_f^{n \! - \! 1}
    \nonumber
    &
    \\
    =&
    B \! + \! \frac{B p_f }{1 \! - \! p_f} \! - \! U_i(0)
    = B \! + \! 2^{-C_2}\frac{1 \! - \! 2^{-NC_2}}{1 \! - \! 2^{-C_2}}B -U_i(0)
    \label{eq: U' and U} 
    \, .&
\end{flalign}
\fi
Then, the value in equation \eqref{eq: U' and U} replaces the inner sum in the left side of \eqref{eq: C2 pf and EU} to obtain:
\ifCLASSOPTIONonecolumn
\begin{align}
    \frac{1}{C}\sum_{i=1}^M  \Pr(\theta \! = \! i) \sum_{n=1}^\infty  \E[U'_i(T'_n) \! - \! U'_i(t_0^{(n)}) \! \mid \! \theta \! = \! i]
    &\le \frac{1}{C}\sum_{i=1}^M  \Pr(\theta \! = \! i) \left(U_i(0) \! + \! B \! + \! 2^{-C_2}\frac{1 \! - \! 2^{-NC_2}}{1-2^{-C_2}} B\right)&
    \nonumber
    \\
    &= \frac{B}{C} \left(1 \! + \! 2^{-C_2}\frac{1 \! - \! 2^{-NC_2}}{1 \! - \! 2^{-C_2}}\right) -\frac{\E[U_i(0)]}{C} \, .&
\end{align}
\else
\begin{flalign}
    \frac{1}{C}&\sum_{i=1}^M  \Pr(\theta \! = \! i) \sum_{n=1}^\infty  \E[U'_i(T'_n) \! - \! U'_i(t_0^{(n)}) \! \mid \! \theta \! = \! i]& \nonumber
    \\
    &\le \frac{1}{C}\sum_{i=1}^M  \Pr(\theta \! = \! i) \left(U_i(0) \! + \! B \! + \! 2^{-C_2}\frac{1 \! - \! 2^{-NC_2}}{1-2^{-C_2}} B\right)&
    \nonumber
    \\
    &= \frac{B}{C} \left(1 \! + \! 2^{-C_2}\frac{1 \! - \! 2^{-NC_2}}{1 \! - \! 2^{-C_2}}\right) -\frac{\E[U_i(0)]}{C} \, .&
\end{flalign}
\fi
The proof is complete.
\end{IEEEproof}

\begin{IEEEproof}
[Proof of Thm. \ref{theorem: simple rule}]

When $U_i(t) \ge 0$ for some $i$, constraint \eqref{eq: simple partitioning 2} is the same as the SED constraint \eqref{eq: sed rule} and therefore the constraints \eqref{eq: phase II step size} and \eqref{eq: phase II V} are satisfied as shown in \cite{Yang2021}. Need to show that constraints \eqref{eq: V ge 0}, \eqref{eq: phase 1 max} and \eqref{eq: C step size} are also satisfied. We start the proof by deriving expressions for $\E[U_i(t \! + \! 1) \! - \! U_i(t) \mid Y^t,  \theta \! = \! i]$ to find bounds in terms of the constraints of the theorem.
The posterior probabilities $\rho_i(y^{t+1})$ are computed according to Bayes' Rule:
\begin{equation}
\begin{aligned}
    \rho_i(y^{t+1}) = \frac{\Pr(\theta = i, Y_{t+1} = y_{t+1} \mid Y^t)}{\Pr(Y_{t+1} = y_{t+1} \mid Y^t)} \, .
    \label{eq: Bayes Update}
\end{aligned}
\end{equation}
The top conditional probability in equation \eqref{eq: Bayes Update} can be split into $P(Y_{t+1} = y_{t+1}\mid \theta = i, Y^t = y^t)
\Pr(\theta = i \mid y^t)$. Since the received history $y^t$ fully characterizes the vector of posterior probabilities $\bm{\rho}_t \triangleq [\rho_1(y^t),\rho_2(y^t),\dots,\rho_{M}(y^t)]$, and the new construction of $S_0$ and $S_1$, then the conditioning event $\{\theta = i\}$ sets the value of the encoding function $X_{t+1} = \text{enc}(i, Y^t)$, via its definition:  $\text{enc}(i, Y^t) = \mathbbm{1}_{i \in S_1}$. We can just write the first probability as $\Pr(Y_{t+1} \mid  \text{enc}(i, Y^t))$, which reduces to $q$ if $Y_{t+1} = \text{enc}(i, Y^t)$ and to $p$ if $Y_{t+1} \neq \text{enc}(i, Y^t)$.
The second probability $\Pr(\theta = i \mid Y^t = y^t)$ is just $\rho_i(y^t)$.

The bottom conditional probability can be written as $\underset{x_{t+1}\in \{0,1\}}{\sum \Pr(Y_{t+1}} = y_{t+1} \mid X_{t+1}, Y^t)P(X_{t+1} = x_{t+1}\mid Y^t)$.
By the channel memoryless property, the next output $Y_{t+1}$ given the input $X_{t+1}$ is independent of the past $Y^t$, that is: \\ $\Pr(Y_{t+1} = y_{t+1} \mid X_{t+1}, Y^t) = \Pr(Y_{t+1} = y_{t+1} \mid X_{t+1})$.
Since $\Pr(X_{t+1} =  x_{t+1}\mid Y^t) =\Pr(\theta \in S_{x_{t+1}})$ which is given by $\underset{i \in S_{x_{t+1}}}{\sum} \rho_i(y^t)$, we write:
\ifCLASSOPTIONonecolumn
\begin{equation}
\begin{aligned}
    \rho_i&(t+1) = \frac{\Pr(Y_{t+1}\mid i)\rho_i(y^t)}{\sum_{j\in \Omega}\Pr(Y_{t+1}\mid j)\rho_j(y^t)}
    &= \frac{\Pr(Y_{t+1}\mid i)\rho_i(y^t)}{q \sum_{j\in S_{y_{t+1}}}\rho_j(y^t)+p\sum_{j\in \Omega \setminus S_{y_{t+1}}}\rho_j(y^t)} \, .
    \label{eq: update sum}
\end{aligned}
\end{equation}
\else
\begin{equation}
\begin{aligned}
    \rho_i&(t+1) = \frac{\Pr(Y_{t+1}\mid i)\rho_i(y^t)}{\sum_{j\in \Omega}\Pr(Y_{t+1}\mid j)\rho_j(y^t)}\\
    &= \frac{\Pr(Y_{t+1}\mid i)\rho_i(y^t)}{q \sum_{j\in S_{y_{t+1}}}\rho_j(y^t)+p\sum_{j\in \Omega \setminus S_{y_{t+1}}}\rho_j(y^t)} \, .
    \label{eq: update sum}
\end{aligned}
\end{equation}
\fi
For $\{i=\theta\}$ the encoding function $X_{t+1} = \mathbbm{1}_{\theta \in S_1}$ dictates that $X_{t+1} = \mathbbm{1}_{i \in S_1}$.
Thus $\Pr(Y_{t+1}=\mathbbm{1}_{i \in S_1}\mid i = \theta) = \Pr(Y_{t+1}= X_{t+1}) = q$, and $\Pr(Y_{t+1}=\mathbbm{1}_{i \notin S_1}\mid i = \theta)  = \Pr(Y_{t+1} = X_{t+1}\oplus 1) = p$. 
Let $P_0 = \sum_{j\in S_{0}}\rho_j(y^t)$ and $P_1 = \sum_{j\in S_{1}}\rho_j(y^t)$ and let $ \Delta \triangleq P_0 - P_1$, so that $P_0 = \frac{1}{2} + \frac{\Delta}{2}$ and $P_1 = \frac{1}{2} - \frac{\Delta}{2}$. The value of $U_i(t+1)$ for each $Y_{t+1} \in \{0,1\}$ can be obtained from equation \ref{eq: update sum}.

Assume first that $i \in S_0$ to obtain the value of $E[U_i(t+1)-U_i(t)\mid Y^t = y^t,\theta = i]$.
\ifCLASSOPTIONonecolumn
\begin{align*}
    \E[U_i(t+1) \mid Y^t = y^t, \theta = i]
    &= q \log_2\frac{\frac{\rho_i(y^t) q}{P_0 q + P_1 p}}{1-\frac{\rho_i(y^t) q}{P_0 q + P_1 p}} +
    p \log_2\frac{\frac{\rho_i(y^t)p}{P_0 p + P_1 q}}{1-\frac{\rho_i(y^t)p}{P_0 p + P_1 q}}
    \\
    &=
    q \log_2\dfrac{\rho_i(y^t) q}{\frac{1}{2} \! + \! \frac{\Delta(q-p)}{2} \! - \! \rho_i(y^t)q} + 
        p \log_2\cfrac{\rho_i(y^t)p}{\frac{1}{2} \! - \! \frac{\Delta(q-p)}{2} \! - \! \rho_i(y^t)p} \, .
\end{align*}
\else
\begin{align*}
    \E[&U_i(t+1) \mid Y^t = y^t, \theta = i]
    \\
    &= q \log_2\frac{\frac{\rho_i(y^t) q}{P_0 q + P_1 p}}{1-\frac{\rho_i(y^t) q}{P_0 q + P_1 p}} +
    p \log_2\frac{\frac{\rho_i(y^t)p}{P_0 p + P_1 q}}{1-\frac{\rho_i(y^t)p}{P_0 p + P_1 q}}
    \\
    &=
    q \log_2\dfrac{\rho_i(y^t) q}{\frac{1}{2} \! + \! \frac{\Delta(q-p)}{2} \! - \! \rho_i(y^t)q} + 
        p \log_2\cfrac{\rho_i(y^t)p}{\frac{1}{2} \! - \! \frac{\Delta(q-p)}{2} \! - \! \rho_i(y^t)p} \, .
\end{align*}
\fi
For $i \in S_1$ the only difference is the sign of the term with $\Delta$. Let $\iota_i = \mathbbm{1}_{i\in S_0}-\mathbbm{1}_{i\in S_1}$, that is $1$ if $i \in S_0$ and $-1$ if $i \in S_1$ and add a coefficient $\iota_i$ to each $\Delta$ for a general expression. Multiply by $2$ both terms of the fraction inside the logarithm and expand it to obtain:
\ifCLASSOPTIONonecolumn
\begin{align}
    \E[U_i(t \! + \! 1) \! - \! & U_i(t) \mid Y^t,  \theta \! = \! i] 
    = \log_2(\rho_i(y^t)) 
    + q\left(\log_2(2q)\! - \! \log_2\left(1 \! - \! \rho_i(y^t) \! + \! (q \! - \! p)(\iota_i\Delta \! - \! \rho_i(y^t)\right)\right) \label{eq: U t difference}&
    \\
    & \quad \quad \quad \quad \quad \quad
    + p\left(\log_2(2p) \!-\! \log_2\left(1 \! - \! \rho_i(y^t) \! - \! (q \! - \! p)(\iota_i\Delta \! - \! \rho_i(y^t) \right)\right) \nonumber&
    \\
    = & q\left(\log_2(2q) \! - \! \log_2\left(1 \! + \! (q \! - \! p)\frac{ \iota_i\Delta \!  - \! \rho_i(y^t)}{1-\rho_i(y^t)}\right) \right) + p \left(\log_2(2p) \! - \! \log_2\left(1 \! - \! (q \! - \! p)\frac{\iota_i\Delta \! - \! \rho_i(y^t)}{1 \! - \! \rho_i(y^t)}\right)\right) \label{eq: split 2 rho q}&
    \\
    \ge & C - \log_2\left(1 + (q-p)^2\frac{ \iota_i\Delta
    -\rho_i(y^t)}{1-\rho_i(y^t)}\right) \, .&
    \label{eq: jensen step}
\end{align}
\else
\begin{flalign}
    \E[&U_i(t \! + \! 1) \! - \! U_i(t) \mid Y^t,  \theta \! = \! i]= \log_2(\rho_i(y^t)) \label{eq: U t difference}&
    \\
    &+ q\left(\log_2(2q)\! - \! \log_2\left(1 \! - \! \rho_i(y^t) \! + \! (q \! - \! p)(\iota_i\Delta \! - \! \rho_i(y^t)\right)\right) \nonumber&
    \\
    &+ p\left(\log_2(2p) \!-\! \log_2\left(1 \! - \! \rho_i(y^t) \! - \! (q \! - \! p)(\iota_i\Delta \! - \! \rho_i(y^t) \right)\right) \nonumber&
    \\
    =& q\left(\log_2(2q) \! - \! \log_2\left(1 \! + \! (q \! - \! p)\frac{ \iota_i\Delta \!  - \! \rho_i(y^t)}{1-\rho_i(y^t)}\right) \right) \label{eq: split 2 rho q}&
    \\
    &+ p \left(\log_2(2p) \! - \! \log_2\left(1 \! - \! (q \! - \! p)\frac{\iota_i\Delta \! - \! \rho_i(y^t)}{1 \! - \! \rho_i(y^t)}\right)\right) \nonumber &
    \\
    \ge& C - \log_2\left(1 + (q-p)^2\frac{ \iota_i\Delta
    -\rho_i(y^t)}{1-\rho_i(y^t)}\right) \, .&
    \label{eq: jensen step}
\end{flalign}
\fi
Now subtract the term $\log_2(1-\rho_i(y^t))$, and add it back as a factor in the logarithm, to recover $U_i(t)$ from $\log_2(\rho_i(y^t))$. Note that $2 \rho_i(y^t) q = \rho_i(y^t) + (q-p)\rho_i(y^t)$ and $2 \rho_i(y^t) p = \rho_i(y^t) - (q-p) \rho_i(y^t)$. 
And also note that $q\log_2(2q) + p \log_2(2p)=C$. 

The logarithm $\log_2(1-\rho_i(y^t))$ from \eqref{eq: U t difference} is split into $p \log_2(1-\rho_i(y^t))+q \log_2(1-\rho_i(y^t))$, and $1-\rho_i(y^t)$ divides the arguments of the logarithms in \eqref{eq: split 2 rho q}. Equation \eqref{eq: jensen step} follows from applying Jensen's inequality to the convex function $-\log_2(\cdot)$. Then:
\ifCLASSOPTIONonecolumn
\begin{align}
    \sum_{i = 1}^M  \E[U_i(t+1)- & U_i(t)\mid Y^t, \theta = i] \rho_i(Y^t)
    \ge
    C - \sum_{i=1}^M \rho_i(y^t)\log_2\left(1 +(q-p)^2\frac{ \iota_i\Delta
    -\rho_i(y^t)}{1-\rho_i(y^t)}\right)
    \\
    &= C - \sum_{i \in S_0}  \rho_i(y^t)\log_2\left(1 +(q-p)^2\frac{ \Delta
    -\rho_i(y^t)}{1-\rho_i(y^t)}\right)
    \label{eq: jensen over S0}
    - \sum_{i \in S_1}  \rho_i(y^t)\log_2\left(1 -(q-p)^2\frac{ \Delta
    +\rho_i(y^t)}{1-\rho_i(y^t)}\right) \, .
\end{align}
\else
\begin{align}
    \sum_{i = 1}^M  &\E[U_i(t+1)-U_i(t)\mid Y^t, \theta = i] \rho_i(Y^t) \nonumber
    \\
    \ge C - &\sum_{i=1}^M \rho_i(y^t)\log_2\left(1 +(q-p)^2\frac{ \iota_i\Delta
    -\rho_i(y^t)}{1-\rho_i(y^t)}\right)
    \\
    = C - &\sum_{i \in S_0}  \rho_i(y^t)\log_2\left(1 +(q-p)^2\frac{ \Delta
    -\rho_i(y^t)}{1-\rho_i(y^t)}\right)
    \nonumber
    \\
    - &\sum_{i \in S_1}  \rho_i(y^t)\log_2\left(1 -(q-p)^2\frac{ \Delta
    +\rho_i(y^t)}{1-\rho_i(y^t)}\right) \, .
    \label{eq: jensen over S0}
\end{align}
\fi
By the SEAD constraints, equations \eqref{eq: simple partitioning} and \eqref{eq: simple partitioning 2} if $i\in S_0$, then $\Delta \le \rho_{\min} \le \rho_i(y^t)$. For the case where $\Delta \ge 0$, then $i \in S_0 \implies \Delta -\rho_i(y^t) \le 0$ and $ -\Delta - \rho_i(y^t) < 0$. Then the arguments of the logarithms in \eqref{eq: jensen over S0} are both less than $1$ for every $i$.
This suffices to show that the constraints \eqref{eq: C step size} and \eqref{eq: V ge 0} are satisfied when $P_0 \ge P_1$ for the case that $\Delta \ge 0$. 

It remains to prove that constraints \eqref{eq: C step size} and \eqref{eq: V ge 0} hold in the case where $P_1 > P_0$, or equivalently $\Delta < 0$.
Let $\alpha = -\Delta > 0$, and note that since $0 < \alpha < 1$, then:
\begin{equation}
    \frac{\alpha}{1 \! - \! \rho_{\min}} \ge \alpha = \alpha \frac{1 \! - \! \rho_i(y^t)}{1 \! - \! \rho_i(y^t)} 
    \ge 
    \frac{\alpha \! - \! \rho_i(y^t)}{1 \! - \! \rho_i(y^t)} \, , \label{eq: alpha inequality}
\end{equation}
and $\rho_i \ge \rho_{\min} \implies \alpha+\rho_i \ge \alpha + \rho_{\min}$ and $1-\rho_i < 1-\rho_{\min}$,  therefore:
\begin{align*}
    \log_2 \! \left( \! 1 \! - \! (q \! - \! p)^2\frac{ \alpha
     \! + \! \rho_i(y^t)}{1 \! - \! \rho_i(y^t)}\right) \! 
    &\le \log_2 \! \left( \! 1 \! - \! (q \! - \! p)^2\frac{ \alpha \! + \! \rho_{\min}}{1 \! - \! \rho_{\min}}\right) 
    \\
    \log_2 \! \left( \! 1 \! + \! (q \! - \! p)^2\frac{ \alpha \! 
    - \! \rho_i(y^t)}{1 \! - \! \rho_i(y^t)}\right) \! 
    &\le  \log_2 \! \left( \! 1 \! +(q \! - \! p)^2 \frac{\alpha}{1-\rho_{\min}}
    \right) \, .
\end{align*}
Since this holds for all $i = 1,\dots,M$, then:
\ifCLASSOPTIONonecolumn
\begin{align}
     \sum_{i = 1}^M  \E[U_i(t+1)-U_i(t)\mid Y^t = y^t, \theta = i] \rho_i(y^t)
    & \ge  C  \label{eq: pi step}
    -P_0 \log_2 \! \left( \! 1 \! - \! (q \! - \! p)^2\frac{ \alpha \! + \! \rho_{\min}}{1 \! - \! \rho_{\min}}\right)
    \! - \! P_1  \log_2 \! \left( \! 1 \! +\alpha \frac{(q \! - \! p)^2}{1 \! - \! \rho_{\min}}
    \right) 
    \\
    &\ge C - \log_2\left(1 - \frac{(q-p)^2}{1-\rho_{\min}} [P_0(\alpha+\rho_{\min})-P_1 \alpha]\right) \,. \label{eq: jensen over P0 and P1}
\end{align}
\else
\begin{align}
     &\sum_{i = 1}^M  \E[U_i(t+1)-U_i(t)\mid Y^t = y^t, \theta = i] \rho_i(y^t)
    \ge  C  \label{eq: pi step}
    \\
    &-P_0 \log_2 \! \left( \! 1 \! - \! (q \! - \! p)^2\frac{ \alpha \! + \! \rho_{\min}}{1 \! - \! \rho_{\min}}\right)
    \! - \! P_1  \log_2 \! \left( \! 1 \! +\alpha \frac{(q \! - \! p)^2}{1 \! - \! \rho_{\min}}
    \right) \nonumber
    \\
    &\ge C - \log_2\left(1 - \frac{(q-p)^2}{1-\rho_{\min}} [P_0(\alpha+\rho_{\min})-P_1 \alpha]\right) \,. \label{eq: jensen over P0 and P1}
\end{align}
\fi
To satisfy constraint \eqref{eq: C step size} we only need the logarithm term in \eqref{eq: jensen over P0 and P1} to be non-negative. This only requires that $-\Delta^2 +P_0\rho_{\min} > 0$.
Since $P_0-P_1 = \Delta$, then $P_0(\alpha + \rho_{\min})-P_1 \alpha =  (P_0-P_1)\alpha + P_0 \rho_{\min}=-\Delta^2 +P_0\rho_{\min}$. To satisfy constraint \eqref{eq: C step size} it suffices that $-\Delta^2 +P_0\rho_{\min} > 0$, which is equivalent to:
\begin{equation}
     \Delta^2 \le P_0\rho_{\min} \, .
    \label{eq: delta requirement}
\end{equation}
The SEAD constraints, equations \eqref{eq: simple partitioning} and \eqref{eq: simple partitioning 2}, guarantees that $\Delta^2 \le \rho^2_{\min}$. Since $P_0 \ge \underset{i \in S_0}{\min}\rho_i(y^t) = \rho_{\min}$, then $\Delta^2 \le \rho^2_{\min} \le P_0 \rho_{\min}$, which satisfies inequality \eqref{eq: delta requirement}. Then, the SEAD constraints guarantee that constraint \eqref{eq: C step size} is satisfied, and only restricts the  \eqref{eq: delta requirement} is satisfied, and only restricts the absolute difference between $P_0$ and $P_1$.

To prove that constraint \eqref{eq: V ge 0} is satisfied, note that equation \eqref{eq: simple partitioning} of the SEAD constraints guarantees that if $\rho_j(t) \le \frac{1}{2} \: \forall j = 1,\dots, M$, then $|\Delta| \le \frac{1}{3}$. Starting from equation \eqref{eq: jensen step} note that the worst case scenario is when $\iota_i\Delta = \frac{1}{3}$. Using \eqref{eq: alpha inequality} with $\alpha = \frac{1}{3}$ to go from \eqref{eq: before alpha} to \eqref{eq: after alpha} we find obtain:
\ifCLASSOPTIONonecolumn
\begin{align}
    \E[U_i(t+1) - U_i(t) \mid Y^t,  \theta = i]
    \ge&
    C - \log_2\left(1 + (q-p)^2\frac{ \iota_i\Delta
    -\rho_i(y^t)}{1-\rho_i(y^t)}\right) 
    \label{eq: before alpha}
    \\
    \ge& C - \log_2\left(1 + \frac{(q-p)^2}{3}\right)
    \label{eq: after alpha}
    \\
    \ge& C -  \frac{(q-p)^2}{3} 
    \\
    =& C-\frac{(q-p)^2}{2 \ln(2)}+\frac{3-2 \ln(2)}{6 \ln(2)}  (q-p)^2 \label{eq: C and q-p}
    \\
    \ge& \frac{3-2 \ln(2)}{6 \ln(2)}  (q-p)^2 > \frac{(q-p)^2}{3} \label{eq: q-p over 3}\,.
\end{align}
\else
\begin{align}
    \E[U_i(t+1) & -  U_i(t)  \mid Y^t,  \theta = i]
    \nonumber
    \\ 
    \ge& C - \log_2\left(1 + (q-p)^2\frac{ \iota_i\Delta
    -\rho_i(y^t)}{1-\rho_i(y^t)}\right) 
    \label{eq: before alpha}
    \\
    \ge& C - \log_2\left(1 + \frac{(q-p)^2}{3}\right)
    \label{eq: after alpha}
    \\
    \ge& C -  \frac{(q-p)^2}{3} 
    \\
    =& C-\frac{(q-p)^2}{2 \ln(2)}+\frac{3-2 \ln(2)}{6 \ln(2)}  (q-p)^2 \label{eq: C and q-p}
    \\
    \ge& \frac{3-2 \ln(2)}{6 \ln(2)}  (q-p)^2 > \frac{(q-p)^2}{3} \label{eq: q-p over 3}\,.
\end{align}
\fi
To transition from \eqref{eq: C and q-p} to \eqref{eq: q-p over 3} we need to show that $2\ln(2) C \ge (q-p)^2$. For this we find a small constant $a$ that makes $a C-(q-p)^2$, the difference between 2 convex functions, also convex. Take second derivatives $\frac{d^2}{dp^2}a C = \frac{1}{\ln(2)}\frac{a}{p q}$ and $\frac{d^2}{d p ^2}(q-p)^2 = 8$ and subtract them. The constant $a$ is found by noting that $p q \le \frac{1}{4}$.

The SEAD constraints guarantee that both sets, $S_0$ and $S_1$ are non-empty. Then, since the maximum absolute value difference $\mid U_i(t+1) - U_i(t)\mid $ is $C_2$, constraint \eqref{eq: phase 1 max} is satisfied, see the proof of \emph{Claim \ref{claim: j confirmation singleton}}. 

For the proof of existence of a process $U'_i(t)$, with $B=\frac{1}{q}\log_2(2q)$,  see Appendix \ref{sec: proof of surrogate process}.
\end{IEEEproof}

\section{Extension to Arbitrary Initial Distributions}
\label{sec: arbitrary distribution}
The proof of Thm. \ref{theorem: Main Theorem} only used the uniform input distribution to assert $U_i(0)=U_1(0)$ and replace $\E[U_i(0)]$ by $U_1(0) = \log_2(M-1)$ in equation \eqref{eq: T upper bound}. In Lemma \ref{lemma: final and initial values}, we have required that $U_i(0) < 0 \; \forall i$. However, even with uniform input distribution this is not the case when $\Omega = \{0,1\}$.
 To avoid this requirement, the case where $\exists i: U_i(0) \ge 0$ and therefore $T^{(1)} = 0$ needs to be accounted for. Also, if $U_i(t) \ge C_2$, then the probability that an initial fall back occurs is only upper bounded by $p_f$, which can be inferred from the proof of Lemma \ref{lemma: fall back probability}. Then, to obtain an upper bound on the expected stopping time $\E[\tau]$ for an arbitrary input distribution, it suffices to multiply the terms $\E[U_i(T^{(1)}) \mid T^{(1)} > 0, \theta = i] - U_i(0) $ in the proof of Lemma \ref{lemma: final and initial values}, equation \eqref{eq: T1 and Tn}, by the indicator $\mathbbm{1}_{U_i(0) < 0}$. Then, the bound on Lemma \ref{lemma: final and initial values} becomes:
 \ifCLASSOPTIONonecolumn
 \begin{align}
      \sum_{i=1}^M \sum_{n=1}^\infty  \E[U_i(t_0^{(n)}+T^{(n)}) - U_i(t_0^{(n)}) \mid \theta = i] \Pr(\theta = i)
      \le C_2 \frac{p}{q}\frac{1 \! - \! \left(\frac{p}{q}\right)^{N}}{1 \! - \! \frac{p}{q}} 
     \! + \! \E[\left(C_2 \! - \! U_i(0)\right)\mathbbm{1}_{U_i(0)<0}] \,. \label{eq: C2 and indicator}
\end{align}
 \else
\begin{align}
      \sum_{i=1}^M& \sum_{n=1}^\infty  \E[U_i(t_0^{(n)}+T^{(n)}) - U_i(t_0^{(n)}) \mid \theta = i] \Pr(\theta = i) \nonumber
      \\
      &\le C_2 \frac{p}{q}\frac{1 \! - \! \left(\frac{p}{q}\right)^{N}}{1 \! - \! \frac{p}{q}} 
     \! + \! \E[\left(C_2 \! - \! U_i(0)\right)\mathbbm{1}_{U_i(0)<0}] \,. \label{eq: C2 and indicator}
\end{align}
\fi
By Thm. \ref{theorem: simple rule}, we can replace $C_2$ with $q^{-1}\log_2(2q)$ in \eqref{eq: C2 and indicator}. Using the definition of $p_f$ from \eqref{eq: fall back probability} we obtain the bound:
 \ifCLASSOPTIONonecolumn
 \begin{align}
    \E[T] \le \E[T'] 
    \le  2^{-C_2}\frac{1-2^{-NC_2}}{1-2^{-C_2}}\frac{\log_2(2q)}{q C} 
    +\E\left[\left(\frac{\log_2(2q)}{q} -U_i(0)\right) \frac{\mathbbm{1}_{U_i(0)<0} }{C} \right] \, . \label{eq: ET surrogate bound}
\end{align}
\else
\begin{align}
    \E[T] &\le \E[T'] 
    \le  2^{-C_2}\frac{1-2^{-NC_2}}{1-2^{-C_2}}\frac{\log_2(2q)}{q C} \nonumber
    \\&+\E\left[\left(\frac{\log_2(2q)}{q} -U_i(0)\right) \frac{\mathbbm{1}_{U_i(0)<0} }{C} \right] \, . \label{eq: ET surrogate bound}
\end{align}
\fi

\subsection{Generalized Achievability Bound}

An upper bound on $\E[\tau]$ for a arbitrary initial distribution $\bm{\rho}_0$ is then obtained using this bound \eqref{eq: ET surrogate bound} and the bound on $\E[\tau-T]$ from equation \eqref{eq: Etau-ET} to obtain:
\ifCLASSOPTIONonecolumn
\begin{flalign}
    &\E[\tau] \le
    \sum_{i=1}^M \left(\frac{\log_2\left(\frac{1 \! - \! \rho_i(0)}{\rho_i(0)}\right)}{C}+\frac{\log_2(2q)}{q\cdot C} \right) \rho_i(0)  {\huge \mathds{1}_{\rho_0^{(i)} < 0.5}} 
    + \! \left\lceil\frac{\log_2(\frac{1-\epsilon}{\epsilon})}{C_2}\right\rceil \! \frac{C_2}{C_1}
   \! + \! \left(\frac{\log_2(2q) }{q C}-\frac{C_2}{C_1} \right)\frac{1 \! - \! \frac{\epsilon }{1 \! - \! \epsilon }2^{-C_2}}{1 \! -  \! 2^{-C_2}} 2^{-C_2} \,.
  \label{eq: arbitrary_distribution}&
\end{flalign}
\else
\begin{align}
    &\E[\tau] \le
    \sum_{i=1}^M \left(\frac{\log_2\left(\frac{1-\rho_i(0)}{\rho_i(0)}\right)}{C}+\frac{\log_2(2q)}{q\cdot C} \right) \rho_i(0)  {\huge \mathds{1}_{\rho_0^{(i)} < 0.5}} \nonumber \\
    &+ \! \left\lceil\frac{\log_2(\frac{1-\epsilon}{\epsilon})}{C_2}\right\rceil \! \frac{C_2}{C_1} 
   \! + \! \left(\frac{\log_2(2q) }{q C}-\frac{C_2}{C_1} \right)\frac{1 - \frac{\epsilon }{1 \! - \! \epsilon }2^{-C_2}}{1 \! -  \! 2^{-C_2}} 2^{-C_2} \,.
  \label{eq: arbitrary_distribution}
\end{align}
\fi
For the special case where $\rho_i(0) \ll \frac{1}{2} \quad \forall i = 1,\dots, M$, the log likelyhood ratio can be approximated by $\log_2(\frac{\rho_i(0)}{1-\rho_i(0)}) \lessapprox \log_2(\rho_i(i))$ to obtain a simpler expression of the bound \eqref{eq: arbitrary_distribution}:
\ifCLASSOPTIONonecolumn
\begin{align}
    \E[\tau]  < \frac{\mathcal{H}(\bm{\rho}_0)}{C}
    +\frac{\log_2(2q)}{q\cdot C} +\left\lceil\frac{\log_2(\frac{1-\epsilon}{\epsilon})}{C_2}\right\rceil\frac{C_2}{C_1}
   \phantom{=\,}+ \left(\frac{\log_2(2q) }{q C}-\frac{C_2}{C_1} \right)\frac{1 - \frac{\epsilon }{1 \! - \! \epsilon }2^{-C_2}}{1 \! -  \! 2^{-C_2}} 2^{-C_2} \, ,
\end{align}
\else
\begin{align}
    \E[\tau] & < \frac{\mathcal{H}(\bm{\rho}_0)}{C}
    +\frac{\log_2(2q)}{q\cdot C} \nonumber +\left\lceil\frac{\log_2(\frac{1-\epsilon}{\epsilon})}{C_2}\right\rceil\frac{C_2}{C_1}\\
   &\phantom{=\,}+ \left(\frac{\log_2(2q) }{q C}-\frac{C_2}{C_1} \right)\frac{1 - \frac{\epsilon }{1 \! - \! \epsilon }2^{-C_2}}{1 \! -  \! 2^{-C_2}} 2^{-C_2} \, ,
\end{align}
\fi
where $\mathcal{H}(\bm{\rho}(0))$ is the entropy of the p.d.f. $\bm{\rho}_0$ in bits.
\subsection{Uniform and Binomial Initial Distribution}
\label{sec: binomial distribution}

Using the bound of equation \eqref{eq: arbitrary_distribution}, we can develop a better upper bound on the blocklength for a systematic encoder with uniform input distribution when $\Omega = \{0,1\}^K$. It can be shown that the systematic transmissions transform the uniform distribution into a binomial distribution, see \cite{9174232}. The bound is constructed by adding the $K$ systematic transmissions to the bound in \eqref{eq: arbitrary_distribution} applied to the binomial distribution as follows:
\ifCLASSOPTIONonecolumn
\begin{align}
    \E[\tau] 
    \le 
    K + 
    \sum_{i=0}^K \! \left[\frac{\log_2(\frac{1-p^i q^{K\!-\!i}}{p^i q^{K\!-\!i}})}{C} + \frac{\log_2(2q)}{qC}\right] 
    &\binom{K}{i} p^i q^{K-i}  \mathbbm{ 1}_{(q^{K-i}p^i < 0.5)} 
    \nonumber
    \\
    &+
    \left\lceil\frac{\log_2(\frac{1-\epsilon}{\epsilon})}{C_2}\right\rceil\frac{C_2}{C_1} 
   \! + \! \left(\frac{\log_2(2q) }{q C} \! - \! \frac{C_2}{C_1} \right)\frac{1 - \frac{\epsilon }{1 \! - \! \epsilon }2^{-C_2}}{1 \! -  \! 2^{-C_2}} 2^{-C_2} \,.  \label{eq: binomial tau}
\end{align}
\else
\begin{align}
    &\E[\tau] \le K +  \label{eq: binomial tau}\\
    &\sum_{i=0}^K \! \left[\frac{\log_2(\frac{1-p^i q^{K\!-\!i}}{p^i q^{K\!-\!i}})}{C} \! + \! \frac{\log_2(2q)}{qC}\right] \!\! \binom{K}{i} p^i q^{K-i}  \mathbbm{ 1}_{(q^{K-i}p^i < 0.5)} \nonumber \\
    &+\left\lceil\frac{\log_2(\frac{1-\epsilon}{\epsilon})}{C_2}\right\rceil\frac{C_2}{C_1} \nonumber 
   \! + \! \left(\frac{\log_2(2q) }{q C} \! - \! \frac{C_2}{C_1} \right)\frac{1 - \frac{\epsilon }{1 \! - \! \epsilon }2^{-C_2}}{1 \! -  \! 2^{-C_2}} 2^{-C_2} \,.
\end{align}
\fi
This bound, which assumes SEAD and systematic transmission, is the tightest achievability bound that we have developed for the model.


\section{Algorithm and Implementation}
\label{sec: algorithm and implementation}

In this section we introduce a  systematic posterior matching (SPM) algorithm with partitioning by thresholding of ordered posteriors (TOP), that we call SPM-TOP. The SPM-TOP algorithm guarantees the performance of bound \eqref{eq: stopping time bound} of Thm. \ref{theorem: simple rule} because both systematic encoding and partitioning via TOP enforce the SEAD partitioning constraint in equations \eqref{eq: simple partitioning} and \eqref{eq: simple partitioning 2}. 
The SPM-TOP algorithm also guarantees the performance of the bound \eqref{eq: binomial tau} because it is a systematic algorithm. 

\subsection{Partitioning by Thresholding of Ordered Posteriors (TOP)}
\label{sec: set construction}
The TOP rule is a simple method to construct $S_0$ and $S_1$ at any time $t$ from the vector of posteriors $\bm{\rho}_t$, which enforces the SEAD partitioning constraint of Thm. \ref{theorem: simple rule}.  The rule requires an ordering $\{b_{1},\dots,b_{M}\}$ of the vector of posteriors such that $\rho_{b_1}(t) \ge \rho_{b_2}(t) \ge \cdots \ge \rho_{b_M}(t)$. TOP builds $S_0$ and $S_1$ by finding a threshold $m$ to split $\{b_{1},\dots,b_{M}\}$ into two contiguous segments $\{b_{1},\dots,b_{m}\} =S_0$ and $\{b_{m+1},\dots,b_{M}\}=S_1$. To determine the threshold position, the rule first determines an index $m' \in \{1,\dots,M\}$ such that:
\begin{equation}
    \label{eq: median ro}
    \sum_{j=1}^{m'-1}\rho_{b_i}(y^t) < \frac{1}{2} \le \sum_{j=1}^{m'}\rho_{b_i}(y^t) \,,
\end{equation}
Once $m'$ is found, the rule must select between two possible alternatives: Either $m=m'$ or $m=m'-1$. In other words, all that remains to decide is whether to place $b_{m'}$ in $S_0$ or in $S_1$. We select the choice that guarantees that the absolute difference between $P_0$ and $P_1$ is no larger than the posterior of $b_{m'}$. 
The threshold $m$ is selected from $\{m'-1, m'\}$ as follows:
\begin{align}
    &\textrm{if }
    \sum_{i=1}^{m'}\rho_{b_i}(t)-\frac{1}{2} > \frac{1}{2}\rho_{b_{m'}}(t) \quad \text{then: }
    m = m'-1 
    \label{eq: Set construction S_1}
    \\
    &\textrm{if } \sum_{i=1}^{m'}  \rho_{b_i}(t)-\frac{1}{2} \le \frac{1}{2}\rho_{b_{m'}}(t) \quad \text{then: } 
    m = m' \,.
    \label{eq: Set construction S_0}
\end{align}
Note that since $m \in \{m'-1, m'\}$, then the posterior of $b_{m'}$ is no larger than that of $b_m$, and the posterior of $b_m$ is also the value of $\rho_{\min} = \min_{i \in S_0}\{\rho_i(t)\}$. 

\begin{claim}\label{claim: Non empty rule compliant}
The TOP rule guarantees that the SEAD constraints of Thm. \ref{theorem: simple rule}, given by \eqref{eq: simple partitioning} and \eqref{eq: simple partitioning 2} are satisfied.
\end{claim}
\begin{IEEEproof}
The TOP partitioning rule sets the threshold that  separates $S_0$ and $S_1$ exactly before or exactly after item $b_{m'}$ depending on weather case  \eqref{eq: Set construction S_1} or case \eqref{eq: Set construction S_0} occurs.
To show that the TOP rule guarantees that the SEAD constraints in Thm. \ref{theorem: simple rule} are satisfied note that the threshold lies before item $b_{m'}$ if case \eqref{eq: Set construction S_1} occurs. Then, by the first inequality of \eqref{eq: median ro} and by \eqref{eq: Set construction S_1}:
\begin{align}
    P_0 &= \sum_{i=1}^{m'-1}\rho_{b_i}(y^t) 
    = \sum_{i=1}^{m'}\rho_{b_i}(y^t)-\rho_{b_{m'}}(t) < \frac{1}{2}
    \label{eq: partition case 1}
    \\    
    P_0 & > \frac{1}{2} + \frac{1}{2}\rho_{b_{m'}}(t)-\rho_{b_{m'}}(t) = \frac{1}{2}-\frac{1}{2}\rho_{b_{m'}}(t) \,.
    \label{eq: partition case 2}
\end{align}
When case \eqref{eq: Set construction S_0} occurs, the threshold is set after item $b_{m'}$. Then by the second inequality of \eqref{eq: median ro} and by \eqref{eq: Set construction S_1}:
\begin{align}
    P_0 &= \sum_{i=1}^{m'}\rho_{b_i}(y^t) \ge \frac{1}{2}
    \label{eq: partition case 3}
    \\
    P_0 &\le \frac{1}{2}+\frac{1}{2}\rho_{b_{m'}}(t) \,,
\end{align}
In either case we have:
\begin{equation}
    \frac{1}{2}- \frac{1}{2}\rho_{b_m}(t) \le P_0 \le \frac{1}{2}+\frac{1}{2}\rho_{b_m}(t) \label{eq: P0 bounds}
\end{equation}
By definition, $\Delta = P_0-P_1 = P_0-(1-P_0)=2P_0-1$.
Scale equation \eqref{eq: P0 bounds} by $2$ and subtract $1$, then: $-\rho_{b_{m'}}(t) < 2 P_0 - 1 \le \rho_{b_{m'}}(t)$. 
Then, $\mid \Delta \mid \; \le  \rho_{b_{m'}}(t) \le \rho_{b_m}(t) = \rho_{\min}$. 
This concludes the proof.
\end{IEEEproof}
We have shown that the construction of $S_0$ and $S_1$ can be as simple as finding the threshold item $b_{m'}$ where the c.d.f. induced by the ordered vector of posteriors crosses $\frac{1}{2}$, then, determining whether the threshold should be before or after item $b_{m'}$, and finally allocating all items before the threshold to $S_0$ and all items after the threshold to $S_1$.

\subsection{The Systematic Posterior Matching Algorithm}
\label{sec: spm algorithm}
The SPM-TOP algorithm is a low complexity scheme that implements sequential transmission over the BSC with noiseless feedback with a source message sampled from a uniform distribution. The algorithm has the usual communication and confirmation phases and the communication phase starts with systematic transmission. The systematic transmissions of the communication phase are treated as a separate systematic phase, for a total of three phases that we proceed to describe in detail.

\subsection{Systematic phase}
Let the sampled message be $\theta \in \{0,1\}^K$, with bits $b_i^{(\theta)}$, that is $\theta = \{b_1^{(\theta)},b_2^{(\theta)},\dots,b_{K}^{(\theta)}\}$. For $t=1,\dots,K$ the bits $b_t^{(\theta)}$ are transmitted and the vector $y^K \triangleq \{y_1,\dots,y_{K}\}$ is received. After the $K$-th transmission, both transmitter and receiver initialize a list of $K+1$ groups $\{\mathcal{G}_0,\dots,\mathcal{G}_K\}$, where each $\mathcal{G}_i$ is a tuple $\mathcal{G}_i = [N_i, L_i, h_i, \rho_i(y^t)]$. For each tuple $N_i$ is the count of messages in the group $N_i$; $L_i$ is the index of the first message in the group; $h_i$ is the shared Hamming distance between $y^K$ and any message in the group, that is: $l, s \in \mathcal{G}_i \implies \sum_{j=1}^K b^{(l)}_j \oplus y_j = \sum_{j=1}^K b^{(s)}_j \oplus y_j $; and $\rho_i(y^t)$ is the group's shared posterior. 
At time $t=K$, each group $\mathcal{G}_i, i = 1,\dots, K$ has that $N_i = \binom{K}{i}$, $L_i = 0$, $h_i = i$,  and $\rho_i(K) = p^j q^{K-j}$. 
The groups are sorted in order of decreasing probability, equivalent to increasing Hamming weight, since  $j > l \rightarrow p^l q^{K-l} < p^h q^{K-j}$. 
At the end of the systematic phase, the transmitter finds the index of the group $h_{\theta}$ and the index within the group $n_{\theta}$ corresponding to the sampled message $\theta$. The index $h_{\theta}$ is given by $h_{\theta} = \sum_{j=1}^K b_j^{(\theta)} \oplus y_j$ and the index $n_{\theta} \in \{0,\dots,\binom{K}{h_{\theta}}-1\}$ and is found via algorithm \ref{alg: get_index}.
The systematic phase is described by algorithm \ref{alg: systematic phase}.


\begin{algorithm}
\SetAlgoLined
\KwIn{Tx: $\theta = [b_1^{(\theta)},\dots,b_K^{(\theta)}]$ \Comment{Transmitted message.}}
\For{$t = 1, \dots, K$}{
\nl    channel input: $x_t = b^\theta_t$, output: $y_t$\;
}
Construct $\{\mathcal{G}_0,\dots,\mathcal{G}_K\}$, $\mathcal{G}_i=[N_i, L_i, h_i, \rho_i(K)] $\;
\nl $N_i \gets \binom{K}{i}$ 
\Comment{$N_i \triangleq \mid \mathcal{G}_i\mid$ }\;
\nl  $\rho_i(K) \gets q^{K-i}p^i$ \Comment{$j \in \mathcal{G}_i \rightarrow \rho_j(K) = \rho_i$ }\;
 
\nl $L_i \gets 0$, \quad $h_i \gets i$\;

\nl \textbf{map} $h_{\theta}$, $n_{\theta} \in
\mathcal{G}_{h_{\theta}} \triangleq \mathcal{G}^{(\theta)}$ \Comment{\textbf{Only Tx}, algorithm \ref{alg: get_index}}\;


 \caption{Systematic Phase}
 \label{alg: systematic phase}
\end{algorithm}



\begin{figure}[t]
\centering
\ifCLASSOPTIONonecolumn
\includegraphics[width=0.8\textwidth]{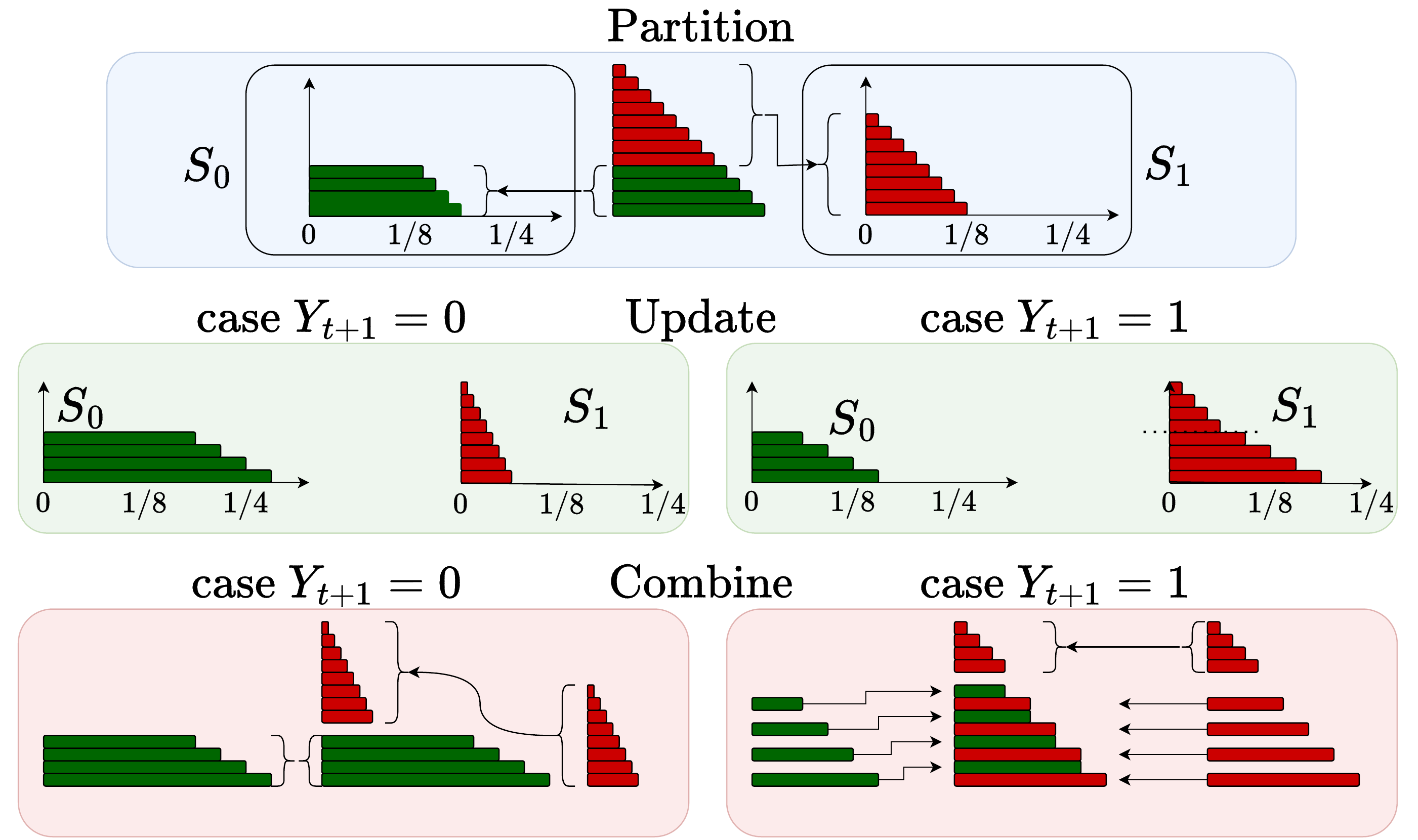}
\else
\includegraphics[width=0.5\textwidth]{combine4.pdf}
\fi
\caption{The two cases for update and combine $S_0$ and $S_1$ after partitioning with the TOP rule of claim \ref{claim: Non empty rule compliant}.}
\label{fig: SPM merge}
\end{figure}
\subsection{Communication Phase}
\label{sec: comm phase}
The communication phase consists of the transmissions after the systematic phase, and while all posteriors are lower than $\frac{1}{2}$. 
During communication phase, the transmitter attempts to boost the posterior of the transmitted message, past the threshold $\frac{1}{2}$, though any other message could cross the threshold instead, due to channel errors. 

The list of groups initialized in the systematic phase is maintained ordered by decreasing common posterior. The list of groups is partitioned into $S_0$ and $S_1$ before each transmission using rule \eqref{eq: median ro}. For this, the group $\mathcal{G}_m$ that contains the threshold item $b_{m}$ is found first, then all groups before $\mathcal{G}_m$ are assigned to $S_0$ and all the groups after $\mathcal{G}_m$ are assigned to $S_1$. To assign group $\mathcal{G}_m$ the index $n_m$ of item $b_m$ that sets the threshold is determined within group $\mathcal{G}_m$. The TOP rule demands that all items $j \in \mathcal{G}_m$ with index $n^{(j)}_m \le n_m$ be assigned to $S_0$ and all items $i \in \mathcal{G}_m$ with index $n^{(i)}_m > n_m$ to $S_1$. For this we split the group $\mathcal{G}_m$ into two by creating an new group with the segment of items past $n_m$ that belongs in $S_1$. However, if the item with index $n_{m}$ is the last item in $\mathcal{G}_m$, then the entire group $\mathcal{G}_m$ belongs in $S_0$ and no splitting is required.


After each transmission $t$, the posterior probabilities of the groups are updated using the received symbol $Y_t$ according to equation \eqref{eq: Bayes Update}. Each posterior is multiplied by a weight update, computed using equation \eqref{eq: update sum}, according to its assignment, $S_0$ or $S_1$. Then, the lists that comprise $S_0$ and $S_1$ are merged into a single sorted list. The process repeats for the next transmission and the communication phase is interrupted to transition to the confirmation phase when the posterior of a candidate message crosses the $\frac{1}{2}$ threshold. The communication phase might resume if the posterior of message $i$ that triggered the confirmation falls below $\frac{1}{2}$ rather than increasing past $1-\epsilon$, and all other posteriors are still below $\frac{1}{2}$. 

Not all groups need to be updated at every transmission. The partitioning method only requires visiting groups $\mathcal{G}_{1}, \mathcal{G}_{2}, \dots, \mathcal{G}_{m}$.
After the symbol $Y_{t+1}$ is received, we need to combine $S_0$ and $S_1$ into a single list, sorted by decreasing order of posteriors once updated. Figure \ref{fig: SPM merge} shows the three operations that are executed during the communication phase: partitioning the list, updating the posteriors of the partitions, and combining the updated partitions into a single sorted list. Note that in both cases, $Y_{t+1}=0$ and $Y_{t+1}=1$, either the entirety or a segment of the partition $S_1$ is just appended at the end of the new sorted list. This segment starts at the first group  $\mathcal{G}_j \in S_1$, such that its posterior $\rho_j(t)$ is smaller than the smallest in $S_0$. We avoid explicit operations on this segments and only saved the ``weight'' update factor as another item in the tuple described in at the beginning of this section. Every subsequent item in the list will share this update coefficient and could be updated latter on if it is encountered at either partitioning the list, updating the posteriors or combining the partitions. If this happens, the ``weight'' update is just ``pushed'' to the next list item. Since most of the groups belong to this ``tail'' segment, we expect to perform explicit operations only for a ``small'' fraction of the groups. This results in a large complexity reduction, which is validated by the simulation data of figure \ref{fig: complexity}.


\begin{algorithm}
\SetAlgoLined
\KwResult{index $\hat\theta$ s.t. $\rho_{\hat\theta}(\tau) \ge \frac{1}{2}$}
\KwIn{List of Groups $\{\mathcal{G}_0,\dots,\mathcal{G}_K\}$} 

\KwIn{ $h_{\theta}$, $n_{\theta} \in
\mathcal{G}_{h_{\theta}}$ \Comment{Tx Only}}

 \While{$\rho_0(t) < \frac{1}{2}$}{
    $m\gets 0, P_0 \gets 0, P_1 \gets 0$, $S_0 \gets \emptyset$\;
    \While{$P_0 + N_m \rho_m < \frac{1}{2}$}{
    $P_0 \gets P_0 + N_m \rho_m(t)$\;
    $m\gets m+1$\;
    }
    $S_0 \gets \{\mathcal{G}_0,\dots,\mathcal{G}_{m-1}\}$, $S_1 \gets \{\mathcal{G}_{m+1},\dots\}$\;
    $n \gets \lceil \frac{0.5-P_0}{\rho_m(t)}\rceil$ \Comment{Initial $n$ value}\;
    \If{$P_0 + n \rho_m(t) > \frac{1}{2}(1+\rho_m(t))$}{
    $n\gets n-1$ \Comment{\textbf{TOP rule}}}
    \eIf{$n > 0$ \&\& $n < N_m$}{
     $\mathcal{G}_{new} \gets
    \textbf{copy} \left(\mathcal{G}_m\right)$\; 
    $N_{new} \gets N_m - n$, $L_{new} \gets L_m+n, N_m \gets n$\;
    $S_0 \gets S_0 \cup \mathcal{G}_m, S_1 \gets \mathcal{G}_{new} \cup S_1$\;
    }{
    \textbf{if } $n==0$ \textbf{then} $S_1 = \mathcal{G}_m \cup S_1$, $m \gets m-1$\;
    \textbf{else then } $S_0 = S_0 \cup \mathcal{G}_m$
    }
    $P_0 \gets P_0 + n \rho_m(t), P_1 \gets 1-P_0$\;
    
    \If{Tx}{
    \If{$\mathcal{G}_m == \mathcal{G}^{(\theta)}$ \&\& $n_{\theta} \le L_m+n$}{$\mathcal{G}^{(\theta)}\gets \mathcal{G}_{new}^0$}
    \textbf{if }$\mathcal{G}^{(\theta)} \in S_0$ \textbf{then}  $x_{t+1} = 0$, \textbf{else }$x_{t+1} = 1$ \textbf{end}\;}  
    $t \gets t+1$ \Comment{Increase time index}\; 
    Update and Merge $S_0$ and $S_1$ via algorithm \ref{alg: update and merge}\;
 }
 \caption{Communication Phase}
 \label{alg: communication phase}
\end{algorithm}

\begin{algorithm}
\SetAlgoLined
\KwData{\textbf{channel input}: $x_{t+1}$, \textbf{output:} $y_{t+1}$}
\KwData{$m$: index of last item in $S_0$}
\KwData{$W_i$: Additional group parameter initialized at $1$: $\mathcal{G}_i=[N_i, L_i, h_i, \rho_i(y^t), W_i]$}

$W_0 \gets \frac{q}{P_{y_{t+1}}(q-p) + p}$ \Comment{Weight update for items in $S_0$} \;
$W_1 \gets \frac{p}{P_{y_{t+1}}(q-p) + p}$ \Comment{Weight Update for items in $S_1$} \;
$n_0 \gets 0$, $n_1 \gets 0$ \Comment{index of first groups in $S_0$ and $S_1$} \;
$W_{n_1} \gets W_{n_1} \cdot W_1$ \Comment{Update weight of first group in $S_1$}\;
\While{$n_0 \le m$}{
\eIf{$\rho_{n_0}(t) < W_{n_1} \cdot \rho_{n_1}(t)$}{
    $\rho_{n_1}(t) \gets \rho_{n_1}(t)\cdot W_{n_1}$\;
    $W_{n_1 + 1} \gets  W_{n_1+1}\cdot W_{n_1}$\Comment{Update Next weight}\; 
    $W_{n_1} \gets 1$  \Comment{Reset weight $W_{n_1}$} \;
    \textbf{insert } $\mathcal{G}_{n_1}$ in $S_0$ before $\mathcal{G}_{n_0}$\;
    $n_1 \gets n_1 + 1$ \Comment{Get next item from $S_1$}\;
    }{
    $\rho_{n_0}(t) \gets \rho_{n_0}(t) \cdot W_0$ \Comment{Update $\rho_{n_0}(t)$} \; 
    $n_0 \gets n_0+1$ \Comment{Get next item from $S_0$}\;
    }
}
 \caption{Simplified Update and Merge Algorithm}
 \label{alg: update and merge}
\end{algorithm}

\begin{algorithm}
\SetAlgoLined
\KwData{Group $\mathcal{G}_i$ with $N_i=1$ and $\rho_i(y^t) \ge \frac{1}{2}$}
$Z(t) \gets 0$, $N = \left \lceil C_2^{-1} \log_2\left(\frac{1-\epsilon}{\epsilon}\right)\right \rceil$\; 
\textbf{if } $U_i(t) + (N-1) C_2 \ge \log_2\left(\frac{1-\epsilon}{\epsilon}\right)$ \textbf{then }$N \gets N-1$\;
\While{$Z(t) \ge 0$ \&\& $Z(t) < N$}{
\textbf{Tx:} $X_t \gets 0$ \textbf{if } $\mathcal{G}^{(\theta)} == \mathcal{G}_i$ \textbf{else} $X_t \gets 1$\;
$Z(t) \gets Z(t) + 1-2 \cdot Y_t$\;
}
\eIf{$Z(t)==-1$}{
\textbf{Run } update and merge \Comment{Algorithm \ref{alg: update and merge}}\;
\textbf{Go to } Communication Phase \Comment{Algorithm \ref{alg: communication phase}}\;
}{
\textbf{Rx:} Get estimate $\hat{\theta}$ \Comment{ Algorithm \ref{alg: recover_estimate}}\;
}
 \caption{Confirmation Phase Algorithm}
 \label{alg: confirmation phase}
\end{algorithm}

\subsection{Confirmation Phase}
\label{sec: conf phase}
The Confirmation Phase is triggered when a candidate $i$ attains a posterior $\rho_i(y^t) \ge \frac{1}{2}$. During this phase the transmitter will attempt to boost $\rho_i(y^t)$, the posterior of candidate $i$, past the $1-\epsilon$ threshold, if it is the true message $\theta$. Otherwise it will attempt to drive its posterior below $\frac{1}{2}$. Clearly, the randomness of the channel could allow the posterior $\rho_i(y^t)$ to grow past $1-\epsilon$, even if it is the wrong message, resulting in a decoding error. Alternatively, the right message could still fall back to the communication phase, also due to channel errors. 
The confirmation phase lasts for as long as the posterior of the message that triggered its start stays between $\frac{1}{2}$ and $1-\epsilon$ or equivalently $U_i(t)$ stays between $0$ and $ \epsilon_U \triangleq \log_2\left(\frac{1-\epsilon}{\epsilon}\right)$.

There are no partitioning, update, or combining operations during the confirmation phase. If $j$ is the message in the confirmation phase, then the partitioning is just $S_0 = \{j\}$, $S_1 = \Omega \setminus \{j\}$.  A single update is executed if a fallback occurs, letting $\rho_i(y^{t})=\rho_i(y^{T_n}) \quad \forall i = 1,\dots,M$, where $n$ is the index of the confirmation phase round that just ended and $T_n$ is the time at which it started. This is because every negative update that follows a positive update results in every $\rho_i(y^{t})$ returning to the state it was at time $t-2$. This is summarized in claim \ref{claim: fix state space} that follows. During the confirmation phase it suffices to check if $U_j(t) \ge \epsilon_U$, in which case the process should terminate, or if $U_j(t) < 0$, in which case a fall back occurs. 

\begin{claim}[Confirmation Phase is a Discrete Markov Chain]
\label{claim: fix state space}
Let the partitioning of $\Omega$ at time $t=s$ be $S_0 = \{j\}$, $S_1 = \Omega \setminus \{j\}$, and suppose $Y_{s+1} = 0$. If the partitioning at time $t=s+1$ is also $S_0 = \{j\}$, $S_1 = \Omega \setminus \{j\}$, the same partitioning of time $s$, and $Y_{t+1}=1$, then for all $i=1,\dots,M$, $\rho_i(y^{t+1})=\rho_i(y^{t-1})$, that is $\rho_i(y^{s})=\rho_i(y^{s+2})$. 
\begin{IEEEproof}
See appendix \ref{sec: proof of fix state space}
\end{IEEEproof}
\end{claim}

During the confirmation phase we only need to count the difference between boosting updates and attenuating updates. 
Since the $U_i(t)$ changes in steps with magnitude $C_2$, then there is a unique number $N$ such that $U_i(T_n)+N C_2 \ge \epsilon_U$ and $U_i(T_n) + (N-1) C_2 < \epsilon_U$. Starting at time $t=T_n$, , since $S_0 = \{j\}$, any event $Y_{t+1}=0$ is a boosting update that results in $U_i(t+1)=U_i(t)+C_2$ and any event $Y_{t+1} = 1$ is an attenuating update that results in $U_i(t+1) = U_i(t)-C_2$. A net of $N$ boosting updates are needed to reach $U_i(T_n)+N C_2$.  Let the difference between boosting and attenuating updates  be $Z(t) \triangleq \sum_{s = T_n+1}^t (1-2 Y_{s})$.
The transmission terminates the first time $\tau$ where $Z(\tau)=N$. However, a fall back occurs if $Z(t)$ ever reaches $-1$ before reaching $N$.
The value of $N$ can be computed as follows: let $N_1 \triangleq \left \lceil C_2^{-1} \log_2\left(\frac{1-\epsilon}{\epsilon}\right)\right \rceil$ and let $\epsilon_n \triangleq \log_2\left(\frac{1-\epsilon}{\epsilon}\right)-N_1 C_2$. Suppose the confirmation phase starts at some time $t=T_n$, then, $N = N_1$ if $U_i(T_n) \ge \epsilon_N$, otherwise $N = N_1 + 1$.
Once $N$ is computed, all that remains is to track $Z(t)$, where $Z(t+1) = Z(t)+(1-2 Y_{t+1})$, and return to the communication phase if $Z(t)$ reaches $-1$ or terminate the process if $Z(t)$ reaches $N$.

\begin{algorithm}
\caption{map message i to index n and type h}
\label{alg: get_index}

\SetAlgoLined
\KwIn{$i,K$ \Comment{$i$: message, $K$: message length=}} 
\KwIn{\textbf{channel output}: $y_{K}$}
\KwResult{tuple $(h, n)$  \Comment{(type, index)}}
$e^i = i \oplus y^K$\;
$h \gets \sum^K_{j = 0}e^i_j$\;
$n \gets 0$\;
$c \gets h$\;
\For{$j=0, \dots, K$}{
    \If{$c == 0$} 
        {Break}
    \eIf{$e^i_j == 0$}{
        $n \gets n+\binom{K-j-1}{c-1}$\;
    }{
        $c \gets c-1$\;
    }
    
}
\end{algorithm}


\begin{theorem}
\label{thm: systematic}
[from \cite{9174232}] 
Suppose that $\Omega = \{0,1\}^K$ and $\rho_i(0)=2^{-K} \, \forall i \in \Omega$. Then, for $t=1,\dots,K$ the partitioning rule $S_0 = \{i \in \Omega \mid b^{(i)}_{t} = 0\}, S_1 = \{i \in \Omega \mid b^{(i)}_{t}= 1\}$, results in systematic transmission: $x^K = \theta$, and achieves exactly equal partitioning $P_0 = P_1 = \frac{1}{2}$.
\end{theorem}
\begin{IEEEproof}
First note that if $\Omega = \{0,1\}^K$, then for each $t=1,\dots,K$, exactly half of the items in $i \in \Omega$ have bit $b^{(i)}_t=0$ and the other half have bit $b^{(i)}_t = 1$. The theorem holds for $t = 1$, since the partitioning $S_0 = \{i \in \Omega \mid b^{(i)}_1 = 0\}, S_1 = \{i \in \Omega \mid b^{(i)}_{1}= 1\}$ results in half the messages in each partition and all the messages have the same prior. For $t=1,\dots,K-1$ note the partitioning $S_0 = \{i \in \Omega \mid b^{(i)}_{t} = 0\}, S_1 = \{i \in \Omega \mid b^{(i)}_{t}= 1\}$ only considers the first $t$ bits $b^{(i)}_1, \dots, b^{(i)}_t$ of each message $i$. Thus, all item $\{j \in \Omega \mid b^{(j)}_1,\dots,b^{(j)}_t = b_1,\dots,b_t\}$ that share a prefix sequence $b_1,\dots,b_t$ have shared the same partition at times $s = 1,\dots,t$, and therefore share the same posterior. There are exactly $2^{K-t}$ such difference posteriors. Also, exactly half
of the items that share the sequence $b_1,\dots,b_t$ have bit $b_{t+1} = 0$ and are assigned to $S_0$ at time $t+1$ and the other half have bit $b_{t+1} = 1$ and are assigned to $S_1$ at time $t+1$. Then, $S_0$ and $S_1$ will each hold half the items in each posterior group at each next time $t+1$ for $t = 1,\dots,K-1$, and therefore equal partitioning holds also at times $t=2,\dots,K$.
\end{IEEEproof}

\begin{algorithm}
\SetAlgoLined
\KwData{Index $L_i$, Hamming weight $h_i$}
\KwResult{$\hat\theta$ \Comment{Receiver Estimate of $\theta$}}
$\hat\theta \gets y^K$, $n\gets L_i$, $h \gets h_i$\;
\For{$j=0, \dots, K-1$}{
    \If{$h == 0$} 
        {Break}
    \eIf{$n < \binom{K-1-j}{h-1}$}{
        $\hat\theta_j \gets \neg \hat\theta_j$\;
        $h \gets h-1$\;
    }{
        
        $n \gets n-\binom{K-j-1}{h-1}$\;
    }
}
 \caption{get estimate $\hat\theta$ from index n and type h}
 \label{alg: recover_estimate}
\end{algorithm}
\subsection{Complexity of the SPM-TOP Algorithm}
\label{sec: spm complexity}

The memory complexity of the SPM-TOP algorithm is of order $O(K^2)$ because we use a triangular array of all combinations of the form $\binom{K}{i} \quad i \in \{0,\dots,K\}$. The algorithm itself stores a list of groups that grow linearly with $K$, since the list size is bounded by the decoding time  $\tau$. 

The time complexity of the SPM-TOP algorithm is of order $O(K^2)$. To obtain this result note that
the total number of items that the system tracks is bounded by the transmission index $t$. At each transmission $t$, partitioning, update and combine operations require visiting every item at most once. Furthermore, because of the complexity reduction described in Sec. \ref{sec: comm phase}, the system executes operations for only a fraction of all the items that are stored. The time complexity at each transmission is then of order $O(K)$, with a small constant coefficient. The number of transmissions required is approximately $K/C$ as the scheme approaches capacity. A linear number of transmissions, each of which requires a linear number of operations, results in an overall quadratic complexity, that is, order $O(K^2)$, for fixed channel capacity $C$.

The $K$ systematic transmissions only require storing the bits, and in the confirmation phase we just add each symbol $Y_t$ to the running sum. The complexity of this phase is then of order $O(K)$. Therefore, the complexity of $O(K^2)$ is only for the communication phase.

\section{SPM-TOP simulation results}
\label{sec: simulation results}
\begin{figure}[t]
\centering
\ifCLASSOPTIONonecolumn
\includegraphics[width=0.8\textwidth]{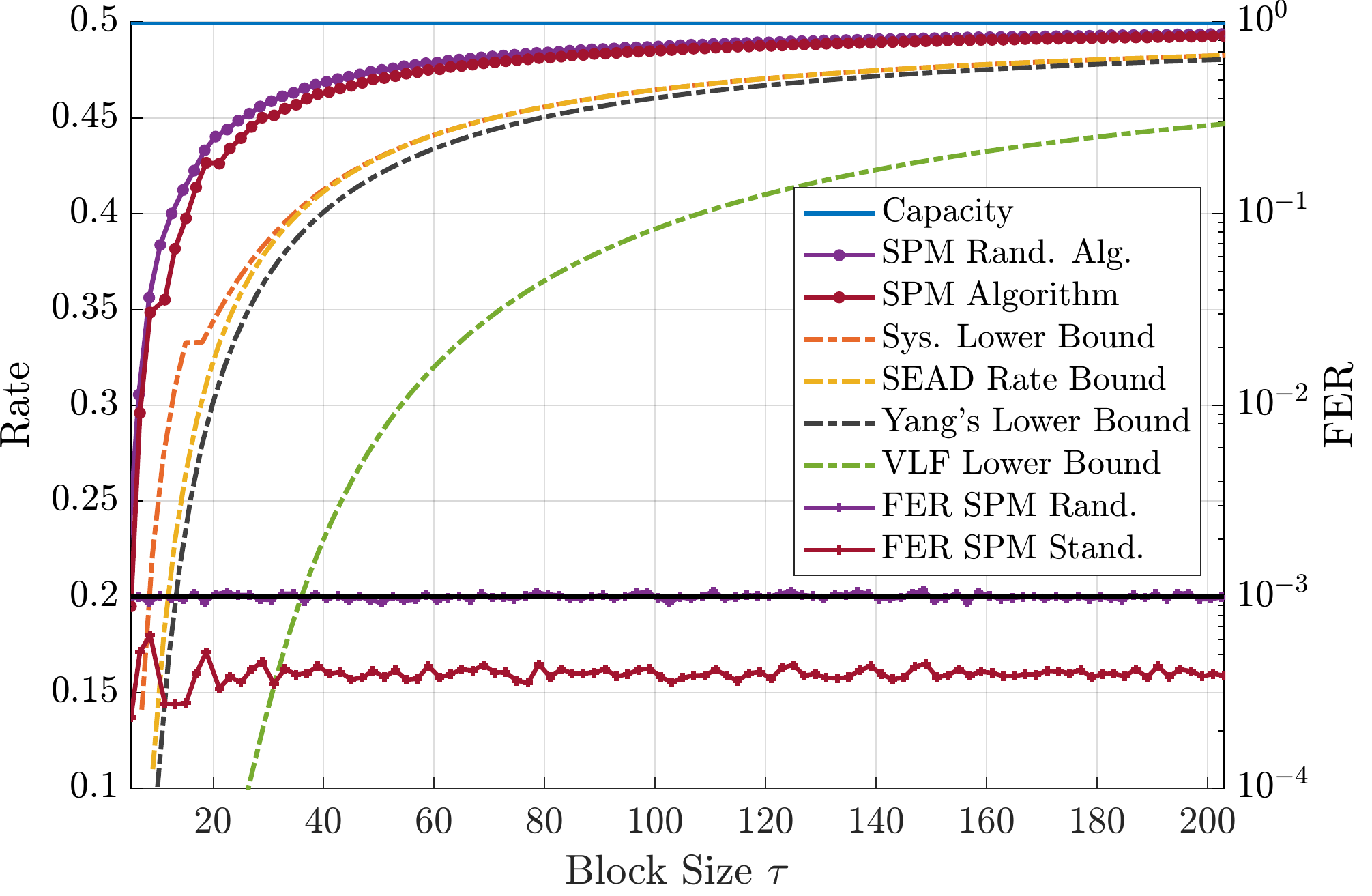}
\else
\includegraphics[width=0.47\textwidth]{allBounds.pdf}
\fi
\caption{SPM-TOP Rate and FER performance as a function of average blocklength. The red lines are rate and FER for the standard version of the SPM-TOP algorithm. The purple lines are rate and FER for a randomized version of the SPM-TOP algorithm.
The orange dash-dot line $-\cdot$ shows the bound defined in \ref{sec: binomial distribution} for a systematic encoder; the yellow dash-dot line $-\cdot$ shows the new bound introduced in Thm. \ref{theorem: simple rule} for system that enforces the SEAD constraints an the initial distribution is uniform; the black dash-dot line $-\cdot$ shows the SED lower bound by Yang \emph{et al.} \cite{Yang2021}, equation \eqref{eq: general bound} sec. \ref{sec: Achievable Rate}; and the green dash-dot line  $-\cdot$ shows Polyanskiy's VLF lower bound for a stop feedback system. The simulation consists of 1M Monte Carlo trials and for all curves the channel is one with capacity $C=0.50$, where $p \approx 0.110$, and a decoding threshold $\epsilon=10^{-3}$. }
\label{fig: rate_vs_blocklength}
\end{figure}

We validate the theoretical achievability bounds in Sec. \ref{sec: Main Section} and Sec. \ref{sec: arbitrary distribution} via simulations of the SPM-TOP algorithm. Figure \ref{fig: rate_vs_blocklength} shows
Simulated rate vs. blocklength performance curves of the SPM-TOP algorithm and the corresponding frame error rate (FER).
The plots show simulated rate curves for the standard SPM-TOP algorithm and a randomized version, as well as their associated error rate. The rate for the standard SPM-TOP algorithm is shown with the red curve with dots at the top and the corresponding FER is the red, jagged curve at the bottom. The standard SPM-TOP algorithm stops when a message $i$ attains $U_i(t)\ge \log_2 \left( (1-\epsilon)/\epsilon\right)$. Note that the FER is well below the threshold $\epsilon$. 
The randomized version of the SPM-TOP algorithm, 
implements a stopping rule that randomly alternates between the standard rule, which is stopping when a message $i$ attains $U_i(t)\ge \log_2 \left( (1-\epsilon)/\epsilon\right)$, and stopping when a message attains $U_i(t)\ge \log_2 \left( (1-\epsilon)/\epsilon\right)-C_2$, which requires one less correctly received transmission. The simulated rate of the randomized SPM-TOP version is the purple curve above the standard version. This randomized version aims to obtain a higher rate by forcing the FER to be close to the threshold $\epsilon$ rather than upper bounded  bounded by $\epsilon$. Note that
the corresponding FER, the horizontal purple curve with dots, is very close to the threshold $\epsilon=10^{-3}$, but not necessarily bounded above by the threshold. The simulation consisted of $10^6$ trials for each value of $K=1,\dots,100$ and for a decoding threshold $\epsilon=10^{-3}$ and a channel with capacity $C=0.50$. 
The simulated rate curves attain an average rate that approaches capacity rapidly and stay above all these theoretical bounds, also shown in Figure \ref{fig: rate_vs_blocklength} that we describe next. 

The two rate lower bounds introduced in this paper are shown in Figure \ref{fig: rate_vs_blocklength} for the same channel, capacity $C=0.50$ and threshold $\epsilon=10^{-3}$, used in the simulations.
The highest lower bound is labeled Sys. Lower Bound, and is the bound developed in Sec. \ref{sec: binomial distribution} for a system that uses a systematic phase to turn a uniform initial distribution into a binomial distribution and then enforces the SEAD constraints. The next highest bound is labeled SEAD Rate Bound and is the lower bound 
\eqref{eq: stopping time optimized bound} introduced in Thm. \ref{theorem: simple rule} for a system that enforces the SEAD constraints. This bound is a slight improvement from the SED lower bound by Yang \emph{et al.} \cite{Yang2021} that we show for comparison and is labeled Yang's Lower Bound. Also for comparison, we show Polyanskiy's VLF lower bound developed for a stop feedback system. Since a stop feedback system is less capable than a full feedback system, we expect that the lower bound for full feedback system approaches capacity faster than the VLF bound, which is what the previous $3$ bounds achieve.

\begin{figure}[t]
\centering
\ifCLASSOPTIONonecolumn
\includegraphics[width=0.8\textwidth]{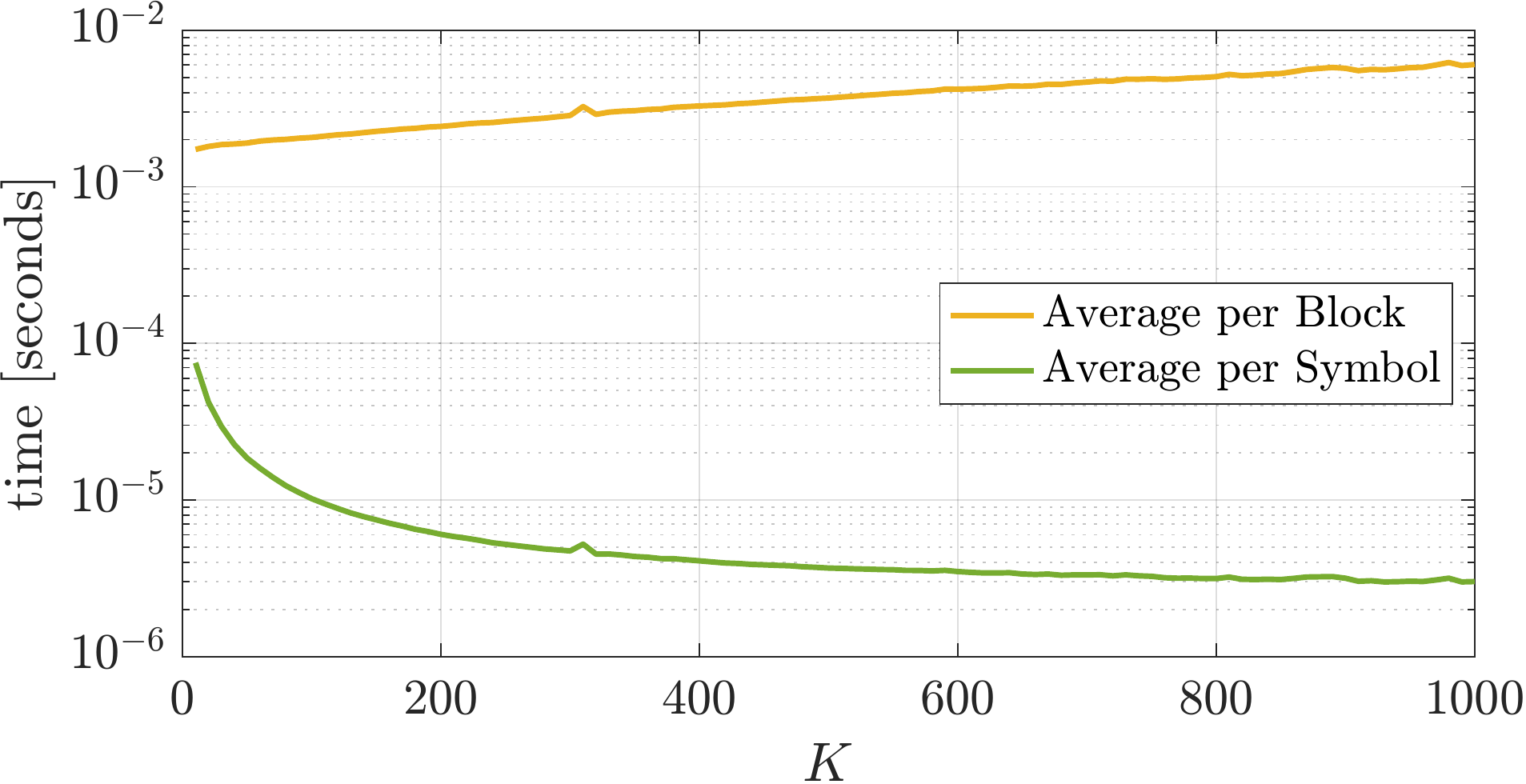}
\else
\includegraphics[width=0.47\textwidth]{logtimesec.pdf}
\fi
\caption{Shows average time as a function of $K$ for values of $K = 10,20,\dots, 1000$. The yellow line shows the time in milliseconds per frame transmission and the yellow line, shows microseconds per symbol transmission. The number of trials use to obtain this data is $100,000$ and the channel capacity is $C=0.50$. 
}
\label{fig: time complexity}
\end{figure}

\begin{figure}[t]
\centering
\ifCLASSOPTIONonecolumn
\includegraphics[width=0.7\textwidth]{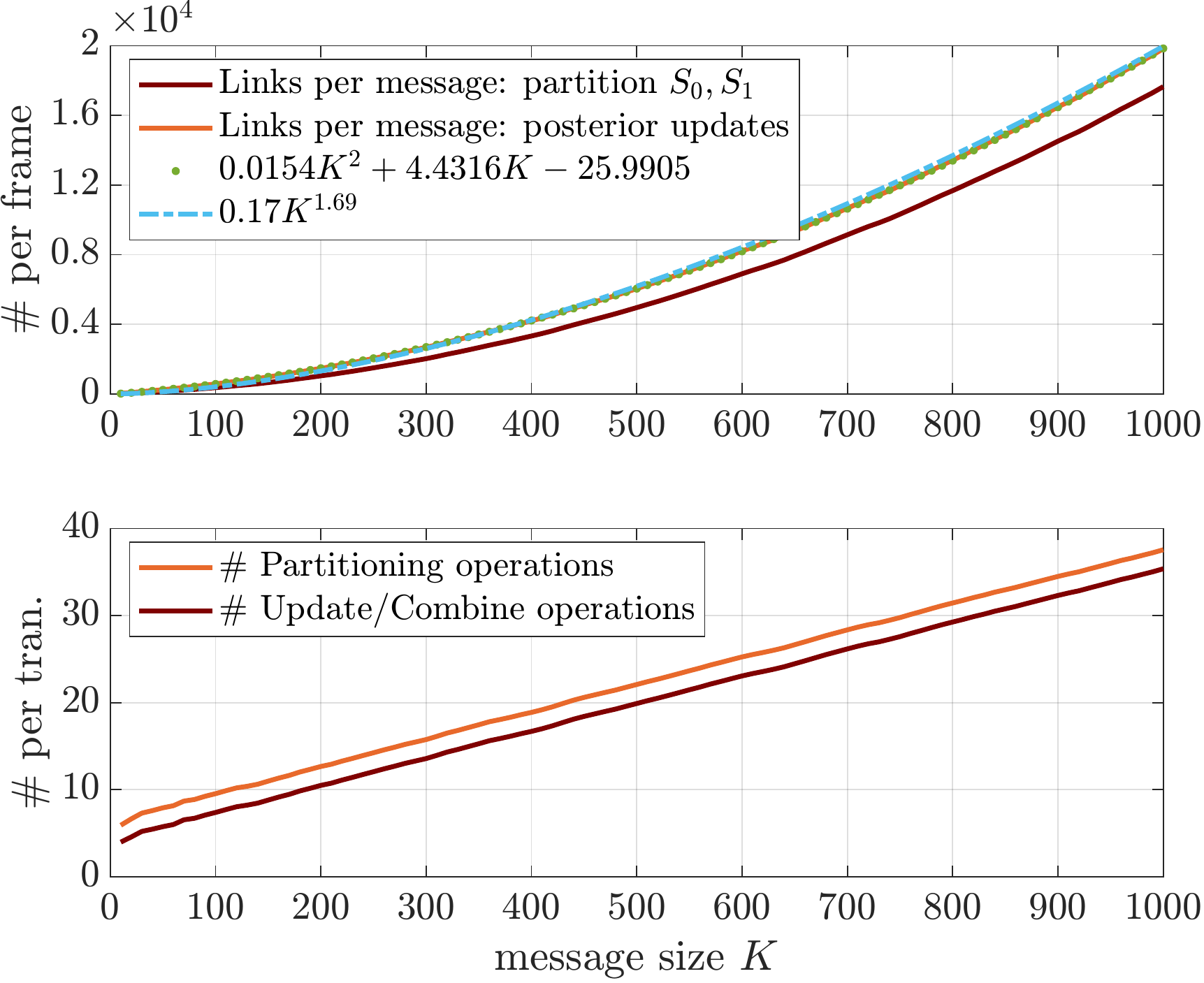}
\else
\includegraphics[width=0.47\textwidth]{operations2.pdf}
\fi
\caption{The plots shows the average of the number of links visited for update-merge operations, orange solid line, and for partitioning the list of groups into $S_0$ and $S_1$, red solid line, as a function of message size $K$. The top plot shows the average for the entire transmission of a frame, while the bottom plot shows the average for a single transmission. The top plot includes a quadratic line, green line with dots $..$, fitted to the update-merge curve for comparison with the simulation data. Also, the top plot includes the function $0.17K^{1.69}$ as a reference to show that the complexity order of the number of links visited during the transmission of a frame is below quadratic. This data was obtained with a simulation of $100,000$ trials, and for a channel with capacity $C = 0.5$
}
\label{fig: complexity}
\end{figure}

The empirical time complexity results of the SPM-TOP algorithm simulations vs. message size $K$ is shown in Figures \ref{fig: time complexity} and Figure \ref{fig: complexity}.
All simulations were performed on a 2019 MacBook Pro laptop with a $2.4$ GHz, $8$-core $i9$ processor and $16$ GB of RAM, and with transmitter and receiver operating alternatively on the same processor. 
First we show in Figure \ref{fig: time complexity} the average time, in milliseconds, taken per message, yellow line, and per transmitted symbol, green line. 
The average time per symbol drops very fast as the message size grows from $10$ to $200$ and then slowly stabilizes. This drop could be explained by the initialization time needed for each new message. However, the computer temperature and other processes managed by the computer's OS could also play a role in time measurements. For a more accurate characterization of the complexity's evolution as a function of message size $K$, we count the number of operations executed during the transmission of each symbol and each message, which are probability checks for partitioning before transmitting a symbol and probability updates after the transmission of a symbol.

The average number of probability checks and probability update operations per message vs message size $K$ are shown in the top of Figure \ref{fig: complexity}. To compare the data with a quadratic line, we fitted the parabola $0.0154 K ^{2}+4.4316 K-25.9905$ to the update-merge simulated data. Also for reference, the blue line shows the function $0.17 K^{1.69}$ to highlight that the complexity per message is below quadratic in the region of interest. 
The average number of probability checks and probability update operations per transmitted symbol are shown in the bottom of Figure \ref{fig: complexity}. The number of operations per symbol falls well below $K$, even when the number of probabilities that the system tracks is larger than $K$. This number is $K+1$ at $t=K$ and only increases with $t$. Note that for $K=1000$ both averages are below $40$.  
These results show that complexity of the SPM-TOP algorithm allows for fast execution time and validate the theory that the complexity order as a function of $K$ is linear for each transmission and quadratic for the whole message.

\section{Conclusion}
\label{sec: conclusion}

Naghshvar \emph{et al.} \cite{Naghshvar2015} established the ``small enough difference'' (SED) rule for posterior matching partitioning and used martingale theory to study asymptotic behavior and also showed how to develop a non-asymptotic lower bound on achievable rate.  Yang \emph{et al.} \cite{Yang2021} significantly improved the non-asymptotic achievable rate bound using martingale theory for the communication phase and a Markov model for the confirmation phase, still maintaining the SED rule.  However, partitioning algorithms that enforce the SED rule require a complex process of swapping messages back and forth between the two message sets $S_0$ and $S_1$ and updating the posteriors.

To reduce complexity, this paper replaces SED with the small enough {\em absolute} difference (SEAD) partitioning constraints. The SEAD constraints are more relaxed than SED, and they admit the TOP partitioning rule. In this way, SEAD allows a low complexity approach that organizes messages according to their type, i.e. their Hamming distance from the received word, orders messages according to their posterior, and partitions the messages with a simple threshold without requiring any swaps.

The main analytical results show that the SEAD constraints suffice to achieve at least the same lower bound that Yang \emph{et al.} \cite{Yang2021} showed to be achievable by SED.  Moreover, the new SEAD analysis establishes achievable rate bounds higher than those found by Yang \emph{et al.} \cite{Yang2021}.  The analysis does not use martingale theory for the communication phase and applies a surrogate channel technique to tighten the results.  An initial systematic transmission further increases the achievable rate bound.

The simplified encoder associated with SEAD has a complexity below order $O(K^2)$ and allows simulations for message sizes of at least 1000 bits.  These simulations reveal actual achievable rates that are enough above our new achievable-rate bounds that further analytical investigation to obtain even tighter achievable-rate bounds is warranted.  From a practical perspective, the simulation results themselves provide new lower bounds on the achievable rates possible for the BSC with full feedback.  For example, with an average block size of 200.97 bits corresponding to $k=99$ message bits, simulation results for a target codeword error rate of $10^{-3}$ show a rate of $R=0.493$ for the channel with capacity 0.5, i.e. $99$\% of capacity.

\section {Acknowledgements}
The authors would like to thank Hengjie Yang and Minghao Pan for their help with this manuscript.

\appendices
\label{appendix: appendices}

\section{Proof of claim \ref{claim: j confirmation singleton}}\label{sec: proof of singleton claim}
Proof that $\mid U_i(t+1)-U_i(t)\mid = C_2$  is equivalent to $S_0 = \{j\}$ ( or $S_1=\{j\}$). First let's prove the converse, if the set containing $j$ is not singleton, then constraint \eqref{eq: phase II step size} does not hold. Without loss of generality, assume $j \in S_0$ and suppose $\exists l \neq j$ s.t. $l \in S_0$. Since $P_0 \ge \rho_j(y^t)+\rho_l(t)$, then, $\Delta = 2 P_0-1 \ge 2 \rho_i(y^t)+2 \rho_j(y^t)-1 \ge 2 \rho_j(y^t)-1 > 0$.
By equation \eqref{eq: split 2 rho q}, when $j \in S_y$, then:
\ifCLASSOPTIONonecolumn
\begin{align}
    U_j(t+1) - U_j(t) 
    &=\log_2(2q) - \log_2\left(1 +(q-p)\frac{ \Delta
    -\rho_j(y^t)}{1-\rho_j(y^t)}\right) \label{eq: Ui > 0}
    \\
    &\le  
    \log_2(2q) \! - \! \log_2\left(1 \! + \! (q \! - \! p)\frac{ 2\rho_j(y^t)+2\rho_l(t)\!-\! 1 \!
    - \! \rho_j(y^t)}{1 \! - \! \rho_j(y^t)}\right) \nonumber
    \\
    &= 
    \log_2(2q) \! - \! \log_2\left(1 \! - \! (q \! - \! p) \! + \! (q \! - \! p)\frac{2\rho_l(t)}{1 \! - \! \rho_j(y^t)}\right)
    \nonumber
    \\
    & <
    \log_2(2q) - \log_2\left(1 -(q-p)\right) = C_2 \,.
\end{align}
\else
\begin{align}
    U_j&(t+1) - U_j(t) \nonumber
    \\
    &=\log_2(2q) - \log_2\left(1 +(q-p)\frac{ \Delta
    -\rho_j(y^t)}{1-\rho_j(y^t)}\right) \label{eq: Ui > 0}
    \\
    &\le  
    \log_2(2q) \! - \! \log_2\left(1 \! + \! (q \! - \! p)\frac{ 2\rho_j(y^t)+2\rho_l(t)\!-\! 1 \! 
    - \! \rho_j(y^t)}{1 \! - \! \rho_j(y^t)}\right) \nonumber
    \\
    &= 
    \log_2(2q) \! - \! \log_2\left(1 \! - \! (q \! - \! p) \! + \! (q \! - \! p)\frac{2\rho_l(t)}{1 \! - \! \rho_j(y^t)}\right)
    \nonumber
    \\
    & <
    \log_2(2q) - \log_2\left(1 -(q-p)\right) = C_2 \,.
\end{align}
\fi
Note from equation \eqref{eq: Ui > 0} that $U_j(t+1)-U_j(t)$ decreases with $\Delta$, therefore, replacing $\Delta$ with a lower bound gives an upper bound of difference \eqref{eq: Ui > 0}. For a lower bound, note that $\Delta \le 1$, and that setting $\Delta=1$ in \eqref{eq: Ui > 0}, results in $U_j(t+1)-U_j(t)=0$. In the case where $Y_{t+1} = X_{t+1}\oplus 1$, or $j \in S_{y^c}$,
by equation \eqref{eq: split 2 rho q}, the difference $U_j(t+1)-U_j(t)$ is:
\ifCLASSOPTIONonecolumn
\begin{align}
    U_j(t+1) - U_j(t) 
    &=\log_2(2p) - \log_2\left(1 -(q-p)\frac{ \Delta
    -\rho_j(y^t)}{1-\rho_j(y^t)}\right) \label{eq: Ui > 0 yc}
    \\
    &\ge 
    \log_2(2p) \! - \! \log_2\left( \! 1 \! -(q \! - \! p)\frac{ 2\rho_j(y^t) \! + \! 2\rho_l(t) \! - \! 1
    - \! \rho_j(y^t)}{1-\rho_j(y^t)}\right) 
    \\
    &= 
    \log_2(2p) \! - \! \log_2\left( \! 1  \! + \! (q \! - \! p)\left(1 \! - \! \frac{2\rho_l(t)
    }{1 \! - \! \rho_j(y^t)}\right)\right)
    \\
    &>
    \log_2(2p) - \log_2\left(2q\right) = -C_2 \, .
\end{align}
\else
\begin{flalign}
    U_j&(t+1) - U_j(t) & \nonumber
    \\
    &=\log_2(2p) - \log_2\left(1 -(q-p)\frac{ \Delta
    -\rho_j(y^t)}{1-\rho_j(y^t)}\right) \label{eq: Ui > 0 yc}&
    \\
    &\ge 
    \log_2(2p) \! - \! \log_2\left( \! 1 \! -(q \! - \! p)\frac{ 2\rho_j(y^t) \! + \! 2\rho_l(t) \! - \! 1
    - \! \rho_j(y^t)}{1-\rho_j(y^t)}\right) &
    \\
    &= 
    \log_2(2p) \! - \! \log_2\left( \! 1  \! + \! (q \! - \! p)\left(1 \! - \! \frac{2\rho_l(t)
    }{1 \! - \! \rho_j(y^t)}\right)\right)&
    \\
    &>
    \log_2(2p) - \log_2\left(2q\right) = -C_2 \, .&
\end{flalign}
\fi
To prove that if the set containing $j$ is singleton, then $|U_j(t+1)-U_j(t)|=C_2$, note that $S_0=\{j\} \implies \Delta = 2 \rho_j(y^t)-1$. The inequalities, therefore, become equalities and equations \eqref{eq: Ui > 0} and \eqref{eq: Ui > 0 yc} become $C_2$ and $-C_2$ respectively.



\section{Proof of existence of $U'_i(t)$ in Thm. \ref{theorem: surrogate martingale}}
\label{sec: proof of surrogate process}
The proof that a process like the one described in Thm. \ref{theorem: surrogate martingale} exists, consists of constructing one such process. Define the process $U'_i(t)$ by $U'_i(t) = U_i(t)$ if $Y^t \in \Bein$ and define $U'_i(t+1) \triangleq U'_i(t)+W'_i(t+1)$, where, to enforce constraints \eqref{eq: V ge 0} and \eqref{eq: C step size}, if $Y^t \!  \in \Yein$, the update $W'_i(t+1)$ is defined by \eqref{eq: W' definition}, and when $Y^t \!  \notin \Yein$, in which case $U'_i(t) \ge 0$, to enforce constraints \eqref{eq: martingale C1} and \eqref{eq: phase II step size} the update $W'_i(t+1)$ is given by $C_2(\mathbbm{1}_{(Y_{t+1}=0)}-\mathbbm{1}_{(Y_{t+1}=1)})$.


Denote the transmitted symbol $X_{t+1}$ by $X$, and the received symbol $Y_{t+1}$ by $Y$ and let $X^c = X\oplus 1$. The symbol $Y$ could either be $X$ or $X^c$. Also, we could have $\{i\in S_0\}$ or $\{i \in S_1\}$. These cases combine to $4$ possible events.
Define $W_i(t+1) \triangleq U_i(t+1)-U_i(t)$, and note that $W_i(t+1)$ can be derived from equation \eqref{eq: split 2 rho q} as follows:
\begin{equation}
    W_i(t \! + \! 1) =
    \begin{cases}
    \log_2(2q) \! + \! a_i \quad \text{if } i \in S_0, Y = X
    \\
    \log_2(2p) \! + \! b_i \quad \text{if } i \in S_0, Y = X^c
    \\
    \log_2(2p) \! + \! c_i \quad \text{if } i \in S_1, Y = X^c
    \\
    \log_2(2q) \! + \! d_i \quad \text{if } i \in S_1, Y = X
    \end{cases} \,,
    \label{eq: W definition}
\end{equation}
where $a_i$, $b_i$, $c_i$ and $d_i$ are given by:
\begin{align}
    a_i &= - \log_2\left(1-(q-p)\frac{\rho_j(y^t)-\Delta}{1-\rho_j(y^t)}\right) \label{eq: a_i}
    \\
    b_i &= -  \log_2\left(1+(q-p)\frac{\rho_j(y^t)-\Delta}{1-\rho_j(y^t)}\right)
    \\
    c_i &= - \log_2\left(1+(q-p)\frac{\rho_j(y^t)+\Delta}{1-\rho_j(y^t)}\right)
    \\
    d_i &= - \log_2\left(1-(q-p)\frac{\rho_j(y^t)+\Delta}{1-\rho_j(y^t)}\right) \,. \label{eq: d_i}
\end{align}

Let $a'_i$ and $d'_i$ be defined by:
\begin{align}
    a'_i &\triangleq \mathbb{1}_{(\Delta < 0)}\frac{1-\Delta}{1+\Delta}\log_2\left(1-(q-p)\Delta\right)-\frac{p}{q}b_i, \label{eq: a'_i}
    \\
    d'_i &\triangleq \mathbb{1}_{(d_i < 0)}d_i-\mathbb{1}_{(d_i \ge 0)}\frac{p}{q}c_i
    \, ,
    \label{eq: d'_i}
\end{align}
then, define the update $W'_i(t+1)$ by:
\begin{equation}
    W'_i(t \! + \! 1) =
    \begin{cases}
    \log_2(2q) \! + \! a'_i \quad \text{if } i \in S_0, Y = X
    \\
    \log_2(2p) \! + \! b_i \quad \text{if } i \in S_0, Y = X^c
    \\
    \log_2(2p) \! + \! c_i \quad \text{if } i \in S_1, Y = X^c
    \\
    \log_2(2q) \! + \! d'_i \quad \text{if } i \in S_1, Y = X 
    \end{cases}
    \,.
    \label{eq: W' definition}
\end{equation}
Need to show that the constraints \eqref{eq: C step size}-\eqref{eq: phase II step size} of Thm. \ref{theorem: Main Theorem}, and constraints \eqref{eq: U' smaller} and \eqref{eq: U' B bound}
are satisfied. When $U'_i(t) \ge 0$, since $W'_i(t+1)$ is defined in the same manner as $W_i(t+1)$, then constraints \eqref{eq: phase II V} and \eqref{eq: phase II step size} are satisfied. 

The proof that $U'_i(t)$ satisfies constraints \eqref{eq: C step size}, \eqref{eq: V ge 0} and \eqref{eq: U' smaller} is split into the case where $\Delta \ge 0$ and the case where $\Delta < 0$.

\subsection{Case $\Delta \ge 0$.}\label{sec: case non-negative Delta}

It suffices to show that for all $y^t \in \Yein$ and for all $i = 1,\dots,M$, if $\Delta \ge 0$, then  $\E[W'_i(t+1)\mid \theta = i, Y^t = y^t] = C$, Since $C > 0$ (constraint \eqref{eq: V ge 0}), and any weighted average would add up to $C$ (constraint \eqref{eq: C step size}).

When $\Delta \ge 0$, then $\rho_i(y^t)+\Delta > 0$, and therefore, $a'_i = - \frac{p}{q}b_j$ since $d_i > 0$. The expectation $\E[W'_j(t+1)\mid \theta = i, Y^t = y^t]$ can be computed from \eqref{eq: split 2 rho q}, where $\iota_i$ depends on whether $i \in S_0$ or $i\in S_1$. The expectation is given by either \eqref{eq: case S0 dif} or \eqref{eq: case S1 dif} respectively:
\begin{align}
    &q \log_2(2q) - p b_i
    + p  \log_2(2p) + p b_i = C\; \text{if } i \in S_0
    \label{eq: case S0 dif}
    \\
    &q \log_2(2q) - p c_i
    + p \log_2(2p) + p c_i = C \; \text{if } i \in S_1 \,.
    \label{eq: case S1 dif}
\end{align}
This proofs constraints \eqref{eq: C step size} and \eqref{eq: V ge 0} satisfied. To proof that constraint \eqref{eq: U' smaller} is satisfied, need to show that $W'_i(t+1)\le W_i(t+1)$. If suffices to compare the cases where $W'_i(t+1) \neq W_i(t+1)$, when $Y = X$. It suffices to compare the terms in which the pairs differ, that is, that $a'_i \le a_i$ and $d'_i \le d_i$. For this comparison, express $a_i$ and $d_i$ as positive logarithms as follows:
\begin{align}
    a_i &= \log_2\left(1 +                      \frac{(q-p)(\rho_i(y^t)-\Delta)}{1-\rho_i(y^t)-(q-p)(\rho_i(y^t)-\Delta)}\right) \label{eq: a_i Delta greater}
    \\
    a'_i &=
    \frac{p}{q} \log_2\left(1+\frac{(q-p)(\rho_i(y^t)-\Delta)}{1-\rho_i(y^t)}\right) \label{eq: a'_i Delta greater}
    \\
    d_i &= \log_2\left(1+\frac{(q-p)(\Delta+\rho_i(y^t))}{1 -\rho_i(y^t) -(q-p)(\Delta+\rho_i(y^t))}\right) \label{eq: d_i Delta greater}
    \\
    d'_i &= 
    \frac{p}{q}
    \log_2\left(1+\frac{(q-p)(\rho_i(y^t)+\Delta)}{1-\rho_i(y^t)}\right) \,.
    \label{eq: d'_i Delta greater}
\end{align}
Since $p \le \frac{1}{2}\rightarrow \frac{p}{q} \le 1$, then, we only need to show that the arguments of the logarithm in \eqref{eq: a_i Delta greater} is greater than that of \eqref{eq: a'_i Delta greater}, and similarly for \eqref{eq: d_i Delta greater} and \eqref{eq: d'_i Delta greater}. All arguments share the term $1$, then only inequalities \eqref{eq: arg a_i and a'_i} and \eqref{eq: arg d_i and d'_i} to hold:

\begin{align}
    \frac{(q-p)(\rho_i(y^t)-\Delta)}{1 \! - \!\rho_i(y^t) \!- \!(q \!- \!p)(\rho_i(y^t) \!- \!\Delta)} 
    &\ge
    \frac{(q-p)(\rho_i(y^t)-\Delta)}{1 \!- \!\rho_i(y^t)}
    \label{eq: arg a_i and a'_i}
    \\
    \frac{(q-p)(\Delta+\rho_i(y^t))}{1 \! - \!\rho_i(y^t) \!- \!(q \!- \!p)(\rho_i(y^t) \!+ \!\Delta)}  
    &\ge
    \frac{(q-p)(\rho_i(y^t)+\Delta)}{1 \!- \!\rho_i(y^t)} 
    \,. 
    \label{eq: arg d_i and d'_i}
\end{align}
The numerators on both inequalities are the same, and positive, since that $q-p > 0$ and if $i \in S_0$ then $\rho_i(y^t)-\Delta \ge 0$ and for all cases  $\rho_i(y^t)+\Delta \ge 0$ regardless. Both denominators on the left hand side are smaller than those in the right side, by exactly the numerators, and therefore, the inequalities hold.






\subsection{Case $\Delta < 0$.}
Next we show that constraints \eqref{eq: C step size}, \eqref{eq: V ge 0} and \eqref{eq: U' smaller} are satisfied when $\Delta < 0$. In the case where $\rho_i(y^t)+\Delta > 0$, then $d'_i$ is still $-\frac{p}{q}c_i \le d_i$ by equation \eqref{eq: d_i Delta greater}. 
However, whenever $\Delta < 0$, the term $\frac{1-\Delta}{1+\Delta}\log_2\left(1-(q-p)\Delta\right) \ge 0$ is added to $a'_i$. 
To show that constraint \eqref{eq: C step size}  holds, recall from \eqref{eq: jensen over S0} that:
\ifCLASSOPTIONonecolumn
\begin{align}
    \sum_{i=1}^M \rho_i \E[W_i(t+1)\mid \theta = i, Y^t=y^t]-C
    =& \sum_{i\in S_0}\rho_i(y^t) 
    \left(q a_i + \!  p b_i\right) + \! \sum_{i\in S_0}\rho_i(y^t) 
    \left(q d_i + \!  p c_i\right)
    \\
    \ge & \sum_{i\in S_0}\rho_i(y^t) 
    \left(q a_i + \!  p b_i\right)
    - \sum_{i \in S_1}  \rho_i(y^t)\log_2\left(1 -(q-p)^2\frac{ \rho_i(y^t)-\alpha}{1-\rho_i(y^t)}\right)
     \\
   \ge & \sum_{i\in S_0} \! \rho_i(y^t) 
    \left(q a_i \! + \!  p b_i\right) \! - \! \frac{1 \! + \! \alpha}{2}\log_2\left(1 \! + \! (q \! - \! p)^2\alpha\right).
    \label{eq: only P1}
\end{align}
\else
\begin{align}
    \sum_{i=1}^M &\rho_i \E[W_i(t+1)\mid \theta = i, Y^t=y^t]-C \nonumber
    \\ 
    =& \sum_{i\in S_0}\rho_i(y^t) 
    \left(q a_i + \!  p b_i\right) + \! \sum_{i\in S_0}\rho_i(y^t) 
    \left(q d_i + \!  p c_i\right)
    \\
    \ge & \sum_{i\in S_0}\rho_i(y^t) 
    \left(q a_i + \!  p b_i\right) \nonumber
    \\
    &- \sum_{i \in S_1}  \rho_i(y^t)\log_2\left(1 -(q-p)^2\frac{ \rho_i(y^t)-\alpha}{1-\rho_i(y^t)}\right)
     \\
   \ge & \sum_{i\in S_0} \! \rho_i(y^t) 
    \left(q a_i \! + \!  p b_i\right) \! - \! \frac{1 \! + \! \alpha}{2}\log_2\left(1 \! + \! (q \! - \! p)^2\alpha\right).
    \label{eq: only P1}
\end{align}
\fi
To obtain $\E[W'_i(t+1)\mid \mathcal{F}_t, \theta = i]$, replace $a_i$ by $a'_i$ in equation \eqref{eq: only P1}, and let $e_i \triangleq \frac{1+\alpha}{1-\alpha}\log_2\left(1+(q-p)^2\alpha\right)$. Then $a'_i = e_i - \frac{p}{q}b_i$ and $q a'_i + p b_i = q e_i$,  and replace to obtain:
\ifCLASSOPTIONonecolumn
\begin{align}
    \sum_{i=1}^M \rho_i \E[W'_i(t+1) \mid \theta = i, Y^t=y^t] - C
    =& \sum_{i\in S_0} \! \rho_i(y^t) 
    q e_i
    \! + \! \sum_{i\in S_1} \! \rho_i(y^t)\left(
    q d_i + p c_i\right) \label{eq: expected W' Delta smaller}
    \\
    \ge& q e_1\sum_{i\in S_0}\rho_i(y^t) 
    - \frac{1+\alpha}{2}\log_2\left(1+ (q-p)^2\alpha \right)
    \label{eq: only S1 prime}
    \\
    =& \frac{1-\alpha}{2} e_1  - \frac{1+\alpha}{2}\log_2\left(1+ (q-p)^2\alpha \right)
    \\
    =& \left(\frac{1 -  \alpha}{2}\frac{1 +  \alpha}{1 - \alpha} - \frac{1 + \alpha}{2}\right)\log_2\left(1+(q-p)^2\alpha\right) 
    \\
    =& \; 0 \,.
\end{align}
\else
\begin{align}
    \sum_{i=1}^M &\rho_i \E[W'_i(t+1) \mid \theta = i, Y^t=y^t] - C \nonumber
    \\
    =& \sum_{i\in S_0} \! \rho_i(y^t) 
    q e_i
    \! + \! \sum_{i\in S_1} \! \rho_i(y^t)\left(
    q d_i + p c_i\right) \label{eq: expected W' Delta smaller}
    \\
    \ge& q e_1\sum_{i\in S_0}\rho_i(y^t) 
    - \frac{1+\alpha}{2}\log_2\left(1+ (q-p)^2\alpha \right)
    \label{eq: only S1 prime}
    \\
    =& \frac{1-\alpha}{2} e_1  - \frac{1+\alpha}{2}\log_2\left(1+ (q-p)^2\alpha \right)
    \\
    =& \left(\frac{1 -  \alpha}{2}\frac{1 +  \alpha}{1 - \alpha} - \frac{1 + \alpha}{2}\right)\log_2\left(1+(q-p)^2\alpha\right) 
    \\
    =& \; 0 \,.
\end{align}
\fi
To show that $\E[W'_i(t+1)\mid \theta = i, Y^t = y^t] \ge 0$, (constraint \eqref{eq: V ge 0}), note that $d'_i$ is either unchanged from $d_i$, or the same as when $\Delta \ge 0$ and therefore, it holds for $i \in S_1$. For $i \in S_0$, note from the first term of equation \eqref{eq: expected W' Delta smaller}, that $\E[W'_i(t+1)\mid \theta = i, Y^t = y^t]-0 = \rho_i(y^t) q e_i$. Since $e_i \ge 0$, then the expectation is either $C$ or greater. 

Need to show, $W'_i(t+1) \le W_i(t+1)$ (constraint \eqref{eq: U' smaller}). It suffices to show that $a'_i \le a_i$ and $d'_i \le d_i$. Again, since $d'_i$ is either $d_i$ or the same as when $\Delta \ge 0$, we only need to show that $a'_i = e_i-\frac{p}{q}b_i \le a_i$. 
It suffices to show that for a positive scalar $\gamma$:
\begin{align}
    \gamma \left(q \left(e_i-\frac{p}{q}b_i\right) +  p b_i \right)
    &\le \gamma \left( q a_i + p b_i \right) \,.
    \label{eq: W' scaled smaller}
\end{align}
When $\Delta < 0$, then $e_i > 0$. We have that:
\begin{align}
    &q \left(e_i-\frac{p}{q}b_i\right) + \!  p b_i = \frac{1+\alpha}{1-\alpha}\log_2\left(1+(q-p)^2\alpha\right) \,.
    \label{eq: case 1 bound}
\end{align}
Recall from equation \eqref{eq: jensen over P0 and P1} that:
\begin{align}
    &q a_i + p b_i \ge - \log_2\left(1 -(q-p)^2\frac{\rho_{\min}+\alpha}{1-\rho_{\min}}\right)\,, \label{eq: qa + pb bound}
\end{align}
and let $\gamma = \frac{1-\alpha}{2}$, then, the scaled difference between left and right terms in \eqref{eq: W' scaled smaller} is given by:
\ifCLASSOPTIONonecolumn
\begin{align}
    \frac{1-\alpha}{2}\left(q a_i + p b_i\right) 
    &
    -\frac{1+\alpha}{2}\log_2(1+(q-p)^2\alpha) 
    \\
    \ge
    & - \frac{1-\alpha}{2} \log_2\left(1 -(q-p)^2\frac{\rho_{\min}+\alpha}{1-\rho_{\min}}\right)
    -\frac{1+\alpha}{2}\log_2\left(1-(q-p)^2\alpha\frac{\rho_{\min}-1}{1-\rho_{\min}}\right) \label{eq: alpha fraction}
    \\
    \ge & - \log_2\left( \! 1 \! - \!  \frac{(q \! - \! p)^2}{1 \! - \! \rho_{\min}} \left(\rho_{\min}\frac{1 \! + \! \alpha^2}{2}
    \! - \! \alpha^2\right)\right) \label{eq: jensen over U'}
    \\
    \ge & - \log_2\left( \! 1 \! - \!  \frac{(q \! - \! p)^2}{1 \! - \! \rho_{\min}} \left(\rho_{\min}\frac{1 \! + \! \alpha^2}{2}
    \! - \!  \rho_{\min} \alpha\right)\right) \label{eq: alpha and rho min}
    \\
    = & - \log_2\left( \! 1 \! - \!  \frac{(q \! - \! p)^2}{1 \! - \! \rho_{\min}} \left(\rho_{\min}\frac{(1 \! - \!  \alpha)^2}{2}\right)\right)  \ge 0
    \,.
    \label{eq: perfect square}
\end{align}
\else
\begin{align}
    \frac{1-\alpha}{2}&\left(q a_i + p b_i\right) -\frac{1+\alpha}{2}\log_2(1+(q-p)^2\alpha) 
    \nonumber
    \\
    \ge & - \frac{1-\alpha}{2} \log_2\left(1 -(q-p)^2\frac{\rho_{\min}+\alpha}{1-\rho_{\min}}\right) \nonumber
    \\
    &-\frac{1+\alpha}{2}\log_2\left(1-(q-p)^2\alpha\frac{\rho_{\min}-1}{1-\rho_{\min}}\right) \label{eq: alpha fraction}
    \\
    \ge & - \log_2\left( \! 1 \! - \!  \frac{(q \! - \! p)^2}{1 \! - \! \rho_{\min}} \left(\rho_{\min}\frac{1 \! + \! \alpha^2}{2}
    \! - \! \alpha^2\right)\right) \label{eq: jensen over U'}
    \\
    \ge & - \log_2\left( \! 1 \! - \!  \frac{(q \! - \! p)^2}{1 \! - \! \rho_{\min}} \left(\rho_{\min}\frac{1 \! + \! \alpha^2}{2}
    \! - \!  \rho_{\min} \alpha\right)\right) \label{eq: alpha and rho min}
    \\
    = & - \log_2\left( \! 1 \! - \!  \frac{(q \! - \! p)^2}{1 \! - \! \rho_{\min}} \left(\rho_{\min}\frac{(1 \! - \!  \alpha)^2}{2}\right)\right)  \ge 0
    \,.
    \label{eq: perfect square}
\end{align}
\fi
Equation \eqref{eq: alpha fraction} follows from \eqref{eq: qa + pb bound}. In \eqref{eq: jensen over U'}, Jensen's inequality is used, where: \\ 
$\frac{1-\alpha}{2}(\rho_{\min}+\alpha)+\frac{1+\alpha}{2}\alpha(\rho_{\min}-1) = -\alpha ^2 +\rho_{\min}\frac{1-\alpha+\alpha+\alpha^2}{2} $. In \eqref{eq: alpha and rho min} note that $\alpha \le \rho_{\min} \implies \alpha^2 \le  \rho_{\min}\alpha$. Finally $1-2\alpha +\alpha^2 = (1-\alpha)^2 \ge 0$.
We conclude that $e_i-\frac{p}{q}b_i \le a_i$, and therefore $W'_i(t+1) \le W_i(t+1)$.

\subsection{Proof of constraint \eqref{eq: U' B bound}}
Finally need to show that constraint  \eqref{eq: U' B bound} is satisfied, that is: $U'_i(T_{n+1}) \! - \! \frac{p}{q}(U_i(T_n)  \! - \! C_2) \le \frac{1}{q}\log_2(2q)$.
We have shown that the update $W'_i(t)$ allows the process $U'_i(t)$ to meet constraints \eqref{eq: V ge 0}-\eqref{eq: phase II step size} of Thm. \eqref{theorem: Main Theorem} and constraints \eqref{eq: U' smaller} and \eqref{eq: U' smaller}. Note that by the definition of $T'_n$ in Thm. \ref{theorem: surrogate martingale} it is possible that the process $U'_i(t)$ restarts when $U_i(t)$ falls from confirmation, at a time $t_0^{(n+1)}$, without ever attaining $U'_i(t) \ge 0$. We could construct a third process $U''_i(t)$ that preserves all the properties of $U'_i(t)$, and with $U''_i(t) \ge 0$ if $U_i(t) \ge 0$. The process $U''_i(t)$ could be initialized by $U''_i(t_0^{(n)}) = U_i(t_0^{(n)})$ and then letting $U''_i(t+1) = U''_i(t) + W''_i(t+1)$ with step size $W''_i(t)$ defined by:
$W''_i(t+1) = \max\{\min\{W_i(t+1), -U_i(t)\}, W'_i(t+1)\}$. The inner minimum guarantees that $U''_i(t)$ reaches $0$ if $U_i(t)$ does, and the outer maximum guarantees that the step size is at least that of $U'_i(t)$. 
Then the processes $U_i(t)$ and $U''_i(t)$ cross $0$ at the same time, and share the same values when $U_i(t) < 0$, that is:
\begin{alignat}{3}
    &U_i(t \! + \! 1) \ge \! 0  \quad &\implies& \quad & U''_i(t+1) \ge 0 \label{eq: same crossing time}
    \\
    &U_i(t) \le 0 &\implies& &U''_i(t) = U_i(t) \,. \label{eq: same smaller values}
\end{alignat}
Using the process $U''_i(t)$ and equation \eqref{eq: same smaller values}, the expression in constraint \eqref{eq: U' B bound} becomes:
\ifCLASSOPTIONonecolumn
\begin{align}
    U''_i(t+1)-\frac{p}{q}\left(U_i(t+1)-C_2\right) U_i(t) \! \left( \! 1 \! - \! \frac{p}{q}\right) \! + \! W''_i(t \! + \! 1) \!
    - \! \frac{p}{q} \left(W_i(t \! + \! 1) \! - \! C_2\right).\label{eq: U W' and W}
\end{align}
\else
\begin{flalign}
    U''_i(&t+1)-\frac{p}{q}\left(U_i(t+1)-C_2\right)
    \nonumber &
    \\
    =& U_i(t) \! \left( \! 1 \! - \! \frac{p}{q}\right) \! + \! W''_i(t \! + \! 1) \!
    - \! \frac{p}{q} \left(W_i(t \! + \! 1) \! - \! C_2\right).\label{eq: U W' and W}&
\end{flalign}
\fi
In the case where $W''_i(t+1) = -U_i(t)$, we have that \\
$U''_i(t+1) = 0$ and $U_i(t+1) \in [0, C_2]$, then:
\ifCLASSOPTIONonecolumn
\begin{align}
    U''_i(t+1)  \! - \! \frac{p}{q}(U_i(t \! + \! 1)  \! - \! C_2) 
    &= -\frac{p}{q}(U_i(t)+W_i(t+1)-C_2)
    \\
    &= \frac{p}{q}C_2 - \frac{p}{q}(U_i(t)+W_i(t+1)) \label{eq: U + W}
    \\
    &\le \frac{p}{q}C_2 \le \frac{1}{q}\log_2(2q) \,.
    \label{eq: case U'Tn = 0}
\end{align}
\else
\begin{align}
    U''_i(t+1) & \! - \! \frac{p}{q}(U_i(t \! + \! 1)  \! - \! C_2) 
    \nonumber
    \\
    &= -\frac{p}{q}(U_i(t)+W_i(t+1)-C_2)
    \\
    &= \frac{p}{q}C_2 - \frac{p}{q}(U_i(t)+W_i(t+1)) \label{eq: U + W}
    \\
    &\le \frac{p}{q}C_2 \le \frac{1}{q}\log_2(2q) \,.
    \label{eq: case U'Tn = 0}
\end{align}
\fi
The first inequality in \eqref{eq: case U'Tn = 0} follows since $U''_i(t)=0 \implies W_i(t+1)\ge -U_i(t)$ and the second inequality holds because:
\ifCLASSOPTIONonecolumn
\begin{align}
    \log_2(2q)-p C_2
    &= 1+ (1-p)\log_2(1-p) +p \log_2(p) \nonumber
    = C \ge 0 \,.
\end{align}
\else
\begin{align}
    \log_2(2q)-p C_2
    &= 1+ (1-p)\log_2(1-p) +p \log_2(p) \nonumber
    \\
    &= C \ge 0 \,.
\end{align}
\fi
For the case where $W''_i(t) > -U_i(t)$ we solve a constraint maximization of expression \eqref{eq: U W' and W}, where the constraint is $U_i(t) < 0$ (or $\rho_i(y^t) < \frac{1}{2}$). 
For simplicity we subtract the constant $\frac{1}{q}\log_2(2q)$ from \eqref{eq: U W' and W}.

Let $i\in \{1,\dots,M\}$ be arbitrary and let $\rho \triangleq \rho_i(y^t)$, and $\alpha \triangleq |\Delta|$. Using the definitions of $W_i(t)$ and $W''_i(t)$ in \eqref{eq: W definition}, \eqref{eq: W' definition} and \eqref{eq: a'_i}-\eqref{eq: d'_i}, we explicitly find expressions for
$U''_i(t+1) \! - \! \frac{p}{q}(U_i(t \! + \! 1)  \! - \! C_2)$ in terms of $\rho$, $p$, $q$ and $\alpha$. 

When $\{i \in S_0\}\cap \{\Delta < 0\}$ or $\{i \in S_1\}\cap \{\Delta \ge 0\}$ the expression is given by:
\ifCLASSOPTIONonecolumn
\begin{align}
    \log_2\left(\frac{\rho}{1 \! - \! \rho}\right)
    \left(1 \! - \! \frac{p}{q}\right)
    - & \frac{p}{q}\log_2(2q)
    -  \frac{p}{q}\log_2(2p) 
    \label{eq: Delta less than 0}
    \\
    & + \mathbbm{1}_{\Delta < 0}\frac{1+\alpha}{1-\alpha}\log_2(1+(q-p)^2\alpha) 
    + \frac{p}{q}\log_2\left(1 \! + \! (q \! - \! p) \frac{\rho \! + \! \alpha}{1 \! - \! \rho}\right) 
    +
    \frac{p}{q}\log_2\left(1 \! - \! (q \! - \! p) \frac{\rho \! + \! \alpha}{1 \! - \! \rho}\right) \, , 
    \nonumber
\end{align}
\else
\begin{align}
    &\log_2\left(\frac{\rho}{1 \! - \! \rho}\right)
    \left(1 \! - \! \frac{p}{q}\right)
    - \frac{p}{q}\log_2(2q)
    - \frac{p}{q}\log_2(2p) \nonumber
    \\
    &+\mathbbm{1}_{\Delta < 0}\frac{1+\alpha}{1-\alpha}\log_2(1+(q-p)^2\alpha) \label{eq: Delta less than 0}
    \\
    &+ \frac{p}{q}\log_2\left(1 \! + \! (q \! - \! p) \frac{\rho \! + \! \alpha}{1 \! - \! \rho}\right)
    \! + \!
    \frac{p}{q}\log_2\left(1 \! - \! (q \! - \! p) \frac{\rho \! + \! \alpha}{1 \! - \! \rho}\right) \, , \nonumber
\end{align}
\fi
and when $\{i \in S_0 \}\cap \{\Delta \ge 0\}$ or $ \{i \in S_1 \} \cap \{ \Delta < 0\}$ it is given by:
\ifCLASSOPTIONonecolumn
\begin{align}
    \log_2\left(\frac{\rho}{1 \! - \! \rho}\right)
    \left(1 \! - \! \frac{p}{q}\right)
    - \frac{p}{q}\log_2(2q)
    - \frac{p}{q}\log_2(2p) \label{eq: Delta greater than 0}
    +\frac{p}{q}\log_2\left(1 \! + \! (q \! - \! p) \frac{\rho \! - \! \alpha}{1 \! - \! \rho}\right)
    +\frac{p}{q} \log_2\left(1 \! - \! (q \! - \! p) \frac{\rho \! - \! \alpha}{1 \! - \! \rho}\right) \, .
\end{align}
\else
\begin{align}
    &\log_2\left(\frac{\rho}{1 \! - \! \rho}\right)
    \left(1 \! - \! \frac{p}{q}\right)
    - \frac{p}{q}\log_2(2q)
    - \frac{p}{q}\log_2(2p) \label{eq: Delta greater than 0}
    \\
    &+\frac{p}{q}\log_2\left(1 \! + \! (q \! - \! p) \frac{\rho \! - \! \alpha}{1 \! - \! \rho}\right)
    +\frac{p}{q} \log_2\left(1 \! - \! (q \! - \! p) \frac{\rho \! - \! \alpha}{1 \! - \! \rho}\right) \,.\nonumber
\end{align}
\fi

\subsection{Maximizing \eqref{eq: Delta less than 0}}
The maximum of \eqref{eq: Delta less than 0} happens when $\Delta > 0$, since the term with the indicator function is non-negative.  Since $\alpha \le \frac{1}{3}$, then $\frac{1+\alpha}{1-\alpha} \le 2$, and we proceed to solve:

\begin{align}
    &\textbf{maximize } f(\rho, \alpha)
    \\
    &\textbf{subject to }  \rho \le \frac{1}{2}, \alpha \le 1-2\rho \,,
\end{align}
where $f(\rho,\alpha)$ is defined by:
\ifCLASSOPTIONonecolumn
\begin{align}
    f(\rho, \alpha) 
    \triangleq & \log_2\left(\frac{\rho}{1 \! - \! \rho}\right)
    \left(1 \! - \! \frac{p}{q}\right)
    +2\log_2(1+(q-p)^2\alpha) 
    \nonumber
    \\
    &
    - \frac{p}{q}\log_2(2q)
    - \frac{p}{q}\log_2(2p)+ \frac{p}{q}\log_2\left(1 \! + \! (q \! - \! p) \frac{\rho \! + \! \alpha}{1 \! - \! \rho}\right)
    \! + \!
    \frac{p}{q}\log_2\left(1 \! - \! (q \! - \! p) \frac{\rho \! + \! \alpha}{1 \! - \! \rho}\right) \,. 
    \label{eq: first maximization}
\end{align}
\else
\begin{align}
    &f(\rho, \alpha) \triangleq \log_2\left(\frac{\rho}{1 \! - \! \rho}\right)
    \left(1 \! - \! \frac{p}{q}\right)
    \label{eq: first maximization}
    \\
    & +2\log_2(1+(q-p)^2\alpha) - \frac{p}{q}\log_2(2q)
    - \frac{p}{q}\log_2(2p)
    \nonumber
    \\
    &+ \frac{p}{q}\log_2\left(1 \! + \! (q \! - \! p) \frac{\rho \! + \! \alpha}{1 \! - \! \rho}\right)
    \! + \!
    \frac{p}{q}\log_2\left(1 \! - \! (q \! - \! p) \frac{\rho \! + \! \alpha}{1 \! - \! \rho}\right) \,. \nonumber
\end{align}
\fi
First we show that $f$ is increasing in $\rho$, by showing $\frac{d}{d\rho}f \ge 0$. Note that $\frac{d}{d \rho}\frac{\rho}{1-\rho} = \frac{1}{(1-\rho)^2}$ and $\frac{d}{d\rho}\frac{\rho+\alpha}{1-\rho} = \frac{1}{(1-\rho)^2} + \frac{\alpha}{(1-\rho)^2}$
\ifCLASSOPTIONonecolumn
\begin{align}
    \frac{\partial}{\partial \rho} \ln(2)f(\rho,\alpha) 
    = \frac{q-p}{q}\frac{1}{(1-\rho)\rho}+
     \frac{p}{q}\frac{1 \! + \! \alpha}{(1 \! - \! \rho)^2}\left(
    \frac{(q-p)}{1+(q \! - \! p)\frac{\rho+\alpha}{1-\rho}}
    -
    \frac{(q-p)}{1-(q \! - \! p)\frac{\rho+\alpha}{1-\rho}}\right) \,.
\end{align}
\else
\begin{align}
    \frac{\partial}{\partial \rho}& \ln(2)f(\rho,\alpha) 
    = \frac{q-p}{q}\frac{1}{(1-\rho)\rho}+
    \nonumber
    \\
    & \frac{p}{q}\frac{1 \! + \! \alpha}{(1 \! - \! \rho)^2}\left(
    \frac{(q-p)}{1+(q \! - \! p)\frac{\rho+\alpha}{1-\rho}}
    -
    \frac{(q-p)}{1-(q \! - \! p)\frac{\rho+\alpha}{1-\rho}}\right) \,.
\end{align}
\fi
Factor out the positive constant $\frac{1}{q}\frac{q-p}{1-\rho}$, to obtain:
\ifCLASSOPTIONonecolumn
\begin{align}
    \frac{1}{\rho}+ p\frac{1 \! + \! \alpha}{1 \! - \! \rho}
    &
    \left(
    \frac{1}{1+(q \! - \! p)\frac{\rho+\alpha}{1-\rho}}
    -
    \frac{1}{1-(q \! - \! p)\frac{\rho+\alpha}{1-\rho}}
    \right) 
    =
    \frac{1}{\rho}+ p\frac{1 \! + \! \alpha}{1 \! - \! \rho}\frac{\left(1 \! - \! (q \! - \! p)\frac{\rho \! + \! \alpha}{1 \! - \! \rho}\right)
    -\left(1 \! + \! (q \!- \! p)\frac{\rho \!+ \! \alpha}{1 \! - \! \rho}\right)}{1-(q-p)^2\left(\frac{\rho+\alpha}{1-\rho}\right)^2} 
    \\
    &=\frac{1}{\rho}- 2\frac{p(1+\alpha)(q-p)(\rho+\alpha)
    }{(1-\rho)^2-(q-p)^2(\rho+\alpha)^2} 
    =
    \frac{(1 \! - \! \rho)^2 \! - \!  (q \! - \! p)^2(\rho \! + \! \alpha)^2 \! - \! 2p\rho(1 \! + \! \alpha)(q \! - \! p)(\rho \! + \! \alpha)
    }{\rho (1-\rho)^2-\rho (q-p)^2(\rho+\alpha)^2} \, . \label{eq: f' numerator}
\end{align}
\else
\begin{align}
    &\frac{1}{\rho}+ p\frac{1 \! + \! \alpha}{1 \! - \! \rho}
    \left(
    \frac{1}{1+(q \! - \! p)\frac{\rho+\alpha}{1-\rho}}
    -
    \frac{1}{1-(q \! - \! p)\frac{\rho+\alpha}{1-\rho}}
    \right) 
    \nonumber
    \\
    &=\frac{1}{\rho}+ p\frac{1 \! + \! \alpha}{1 \! - \! \rho}\frac{\left(1 \! - \! (q \! - \! p)\frac{\rho \! + \! \alpha}{1 \! - \! \rho}\right)
    -\left(1 \! + \! (q \!- \! p)\frac{\rho \!+ \! \alpha}{1 \! - \! \rho}\right)}{1-(q-p)^2\left(\frac{\rho+\alpha}{1-\rho}\right)^2}
    \\
    &=\frac{1}{\rho}- 2\frac{p(1+\alpha)(q-p)(\rho+\alpha)
    }{(1-\rho)^2-(q-p)^2(\rho+\alpha)^2} =
    \nonumber
    \\
    &\frac{(1 \! - \! \rho)^2 \! - \!  (q \! - \! p)^2(\rho \! + \! \alpha)^2 \! - \! 2p\rho(1 \! + \! \alpha)(q \! - \! p)(\rho \! + \! \alpha)
    }{\rho (1-\rho)^2-\rho (q-p)^2(\rho+\alpha)^2} . \label{eq: f' numerator}
\end{align}
\fi
It suffices to show that the top of equation \eqref{eq: f' numerator} is non-negative. Since it decreases when $\alpha \le 1-2\rho$ then: 
\ifCLASSOPTIONonecolumn
\begin{align}
    (1-\rho)^2 - (q-p)^2(\rho+\alpha)^2
    & -2p\rho(1+\alpha)(q-p)(\rho+\alpha)
    \nonumber
    \\
    \ge& 
    (1-\rho)^2 - (q-p)^2(\rho +1-2\rho)^2
    -2p\rho(1+1-2\rho)(q-p)(\rho+1-2\rho)
    \\
    =& 
    (1-\rho)^2 - (q-p)^2(1-\rho)^2
    -2p\rho(2-2\rho)(q-p)(1-\rho)
    \\
    =& 
    (1-\rho)^2(1-(q-p)^2 - (q-p)4p\rho)
    \\
    =&(1-\rho)^2 4 p (q - \rho(q-p)) > 4 p (1-\rho)^2(q-\rho) > 0  \label{eq: q-p square} \,.
\end{align}
\else
\begin{align}
    (1-\rho)^2 &- (q-p)^2(\rho+\alpha)^2-2p\rho(1+\alpha)(q-p)(\rho+\alpha) \nonumber
    \\
    \ge& 
    (1-\rho)^2 - (q-p)^2(\rho +1-2\rho)^2
    \nonumber
    \\
    &-2p\rho(1+1-2\rho)(q-p)(\rho+1-2\rho)
    \\
    =& 
    (1-\rho)^2 - (q-p)^2(1-\rho)^2 \nonumber
    \\
    &-2p\rho(2-2\rho)(q-p)(1-\rho) 
    \\
    =& 
    (1-\rho)^2(1-(q-p)^2 - (q-p)4p\rho)
    \\
    =&(1-\rho)^2 4 p (q - \rho(q-p)) \nonumber
    \\
    &> 4 p (1-\rho)^2(q-\rho) > 0 \label{eq: q-p square}
    \,.
\end{align}
\fi
In equation \eqref{eq: q-p square} we have used $(q-p)^2 = (1-2p)^2 = 1-4p+4p^2=1-4 p q$ and $\rho (q-p) < \rho q < \rho$.
Since $\frac{\partial}{\partial \rho}f > 0$ then $f$ is increasing in $\rho$ and we can replace $\rho$ by $\frac{1-\alpha}{2}$ for an upper bound. Since $\frac{\rho + \alpha}{1-\rho} = \frac{\frac{1-\alpha}{2}+\alpha}{1-\frac{1-\alpha}{2}}=\frac{1-\alpha+2\alpha}{2-1+\alpha} = 1$, then $f(\alpha)\triangleq f\left(\frac{1-\alpha}{2},\alpha\right)$ is given by:
\ifCLASSOPTIONonecolumn
\begin{align}
    f(\alpha) =& \log_2\left(\frac{1 \! - \! \alpha}{1 \! + \! \alpha}\right)\left(1 \! - \! \frac{p}{q}\right) \! + \! 2\log_2(1 \! + \! (q \! - \! p)^2\alpha)
    \\
    &- \frac{p}{q}\log_2(2q)-\frac{p}{q}\log_2(2p) 
    +\frac{p}{q}\log_2(1 \! + \! (q \! - \! p))
    +\frac{p}{q} \log_2(1-(q-p))
    \\
    = & \log_2\left(\frac{1 \! - \! \alpha}{1 \! + \! \alpha}\right) \! \left(1 \! - \! \frac{p}{q}\right) \! + \! 2\log_2(1 \! + \! (q \! - \! p)^2\alpha) .
\end{align}
\else
\begin{align}
    f(\alpha) =& \log_2\left(\frac{1 \! - \! \alpha}{1 \! + \! \alpha}\right)\left(1 \! - \! \frac{p}{q}\right) \! + \! 2\log_2(1 \! + \! (q \! - \! p)^2\alpha)
    \\
    &- \frac{p}{q}\log_2(2q)-\frac{p}{q}\log_2(2p)
    \nonumber
    \\
    &+\frac{p}{q}\log_2(1 \! + \! (q \! - \! p))
    +\frac{p}{q} \log_2(1-(q-p))
    \\
    = & \log_2\left(\frac{1 \! - \! \alpha}{1 \! + \! \alpha}\right) \! \left(1 \! - \! \frac{p}{q}\right) \! + \! 2\log_2(1 \! + \! (q \! - \! p)^2\alpha) .
\end{align}
\fi
To complete the proof, it suffices to show that the last expression decreases in $\alpha$:
\ifCLASSOPTIONonecolumn
\begin{align}
    \frac{d}{d\alpha}\ln(2)f(\alpha)
    =&\left(1 \! - \! \frac{p}{q}\right)
    \frac{1 \! + \! \alpha}{1 \! - \! \alpha}
    \frac{-(1 \! + \! \alpha) \! - \! (1 \! - \! \alpha)}{(1+\alpha)^2}
    \! + \! \frac{2(q-p)^2}{1 \! + \! (q \! - \! p)^2\alpha}
    \\
    =&
    -2\frac{1}{q}
    \frac{q-p}{1-\alpha^2}
    +2 \frac{(q-p)^2}{1+(q-p)^2\alpha}
    \\
    =&
    2(q-p) \left(-\frac{1}{q}
    \frac{1}{1-\alpha^2}
    +\frac{q-p}{1+(q-p)^2\alpha}\right) 
    \\
    \le& -2(q-p)\left(1+\frac{p}{q} - q + p\right)
    = -2 p (q-p)\left(2+\frac{1}{q}\right) < 0
    \,.
\end{align}
\else
\begin{align}
    \frac{d}{d\alpha}&\ln(2)f(\alpha) \nonumber
    \\
    =&\left(1 \! - \! \frac{p}{q}\right)
    \frac{1 \! + \! \alpha}{1 \! - \! \alpha}
    \frac{-(1 \! + \! \alpha) \! - \! (1 \! - \! \alpha)}{(1+\alpha)^2}
    \! + \! \frac{2(q-p)^2}{1 \! + \! (q \! - \! p)^2\alpha}
    \\
    =&
    -2\frac{1}{q}
    \frac{q-p}{1-\alpha^2}
    +2 \frac{(q-p)^2}{1+(q-p)^2\alpha}
    \\
    =&2(q-p) \left(-\frac{1}{q}
    \frac{1}{1-\alpha^2}
    +\frac{q-p}{1+(q-p)^2\alpha}\right) 
    \\
    &\le -2(q-p)\left(1+\frac{p}{q} - q + p\right)
    \nonumber
    \\
    &= -2 p (q-p)\left(2+\frac{1}{q}\right) < 0
    \,.
\end{align}
\fi
Since $f$ is decreasing, then the maximum of equation \eqref{eq: Delta less than 0} is $0$, at $\alpha = 0$.

\subsection{Maximizing \eqref{eq: Delta greater than 0}}
The expression \eqref{eq: Delta greater than 0} is given by $g(\rho, \alpha) -\frac{p}{q}\log_2(2q)-\frac{p}{q}\log_2(2p)$ where $f(\rho, \alpha)$ is defined by:
\ifCLASSOPTIONonecolumn
\begin{align}
    g(\rho,\alpha) \triangleq \log_2\left(\frac{\rho}{1 \! - \! \rho}\right)
    \left(1 \! - \! \frac{p}{q}\right)
    +\frac{p}{q}\log_2\left(1 \! + \! (q \! - \! p) \frac{\rho \! - \! \alpha}{1 \! - \! \rho}\right)
    +\frac{p}{q} \log_2\left(1 \! - \! (q \! - \! p) \frac{\rho \! - \! \alpha}{1 \! - \! \rho}\right) \,. 
\end{align}
\else
\begin{align}
    g&(\rho,\alpha) \triangleq \log_2\left(\frac{\rho}{1 \! - \! \rho}\right)
    \left(1 \! - \! \frac{p}{q}\right)
    \\
    &+\frac{p}{q}\log_2\left(1 \! + \! (q \! - \! p) \frac{\rho \! - \! \alpha}{1 \! - \! \rho}\right)
    +\frac{p}{q} \log_2\left(1 \! - \! (q \! - \! p) \frac{\rho \! - \! \alpha}{1 \! - \! \rho}\right) \,. \nonumber
\end{align}
\fi
We proceed to solve:
\begin{align}
    &\textbf{maximize} & &g(\rho,\alpha)
    \\
    &\textbf{subject to }&  &\rho \le \frac{1}{2}, \alpha \le 1-2\rho \,.
\end{align}
Combining the last two terms we obtain:
\ifCLASSOPTIONonecolumn
\begin{align} 
    g(\rho,\alpha)
    =\log_2
    \left(\frac{\rho}{1-\rho}\right)\left(1-\frac{p}{q}\right)
    +\frac{p}{q}\log_2\left(1-(q-p)^2 \left(\frac{\rho-\alpha}{1-\rho}\right)^2\right)
    \,.
    \label{eq: only U}
\end{align}
\else
\begin{align} 
    g(\rho,\alpha)
    =&\log_2
    \left(\frac{\rho}{1-\rho}\right)\left(1-\frac{p}{q}\right) \nonumber
    \\
    &+\frac{p}{q}\log_2\left(1-(q-p)^2 \left(\frac{\rho-\alpha}{1-\rho}\right)^2\right)
    \,.
    \label{eq: only U}
\end{align}
\fi
The first term increases with $\rho$, and
the second one decreases as the quotient $\left(\frac{\rho-\alpha}{1-\rho}\right)$ increases in absolute value.
For $\rho \le \frac{1}{3}$, it is possible to have $\rho = \alpha$, leaving only \eqref{eq: only U}. However, for $\rho \ge \frac{1}{3}$, the quotient is positive because $\alpha \le 1-2\rho \le \frac{1}{3}$. The smallest value of the quotient is then  $\frac{1-\alpha}{1-\rho} \frac{3\rho-1}{1-\rho} = \frac{2\rho}{1-\rho}-1$ with square $1-\frac{4\rho(1-2\rho)}{(1-\rho)^2}$. Let $g(\rho)$ be defined in equation \eqref{eq: g rho definition}, then the maximum of $g(\rho,\alpha)$ is bounded by the maximum of $g(\rho)$, where:
\ifCLASSOPTIONonecolumn
\begin{align}
    g(\rho) &\triangleq \log_2 \left(\frac{\rho}{1-\rho}\right)\left(1-\frac{p}{q}\right)\label{eq: g rho definition}
    +\frac{p}{q}\log_2\left(1 \! - \! (q \! - \! p) \frac{3\rho \! - \! 1}{1 \! - \! \rho}\right)
    +\frac{p}{q}\log_2\left(1 \! + \! (q \! - \! p) \frac{3\rho \! - \! 1}{1 \! - \! \rho}\right) \,. 
\end{align}
\else
\begin{align}
    g(\rho) &\triangleq \log_2 \left(\frac{\rho}{1-\rho}\right)\left(1-\frac{p}{q}\right)\label{eq: g rho definition}
    \\
    +&\frac{p}{q}\log_2\left(1 \! - \! (q \! - \! p) \frac{3\rho \! - \! 1}{1 \! - \! \rho}\right)
    +\frac{p}{q}\log_2\left(1 \! + \! (q \! - \! p) \frac{3\rho \! - \! 1}{1 \! - \! \rho}\right) \,. \nonumber
\end{align}
\fi
To determine the max, we find the behavior of $g(\rho)$ by taking the first derivative:
\ifCLASSOPTIONonecolumn
\begin{align}
    \frac{d}{d\rho}g(\rho) =
    \frac{q-p}{q\ln(2)} \frac{1-\rho}{\rho} \frac{1}{(1-\rho)^2}
    +\frac{p}{q\ln(2)}\left(\frac{(q-p)\frac{2}{(1-\rho)^2}}{1+(q-p)\frac{3\rho-1}{1-\rho}}
    -\frac{(q-p)\frac{2}{(1-\rho)^2}}{1-(q-p)\frac{3\rho-1}{1-\rho}}\right) \,.
\end{align}
\else
\begin{align}
    \frac{d}{d\rho}&g(\rho) =
    \frac{q-p}{q\ln(2)} \frac{1-\rho}{\rho} \frac{1}{(1-\rho)^2} \nonumber
    \\
    +&\frac{p}{q\ln(2)}\left(\frac{(q-p)\frac{2}{(1-\rho)^2}}{1+(q-p)\frac{3\rho-1}{1-\rho}}
    -\frac{(q-p)\frac{2}{(1-\rho)^2}}{1-(q-p)\frac{3\rho-1}{1-\rho}}\right) \,.
\end{align}
\fi
Then, scale by the positive term $\frac{(1-\rho)^2}{q-p}q \ln(2)$ to obtain:
\ifCLASSOPTIONonecolumn
\begin{align}
    g'(\rho)\frac{(1-\rho)^2}{q-p}q \ln(2) 
    =& \frac{1-\rho}{\rho}
    -p\frac{2}{1-(q-p)\frac{3\rho-1}{1-\rho}}
    +p\frac{2}{1+(q-p)\frac{3\rho-1}{1-\rho}}
    \\
    =& \frac{1-\rho}{\rho}
    +2p\frac{1-(q-p)\frac{3\rho-1}{1-\rho}-1 - (q-p)\frac{3\rho-1}{1-\rho}}{1-(q-p)\frac{3\rho-1}{1-\rho}}
    \\
    =& \frac{1-\rho}{\rho}
    -4p(q-p)\frac{\frac{3\rho-1}{1-\rho}}{1-(q-p)\frac{3\rho-1}{1-\rho}}
    \\
    =& \frac{1-\rho}{\rho}
    -4p(q-p)\frac{3\rho-1}{1-(q-p) (3\rho-1)}
    \\
    =&
    \frac{1-\rho -(q-p)(3\rho-1)(1-\rho+4p\rho)}{\rho-(q-p)\rho(3\rho-1)} \,. \label{eq: scaled g'}
\end{align}
\else
\begin{align}
    g'&(\rho)\frac{(1-\rho)^2}{q-p}q \ln(2) 
    \nonumber
    \\
    =& \frac{1-\rho}{\rho}
    -p\frac{2}{1-(q-p)\frac{3\rho-1}{1-\rho}}
    +p\frac{2}{1+(q-p)\frac{3\rho-1}{1-\rho}}
    \\
    =& \frac{1-\rho}{\rho}
    +2p\frac{1-(q-p)\frac{3\rho-1}{1-\rho}-1 - (q-p)\frac{3\rho-1}{1-\rho}}{1-(q-p)\frac{3\rho-1}{1-\rho}}
    \\
    =& \frac{1-\rho}{\rho}
    -4p(q-p)\frac{\frac{3\rho-1}{1-\rho}}{1-(q-p)\frac{3\rho-1}{1-\rho}}
    \\
    =& \frac{1-\rho}{\rho}
    -4p(q-p)\frac{3\rho-1}{1-(q-p) (3\rho-1)}
    \\
    =&
    \frac{1-\rho -(q-p)(3\rho-1)(1-\rho+4p\rho)}{\rho-(q-p)\rho(3\rho-1)} \,. \label{eq: scaled g'}
\end{align}
\fi
To show that $g' \ge 0$ in $[\frac{1}{3},\frac{1}{2}]$ it suffices to show that the top of equation \eqref{eq: scaled g'} is positive:
\ifCLASSOPTIONonecolumn
\begin{align}
    1-\rho -(q-p)(3\rho-1)(1-\rho+4p\rho)
    \ge&
    1-\frac{1}{2}-(q-p)\left(\frac{3}{2}-1\right)(1-\rho(1-4p))
    \\
    =&
    \frac{1}{2}-\frac{1-2p}{2}(1-\rho(1-4p))
    \\
    =&
    \frac{1}{2}-\frac{1-2p}{2}
    +\frac{1-2p}{2}\rho(1-4p)
    \\
    =&
    \frac{1}{2}-\frac{1}{2}+p
    +\rho \frac{(1-2p)(1-4p)}{2}
    \\
    =& p
    +\rho \frac{(1-2p)(1-4p)}{2} \,.
\end{align}
\else
\begin{align}
    1-\rho& -(q-p)(3\rho-1)(1-\rho+4p\rho)
    \nonumber
    \\
    \ge&
    1-\frac{1}{2}-(q-p)\left(\frac{3}{2}-1\right)(1-\rho(1-4p))
    \\
    =&
    \frac{1}{2}-\frac{1-2p}{2}(1-\rho(1-4p))
    \\
    =&
    \frac{1}{2}-\frac{1-2p}{2}
    +\frac{1-2p}{2}\rho(1-4p)
    \\
    =&
    \frac{1}{2}-\frac{1}{2}+p
    +\rho \frac{(1-2p)(1-4p)}{2}
    \\
    =& p
    +\rho \frac{(1-2p)(1-4p)}{2} \,.
\end{align}
\fi
When $p \le \frac{1}{4}$, then $(1-4p) > 0$ and the second term is non-negative and therefore the derivative is positive. When $p > \frac{1}{4}$ we have $1-4p \ge -1$, $0 \le 1-2p < \frac{1}{2}$, then:
\begin{align}
    & p +\rho  \frac{(1-2p)(1-4p)}{2}
    \ge  \frac{1}{4} - \rho\frac{1}{4} = \frac{1-\rho}{4} \ge \frac{1}{8} > 0 \,, \nonumber
\end{align}
therefore, $g$ is increasing in $\rho$ and the maximum is at $\rho = \frac{1}{2}$ where $\frac{\rho}{1-\rho} = 1$. The maximum is given by:
\ifCLASSOPTIONonecolumn
\begin{align}
    g\left(\frac{1}{2}\right) =& \log_2 \left(1\right)\left(1-\frac{p}{q}\right)
    +\frac{p}{q}\log\left(1 \! + \! (q \! - \! p) \right)
    +\frac{p}{q}\ln\left(1 \! - \! (q \! - \! p)\right) 
    \frac{p}{q}\log_2(2q)+\frac{p}{q}\log_2(2p) \,.
\end{align}
\else
\begin{align}
    g\left(\frac{1}{2}\right) =& \log_2 \left(1\right)\left(1-\frac{p}{q}\right)
    \nonumber
    \\
    &+\frac{p}{q}\log\left(1 \! + \! (q \! - \! p) \right)
    +\frac{p}{q}\ln\left(1 \! - \! (q \! - \! p)\right) \nonumber
    \\
    =& \frac{p}{q}\log_2(2q)+\frac{p}{q}\log_2(2p) \,.
\end{align}
\fi
Then the maximum of $g(\rho,\alpha)-\frac{p}{q}\log(2q)-\frac{p}{q}\log_2(2p)$ is zero. Since the maximum of both equations, \eqref{eq: Delta less than 0} and \eqref{eq: Delta greater than 0} are zero, we conclude that $ U'_i(t+1)-\frac{p}{q}\left(U_i(t+1)-C_2\right) \le \frac{1}{q}\log_2(2q)$.

Finally, we prove the last claim that 
$B = \frac{1}{q}\log_2(2q)$ is the smallest value for a system that enforces the SED constraint. It suffices to note that the surrogate process described in  \cite{Yang2021}, Sec. V \textit{E} is a strict martingale. A process $U'_i(t)$ with a lower $B$ value would not comply with constraint \eqref{eq: submartingale C} and therefore would also fail to meet constraint \eqref{eq: C step size}.


\section{Proof: Confirmation Phase State Space \ref{claim: fix state space}}
\begin{IEEEproof}
\label{sec: proof of fix state space}
Suppose that for times $t=s$ and $t=s+1$ the partitioning is fixed at $S_0=\{j\}$ and $S_1 = \Omega \setminus \{j\}$. Suppose also that $Y_s = 0$, and $Y_{s+1}=1$. Need to show that $\forall i = 1,\dots, M$ we have that $\rho_i(y^{s})= \rho_i(y^{s+2})$. Using the update formula \eqref{eq: update sum}, at time $t=s+1$ we have that for $i\neq j$:
\ifCLASSOPTIONonecolumn
\begin{align}
    \rho_i(y^{s+1}) 
    &= \frac{p\rho_i(y^s)}
    {q \rho_j(y^s) \! + \! p(1 \! - \! \rho_j(y^s))} = \frac{p\rho_i(y^s)}
    {\rho_j(y^s)(q \! - \! p) \! + \! p} 
    \\
    \rho_j(y^{s+1}) 
    &= \frac{q \rho_j(y^s)}
    {q \rho_j(y^s)+p(1-\rho_j(y^s))}
    = \frac{q\rho_j(y^s)}
    {\rho_j(y^s)( q \! - \! p) \! - \! p} 
    \, .
    \label{eq: update t = s+1}
\end{align}
\else
\begin{flalign}
    \rho_i(y^{s+1}) &= \frac{p\rho_i(y^s)}
    {q \rho_j(y^s) \! + \! p(1 \! - \! \rho_j(y^s))} 
    = \frac{p\rho_i(y^s)}
    {\rho_j(y^s)(q \! - \! p) \! + \! p} &
    \\
    \rho_j(y^{s+1}) &= \frac{q \rho_j(y^s)}
    {q \rho_j(y^s)+p(1-\rho_j(y^s))}
    = \frac{q\rho_j(y^s)}
    {\rho_j(y^s)( q \! - \! p) \! - \! p} 
    \, .
    \label{eq: update t = s+1}&
\end{flalign}
\fi
At time $t=s+2$, since $Y_{s+2}=1$, equation \eqref{eq: update sum} for $i\neq j$ results in:
\ifCLASSOPTIONonecolumn
\begin{align}
    \rho_i(y^{s+2})
    &= \frac{q\rho_i(y^{s+1})}
    {(p \! - \! q)\rho_j(y^{s+1}) \! + \! q} 
    = \frac{q \frac{p\rho_i(y^s)}
    {\rho_j(y^s)(q \! - \! p) \! + \! p} }
    {(p \! - \! q) \frac{q\rho_j(y^s)}
    {\rho_j(y^s)( q \! - \! p) \! + \! p} \! + \! q}
    = \frac{q p\rho_i(y^s)}
    {(p \! - \! q) q\rho_j(y^s)\! + \! q
    (\rho_j(y^s)( q \! - \! p) \! + \! p) }
    \\
    &= \frac{q p\rho_i(y^s)}
    {-(q \! - \! p) q\rho_j(y^s)\! + \! ( q \! - \! p) q \rho_j(y^s) \! + \! q p}
    = \frac{q p\rho_i(y^s)}{ q p}=\rho_i(y^s)
    \, .
    \label{eq: update i t = s+2}
\end{align}
\else
\begin{align}
    \rho_i(y^{s+2})
    &= \frac{q\rho_i(y^{s+1})}
    {(p \! - \! q)\rho_j(y^{s+1}) \! + \! q} 
    = \frac{q \frac{p\rho_i(y^s)}
    {\rho_j(y^s)(q \! - \! p) \! + \! p} }
    {(p \! - \! q) \frac{q\rho_j(y^s)}
    {\rho_j(y^s)( q \! - \! p) \! + \! p} \! + \! q}
    \nonumber
    \\
    &= \frac{q p\rho_i(y^s)}
    {(p \! - \! q) q\rho_j(y^s)\! + \! q
    (\rho_j(y^s)( q \! - \! p) \! + \! p) }
    \\
    &= \frac{q p\rho_i(y^s)}
    {-(q \! - \! p) q\rho_j(y^s)\! + \! ( q \! - \! p) q \rho_j(y^s) \! + \! q p}
    \nonumber
    \\
    &= \frac{q p\rho_i(y^s)}{ q p}=\rho_i(y^s)
    \, .
    \label{eq: update i t = s+2}
\end{align}
\fi
And for $i=j$ equation \eqref{eq: update sum} results in:
\ifCLASSOPTIONonecolumn
\begin{align}
    \rho_j(y^{s+2})
    &= \frac{p\rho_j(y^{s+1})}
    {(p \! - \! q)\rho_j(y^{s+1}) \! + \! q} 
    = \frac{p \frac{q\rho_j(y^s)}
    {\rho_j(y^s)(q \! - \! p) \! + \! p} }
    {(p \! - \! q) \frac{q\rho_j(y^s)}
    {\rho_j(y^s)( q \! - \! p) \! + \! p} \! + \! q}
    = \frac{p q\rho_j(y^s)}
    {(p \! - \! q) q\rho_j(y^s)\! + \! q
    (\rho_j(y^s)( q \! - \! p) \! + \! p) }
    \\
    &= \frac{q p\rho_j(y^s)}
    {- \! (q \! - \! p) q\rho_j(y^s)\! + \! ( q \! - \! p) q \rho_j(y^s) \! + \! q p}
    = \frac{q p\rho_j(y^s)}{ q p}=\rho_j(y^s)
    \, .
    \label{eq: update j t = s+2}
\end{align}
\else
\begin{align}
    \rho_j(y^{s+2})
    &= \frac{p\rho_j(y^{s+1})}
    {(p \! - \! q)\rho_j(y^{s+1}) \! + \! q} 
    = \frac{p \frac{q\rho_j(y^s)}
    {\rho_j(y^s)(q \! - \! p) \! + \! p} }
    {(p \! - \! q) \frac{q\rho_j(y^s)}
    {\rho_j(y^s)( q \! - \! p) \! + \! p} \! + \! q} \nonumber
    \\
    &= \frac{p q\rho_j(y^s)}
    {(p \! - \! q) q\rho_j(y^s)\! + \! q
    (\rho_j(y^s)( q \! - \! p) \! + \! p) }
    \\
    &= \frac{q p\rho_j(y^s)}
    {- \! (q \! - \! p) q\rho_j(y^s)\! + \! ( q \! - \! p) q \rho_j(y^s) \! + \! q p}
    \nonumber
    \\
    &= \frac{q p\rho_j(y^s)}{ q p}=\rho_j(y^s)
    \, .
    \label{eq: update j t = s+2}
\end{align}
\fi
Then for all $i = i\dots,M$ each posterior at time $t=s+1$ is given by: $\rho_i(y^{s+2})=\rho_i(y^{s})$. The same equalities hold when $Y_{s+1} = 1$ and $Y_{s+2} = 0$, where the only difference is that $p$ and $q$ are interchanged. By induction, we have that $\rho_i(y^{s+2r})=\rho_i(y^{s})$ for $r = 1,\dots$, if for every $t=s, \dots, s+2 r-1$ the partitions are fixed at $S_0 = \{j\}$ and $S_1 = \Omega \setminus\{j\}$, and $\sum_{k=1}^{2 r} Y_{s+k}=0$.
\end{IEEEproof}



\section{Proof of Inequality \eqref{eq: T n+1 inequality}, Sec. \ref{sec: Main Section}}
\label{appendix: proof of T n+1 inequality}
\begin{IEEEproof}
Need to show that the following inequality holds:
\begin{flalign}
    &\E[U_i(T_n)\mid T^{(n+1)} \! > \! 0, \theta \! = \! i]
    \ge
    \E[U_i(T_n)\mid T^{(n)} \! > \! 0, \theta \! = \! i]&
\end{flalign}
Recall that $\mathcal{C}(t^{(n)}_0)$ is the event that message $i$ enters confirmation after time $t^{(n)}_0$, rather than another message $j\neq i$ ending the process by attaining $U_j(t) \ge \log_2(1-\epsilon)-\log_2(\epsilon)$. This event is defined by $\mathcal{C}(t^{(n)}_0)\triangleq \{\exists t > t^{(n)}_0: U_i(t) \ge 0\}$.
Using Bayes rule, the expectation $\E[U_i(T_n)\mid T^{(n)} > 0, \theta \! = \! i]$ can be expanded as a sum of expectations conditioned on events that are defined in terms of $\{T^{(n+1)} > 0\}$, $\{T^{(n)} > 0\}$ and $\mathcal{C}(t^{(n)}_0)$, and whose union is the full event space to leave only the original conditioning $\{T^{(n)} > 0\}$. These events are $\mathcal{C}(t^{(n)}_0)\cap \{T^{(n+1)}> 0\}$, $ \mathcal{C}(t^{(n)}_0)\cap \{T^{(n+1)} \le 0\}$, $\neg \mathcal{C}(t^{(n)}_0)\cap \{T^{(n+1)}> 0\}$ and $\neg \mathcal{C}(t^{(n)}_0)\cap \{T^{(n+1)} \le 0\}$. Note that $\neg \mathcal{C}(t^{(n)}_0)\implies \{T^{(n+1)} = 0\}$ and therefore the third event vanishes. The expansion is given by:
\ifCLASSOPTIONonecolumn
\begin{align}
    \E[U_i(T_n)\mid T^{(n)} > 0, & \theta = i] =
    \nonumber
    \\
    &\E[U_i(T_n)\mid T^{(n+1) }>0, T^{(n)} > 0, \theta = i]
    \Pr(\mathcal{C}(t^{(n)}_0), T^{(n+1)}>0\mid T^{(n)}> 0, \theta = i) 
    \label{eq: P T and C given T n}
    \\
    &+\E[U_i(T_n)\mid \mathcal{C}(t^{(n)}_0), T^{(n+1) } \! \le \! 0, T^{(n)} \! > \! 0, \theta \! = \! i] \label{eq: U Tn with C}
    \Pr(\mathcal{C}(t^{(n)}_0), T^{(n+1)} \le 0 \mid T^{(n)} \! > \! 0, \theta \! = \! i)
    \\
    &+\E[U_i(T_n)\mid \neg \mathcal{C}(t^{(n)}_0), T^{(n+1) } \! \le \! 0, T^{(n)} \! > \! 0, \theta \! = \! i]
    \Pr(\neg \mathcal{C}(t^{(n)}_0), T^{(n+1)} \le 0 \mid T^{(n)} \! > \! 0, \theta \! = \! i)
    \label{eq: Not C negative}
    \, .
\end{align}
\else
\begin{align}
    \E[&U_i(T_n)\mid T^{(n)} > 0, \theta = i] \nonumber
    \\
    =&\E[U_i(T_n)\mid T^{(n+1) }>0, T^{(n)} > 0, \theta = i]
    \nonumber
    \\
    &\quad \quad \quad \cdot \Pr(\mathcal{C}(t^{(n)}_0), T^{(n+1)}>0\mid T^{(n)}> 0, \theta = i) 
    \label{eq: P T and C given T n}
    \\
    &+\E[U_i(T_n)\mid \mathcal{C}(t^{(n)}_0), T^{(n+1) } \! \le \! 0, T^{(n)} \! > \! 0, \theta \! = \! i] \nonumber
    \\
    & \quad \quad \quad \cdot
    \Pr(\mathcal{C}(t^{(n)}_0), T^{(n+1)} \le 0 \mid T^{(n)} \! > \! 0, \theta \! = \! i)
    \label{eq: U Tn with C}
    \\
    &+\E[U_i(T_n)\mid \neg \mathcal{C}(t^{(n)}_0), T^{(n+1) } \! \le \! 0, T^{(n)} \! > \! 0, \theta \! = \! i]
    \nonumber
    \\
    & \quad \quad \cdot
    \Pr(\neg \mathcal{C}(t^{(n)}_0), T^{(n+1)} \le 0 \mid T^{(n)} \! > \! 0, \theta \! = \! i)
    \label{eq: Not C negative}
    \, .
\end{align}
\fi
Since $\{T^{(n+1)} > 0\} \implies \mathcal{C}(t^{(n)}_0) \cap \{T^{(n)}> 0\}$ , we can omit the conditioning on $\mathcal{C}(t^{(n)}_0)$ and $\{T^{(n)}> 0\}$ when accompanied by $\{T^{(n+1)} > 0\}$. 
By the independence of the confirmation phase from the crossing value $U_i(T_n)$ derived from the fix state count of the Markov Chain we have that:
\ifCLASSOPTIONonecolumn
\begin{align}
    \E[U_i(T_n)&\mid \mathcal{C}(t^{(n)}_0), T^{(n+1) } \! \le \! 0, T^{(n)} \! > \! 0, \theta \! = \! i] 
    =
    \E[U_i(T_n)\mid T^{(n+1) } > 0, T^{(n)} > 0, \theta = i] 
    \label{eq: U Tn with C or with T n+1}
    \, .
\end{align}
\else
\begin{align}
    \E[U_i(T_n)&\mid \mathcal{C}(t^{(n)}_0), T^{(n+1) } \! \le \! 0, T^{(n)} \! > \! 0, \theta \! = \! i]
    \nonumber
    \\
    &=
    \E[U_i(T_n)\mid T^{(n+1) } > 0, T^{(n)} > 0, \theta = i] 
    \label{eq: U Tn with C or with T n+1}
    \, .
\end{align}
\fi
Therefore, we can replace the expectation in \eqref{eq: U Tn with C} by the one in \eqref{eq: P T and C given T n} and then add the probabilities in \eqref{eq: P T and C given T n} and  \eqref{eq: U Tn with C} to obtain $\Pr(\mathcal{C}(t^{(n)}_0)\mid T^{(n)} \! > \! 0, \theta \! = \! i)$. Note that $\neg \mathcal{C}(t^{(n)}_0) \implies \{T^{(n+1) } \le 0\}$, thus the conditioning on $\{T^{(n+1) } \le 0\}$ is redundant with $\neg \mathcal{C}(t^{(n)}_0)$.
Then the expectation in the left of \eqref{eq: P T and C given T n} is also given by:
\ifCLASSOPTIONonecolumn
\begin{align}
    \E[U_i(T_n)\mid T^{(n)} > 0, \theta = i]
    =&\E[U_i(T_n)\mid T^{(n+1) }>0, \theta = i]
    \Pr(\mathcal{C}(t^{(n)}_0) \mid T^{(n)} \! > \! 0, \theta \! = \! i)
    \label{eq: E given Tn}
    \\
    &\quad \quad +\E[U_i(T_n)\mid \neg \mathcal{C}(t^{(n)}_0), T^{(n)} \! > \! 0, \theta \! = \! i]
    \Pr(\neg \mathcal{C}(t^{(n)}_0) \mid T^{(n)} \! > \! 0, \theta \! = \! i) 
    \label{eq: E U given not C}
\end{align}
\else
\begin{flalign}
    \E[U_i(T_n) & \mid T^{(n)} > 0, \theta = i]
    \nonumber
    \\
    &=\E[U_i(T_n)\mid T^{(n+1) }>0, \theta = i]&
    \nonumber
    \\
    & \quad \quad \quad \quad \quad \cdot
    \Pr(\mathcal{C}(t^{(n)}_0) \mid T^{(n)} \! > \! 0, \theta \! = \! i)&
    \label{eq: E given Tn}
    \\
    & \quad +\E[U_i(T_n)\mid \neg \mathcal{C}(t^{(n)}_0), T^{(n)} \! > \! 0, \theta \! = \! i]&
    \nonumber
    \\
    & \quad \quad \quad \quad \quad \cdot
    \Pr(\neg \mathcal{C}(t^{(n)}_0)\mid T^{(n)} \! > \! 0, \theta \! = \! i) \label{eq: E U given not C}&
\end{flalign}
\fi
The event $\neg \mathcal{C}(t^{(n)}_0) \cap \{T^{(n)} > 0\} \cap \{\theta = i\}$ implies that the process decodes in error at the $n$th communication phase round, which results in  $U_i(T_n) < 0$. Therefore, we have that $\E[U_i(T_n)\mid \neg \mathcal{C}(t^{(n)}_0), T^{(n+1) } \! \le \! 0, \theta \! = \! i] < 0$,  
This makes the left side of \eqref{eq: E given Tn} an average of the positive quantity in the right of \eqref{eq: E given Tn} and the negative quantity in \eqref{eq: E U given not C}. Then:
\ifCLASSOPTIONonecolumn
\begin{align}
    \E[U_i(T_n)\mid T^{(n)} > 0, \theta = i]
    &\le \E[U_i(T_n)\mid T^{(n+1) }>0, \theta = i]
    \Pr(\mathcal{C}(t^{(n)}_0) \mid T^{(n)} \! > \! 0, \theta \! = \! i)
    \label{eq: P C given Tn alone}
    \\
    &\le  \E[U_i(T_n)\mid T^{(n+1) }>0,  \theta = i]
    \label{eq: E U T n+1 new alone}
\end{align}
\else
\begin{flalign}
    \E[U_i(T_n)\mid & T^{(n)} \! > \! 0, 
    \theta \! = \! i] \nonumber&
    \\
    &\le \E[U_i(T_n)\mid T^{(n+1) } \! > \! 0, \theta \! = \! i]
    \nonumber&
    \\
    & \quad \quad \quad \quad \quad \cdot
    \Pr(\mathcal{C}(t^{(n)}_0) \mid T^{(n)} \! > \! 0, \theta \! = \! i)
    \label{eq: P C given Tn alone}&
    \\
    &\le  \E[U_i(T_n)\mid T^{(n+1) }>0, \theta = i]
    \label{eq: E U T n+1 new alone}&
\end{flalign}
\fi
The last inequality \eqref{eq: E U T n+1 new alone} follows because the expectation is positive and is multiplied by a probability, $0 \le \Pr(\mathcal{C}(t^{(n)}_0) \mid T^{(n)} \! > \! 0, \theta \! = \! i) \le 1$, in \eqref{eq: P C given Tn alone}. 
\end{IEEEproof}

\ifCLASSOPTIONcaptionsoff
  \newpage
\fi

\bibliographystyle{IEEEtran}
\bibliography{IEEEabrv,references}

\begin{thebibliography}{10}
\providecommand{\url}[1]{#1}
\csname url@samestyle\endcsname
\providecommand{\newblock}{\relax}
\providecommand{\bibinfo}[2]{#2}
\providecommand{\BIBentrySTDinterwordspacing}{\spaceskip=0pt\relax}
\providecommand{\BIBentryALTinterwordstretchfactor}{4}
\providecommand{\BIBentryALTinterwordspacing}{\spaceskip=\fontdimen2\font plus
\BIBentryALTinterwordstretchfactor\fontdimen3\font minus
  \fontdimen4\font\relax}
\providecommand{\BIBforeignlanguage}[2]{{%
\expandafter\ifx\csname l@#1\endcsname\relax
\typeout{** WARNING: IEEEtran.bst: No hyphenation pattern has been}%
\typeout{** loaded for the language `#1'. Using the pattern for}%
\typeout{** the default language instead.}%
\else
\language=\csname l@#1\endcsname
\fi
#2}}
\providecommand{\BIBdecl}{\relax}
\BIBdecl

\bibitem{9174232}
A.~{Antonini}, H.~{Yang}, and R.~D. {Wesel}, ``Low complexity algorithms for
  transmission of short blocks over the bsc with full feedback,'' in \emph{2020
  IEEE International Symposium on Information Theory (ISIT)}, 2020, pp.
  2173--2178.

\bibitem{Yang2021}
\BIBentryALTinterwordspacing
H.~Yang, M.~Pan, A.~Antonini, and R.~D. Wesel, ``Sequential transmission over
  binary asymmetric channels with feedback,'' \emph{IEEE Tran. Inf. Theory},
  2021. [Online]. Available: \url{https://arxiv.org/abs/2111.15042}
\BIBentrySTDinterwordspacing

\bibitem{Shannon1956}
C.~Shannon, ``The zero error capacity of a noisy channel,'' \emph{IRE Trans.
  Inf. Theory}, vol.~2, no.~3, pp. 8--19, September 1956.

\bibitem{Burnashev1976}
M.~V. Burnashev, ``Data transmission over a discrete channel with feedback.
  random transmission time,'' \emph{Problemy Peredachi Inf.}, vol.~12, no.~4,
  pp. 10--30, 1976.

\bibitem{Horstein1963}
M.~Horstein, ``Sequential transmission using noiseless feedback,'' \emph{{IEEE}
  Trans. Inf. Theory}, vol.~9, no.~3, pp. 136--143, July 1963.

\bibitem{Shayevitz2011}
O.~Shayevitz and M.~Feder, ``Optimal feedback communication via posterior
  matching,'' \emph{{IEEE} Trans. Inf. Theory}, vol.~57, no.~3, pp. 1186--1222,
  March 2011.

\bibitem{Gorantla2010}
S.~K. Gorantla and T.~P. Coleman, ``A stochastic control approach to coding
  with feedback over degraded broadcast channels,'' in \emph{49th IEEE
  Conference on Decision and Control (CDC)}, 2010, pp. 1516--1521.

\bibitem{Li2015}
C.~T. Li and A.~E. Gamal, ``An efficient feedback coding scheme with low error
  probability for discrete memoryless channels,'' \emph{{IEEE} Trans. Inf.
  Theory}, vol.~61, no.~6, pp. 2953--2963, June 2015.

\bibitem{Shayevitz2016}
O.~Shayevitz and M.~Feder, ``A simple proof for the optimality of randomized
  posterior matching,'' \emph{{IEEE} Trans. Inf. Theory}, vol.~62, no.~6, pp.
  3410--3418, June 2016.

\bibitem{Naghshvar2015}
M.~Naghshvar, T.~Javidi, and M.~Wigger, ``Extrinsic {J}ensen--{S}hannon
  divergence: Applications to variable-length coding,'' \emph{{IEEE} Trans.
  Inf. Theory}, vol.~61, no.~4, pp. 2148--2164, April 2015.

\bibitem{Bae2010}
J.~H. Bae and A.~Anastasopoulos, ``A posterior matching scheme for finite-state
  channels with feedback,'' in \emph{2010 IEEE Int. Symp. Inf. Theory}, 2010,
  pp. 2338--2342.

\bibitem{Kostina2017}
V.~Kostina, Y.~Polyanskiy, and S.~Verd, ``Joint source-channel coding with
  feedback,'' \emph{IEEE Trans. on Inf. Theory}, vol.~63, no.~6, pp.
  3502--3515, 2017.

\bibitem{Kim2013}
S.~Kim, R.~Ma, D.~Mesa, and T.~P. Coleman, ``Efficient bayesian inference
  methods via convex optimization and optimal transport,'' in \emph{2013 IEEE
  Int. Symp. on Inf. Theory}, 2013, pp. 2259--2263.

\bibitem{Sabag2018}
O.~Sabag, H.~H. Permuter, and N.~Kashyap, ``Feedback capacity and coding for
  the bibo channel with a no-repeated-ones input constraint,'' \emph{IEEE Tran.
  Inf. Theory}, vol.~64, no.~7, pp. 4940--4961, 2018.

\bibitem{Truong2014}
L.~V. Truong, ``Posterior matching scheme for gaussian multiple access channel
  with feedback,'' in \emph{2014 IEEE Information Theory Workshop (ITW 2014)},
  2014, pp. 476--480.

\bibitem{Anastasopoulos2012}
A.~Anastasopoulos, ``A sequential transmission scheme for unifilar finite-state
  channels with feedback based on posterior matching,'' in \emph{2012 IEEE Int.
  Symp. Inf. Theory}, 2012, pp. 2914--2918.

\bibitem{Schalkwijk1971}
J.~Schalkwijk, ``A class of simple and optimal strategies for block coding on
  the binary symmetric channel with noiseless feedback,'' \emph{{IEEE} Trans.
  Inf. Theory}, vol.~17, no.~3, pp. 283--287, May 1971.

\bibitem{Schalkwijk1973}
J.~Schalkwijk and K.~Post, ``On the error probability for a class of binary
  recursive feedback strategies,'' \emph{{IEEE} Trans. Inf. Theory}, vol.~19,
  no.~4, pp. 498--511, July 1973.

\bibitem{Tchamkerten2002}
A.~Tchamkerten and E.~Telatar, ``A feedback strategy for binary symmetric
  channels,'' in \emph{Proc. IEEE Int. Symp. Inf. Theory}, June 2002, pp.
  362--362.

\bibitem{Tchamkerten2006}
A.~Tchamkerten and I.~E. Telatar, ``Variable length coding over an unknown
  channel,'' \emph{{IEEE} Trans. Inf. Theory}, vol.~52, no.~5, pp. 2126--2145,
  May 2006.

\bibitem{Naghshvar2012}
M.~{Naghshvar}, M.~{Wigger}, and T.~{Javidi}, ``Optimal reliability over a
  class of binary-input channels with feedback,'' in \emph{2012 IEEE Inf.
  Theory Workshop}, Sep. 2012, pp. 391--395.

\bibitem{almaOST2019}
R.~Durrett, \emph{\BIBforeignlanguage{eng}{Probability : theory and examples /
  Chapter 4.8, Optional Stopping Theorems}}, 5th~ed., ser. Cambridge series in
  statistical and probabilistic mathematics ; 49.\hskip 1em plus 0.5em minus
  0.4em\relax Cambridge ;: Cambridge University Press, 2019.

\bibitem{almaEM2019}
------, \emph{\BIBforeignlanguage{eng}{Probability : theory and examples /
  Chapter 4.2, Exponential Martingale}}, 5th~ed., ser. Cambridge series in
  statistical and probabilistic mathematics ; 49.\hskip 1em plus 0.5em minus
  0.4em\relax Cambridge ;: Cambridge University Press, 2019.

\end{thebibliography}
\end{document}